\documentclass[colorlinks=true,linkcolor=red,citecolor=blue]{article}
\usepackage{threeparttable}

\usepackage[utf8]{inputenc}
\usepackage[T1]{fontenc}
\usepackage[table]{xcolor}  
\usepackage{graphicx}
\usepackage{amsmath,amssymb,amsfonts,bm}
\usepackage{booktabs,multirow,multicol,makecell,threeparttable}
\usepackage{longtable,rotating,adjustbox}
\usepackage{caption,subcaption,placeins}
\usepackage{enumitem,setspace,microtype,url,doi}
\usepackage{natbib}

\usepackage{arxiv}

\usepackage[utf8]{inputenc} 
\usepackage[T1]{fontenc}    
\usepackage{url}            
\usepackage{booktabs}       
\usepackage{amsfonts}       
\usepackage{nicefrac}       
\usepackage{microtype}      
\usepackage{lipsum}		
\usepackage{subcaption} 
\usepackage{makecell} 
\usepackage{graphicx}
\usepackage{natbib}
\usepackage{doi}
\usepackage{rotating}   
\usepackage{booktabs}
\usepackage{import}
\usepackage{bm,amsmath,multicol,multirow,amssymb,comment}
\usepackage{graphicx}
\usepackage{longtable,setspace}
\usepackage{enumitem}
\usepackage{caption,placeins,subcaption}
\usepackage{adjustbox}

\newcommand{\G}{\mathbb{G}}
\newcommand{\E}{\mathbb{E}}

\newtheorem{theorem}{Theorem}
\newtheorem{proposition}[theorem]{Proposition}

\usepackage{xcolor} 
\usepackage{subcaption}
\usepackage[table]{xcolor}
\definecolor{lightgray}{gray}{0.9}

\title{Calibrated Bayes analysis of cluster-randomized trials}

\author{ {\hspace{1mm} Ruyi Liu} \\
	Department of Biostatistics, Yale School of Public Health, \\
    New Haven, CT, USA\\
	\AND
    {\hspace{1mm} Joshua L. Warren} \\
	Department of Biostatistics, Yale School of Public Health, \\
    New Haven, CT, USA\\
    \AND
    {\hspace{1mm} Yuki Ohnishi} \\
	Department of Biostatistics, Yale School of Public Health, \\
    New Haven, CT, USA\\
	\AND
    {\hspace{1mm} Donna Spiegelman} \\
	Department of Biostatistics, Yale School of Public Health, \\
    New Haven, CT, USA\\
	\AND
    {\hspace{1mm} Liangyuan Hu} \\
	Department of Biostatistics and Epidemiology, Rutgers University, \\
    New Brunswick, NJ, USA\\
	\AND
    {\hspace{1mm} Fan Li}\thanks{Corresponding author: Fan Li, Department of Biostatistics, Yale School of Public Health. Email: \href{mailto:fan.f.li@yale.edu}{fan.f.li@yale.edu}.} \\
    Department of Biostatistics, Yale School of Public Health, \\
    New Haven, CT, USA\\
}

\allowdisplaybreaks 

\begin{document}
\maketitle

\newpage
\begin{abstract}
In cluster-randomized trials (CRTs), entire clusters of individuals are randomized to treatment, and outcomes within a cluster are typically correlated. While frequentist approaches are standard practice for CRT analysis, Bayesian methods have emerged as a strong alternative. Previous work has investigated the use of Bayesian hierarchical models for continuous, binary, and count outcomes in CRTs, but these approaches focus on model-based treatment effect coefficients as the target estimands, which may have ambiguous interpretation under model misspecification and informative cluster size. In this article, we introduce a calibrated Bayesian procedure for estimand-aligned analysis of CRTs even in the presence of potentially misspecified models. We propose estimators targeting both the cluster-average treatment effect (cluster-ATE) and individual-average treatment effect (individual-ATE), particularly in scenarios with informative cluster sizes. We additionally explore strategies for summarizing the posterior samples that can achieve the frequentist coverage even under working model misspecification. We provide simulation evidence to demonstrate the model-robustness property of the proposed estimators in CRTs, and further investigate the impact of covariate adjustment as well as the use of more flexible Bayesian nonparametric working models in the CRT context.
\end{abstract}

\keywords{Bayesian causal inference \and informative cluster size \and estimands \and model-robustness \and covariate adjustment \and Bayesian additive regression trees}

\section{Introduction} \label{sec: Introduction}
In a cluster-randomized trial (CRT), entire social units or groups of individuals, rather than individuals themselves, are randomized to different treatment conditions \citep{donner2000design}. CRTs are often the default when randomizing individuals is infeasible, when studying a cluster-level intervention, or when there are concerns about treatment contamination under individual randomization \citep{murray1998design}. There is a growing number of CRTs conducted to evaluate the effect of new healthcare innovations into the routine practice settings \citep{weinfurt2017pragmatic}. 
However, cluster randomization is typically less efficient than individual randomization because outcomes from individuals within the same cluster tend to be more similar than those from different clusters, requiring the adjustment for this within-cluster correlation when estimating the treatment effect. For this purpose, linear and generalized linear mixed models are a popular class of regression-based methods for analyzing CRTs that account for within-cluster correlations and adjust for baseline covariates \citep{wang2026mixed,wang2024model}.

Although analyses of CRTs have traditionally relied on frequentist methods, a growing literature demonstrates the utility of Bayesian approaches. For example, \citet{spiegelhalter2001bayesian} considered a linear mixed model for analyzing continuous outcomes in CRTs, and assessed the sensitivity under different priors for variance components and intracluster correlation coefficient (ICC). \citet{tong2024hierarchical} extended this approach to further address heterogeneous variances across clusters. \citet{turner2001bayesian} and \citet{thompson2004bayesian} provided extensions to binary outcomes using generalized linear mixed models and possibly an arm-specific random-effects structure. \citet{clark2010bayesian} focused on Bayesian hierarchical models to handle Poisson outcomes. In particular, Bayesian methods offer several appealing features in analyzing CRTs \citep{jones2021bayesian}. First, the analyst has the flexibility in integrating informative priors on variance components and ICCs to inform the data analysis, especially when a limited number of clusters dictates fewer replication at the higher hierarchical level \citep{grantham2022evaluating}. Second, Bayesian inference automatically quantifies uncertainty through the joint posterior distribution of the model parameters; in practice, even under complex models, we can approximate this posterior via Markov chain Monte Carlo (MCMC) to obtain samples for interval estimation and prediction \citep{spiegelhalter2001bayesian}. In the context of CRTs, Bayesian methods naturally facilitate posterior inference on variance components and ICCs, which may not be directly available through standard model fitting procedures in existing software \citep{turner2001bayesian, clark2010bayesian}. Finally, Bayesian nonparametric priors provide a promising avenue for relaxing the restrictive distributional assumptions on fixed effects and random effects inherent in hierarchical models, which can be valuable when cluster-specific heterogeneity departs from standard parametric forms. For instance, Bayesian Additive Regression Trees (BART) \citep{chipman_bart_2010}, an ensemble approach that models outcomes as a sum of regression trees, allows treatment effects to be estimated without relying on strong parametric specifications of the outcome model.

With a few recent exceptions (e.g., \citet{ohnishi2026bayesian} who focused on causal mediation analysis), previous efforts on Bayesian analysis of CRTs have primarily focused on model-based treatment effect estimands. In the context of CRTs, there has been little investigation, to date, on applying Bayesian methods to target nonparameteric treatment effect estimands defined under the potential outcomes framework. Thus, it remains unclear, for example, whether a misspecified Bayesian hierarchical model can still target a clear treatment effect estimand under the potential outcomes framework in CRTs. 
In a parallel-arm CRT, \citet{kahan2023estimands} exemplified two different average treatment effect (ATE) estimands. The cluster-average treatment effect (cluster-ATE) assigns equal weight to each cluster regardless of the cluster size, and the individual-average treatment effect (individual-ATE) assigns equal weight to each individual participant regardless of cluster membership{\color{black}, thereby giving greater weight to larger clusters}. These two estimands differ in magnitude in the presence of informative cluster size, that is, when the outcomes of individuals or the within-cluster treatment effect are marginally associated with cluster size. In the latter case, the cluster size is a marginal effect modifier that requires special consideration in the analysis. {\color{black}For example, larger clinics may serve patients with a different case mix, or may differ in resources and implementation capacity, so that both the outcomes and the treatment effect vary systematically with clinic size. In such settings, the cluster-ATE and individual-ATE can differ, and interpreting a single model coefficient as the treatment effect can be misleading. This is because a model coefficient can sometimes estimate a weighted average of cluster-specific effects with weights driven implicitly by the data rather than by the scientific question, and the probability limit of that coefficient estimator is neither a cluster-average or an individual-average treatment effect \citep{wang2022two,kahan2023demystifying}.} As \citet{greenland1982interpretation} highlighted, summary estimators may exhibit inconsistent behavior depending on the distribution of cluster sizes, underscoring the importance of accounting for possible interactions between treatment effects and cluster size distributions when estimating marginal treatment effects. The concept of informative cluster size can be viewed as a specific manifestation of this broader principle. When the cluster size is non-informative, \citet{wang2026mixed} have proved that the treatment coefficient estimator under an arbitrarily misspecified linear mixed model converges to the ATE (this is the case when cluster-ATE and individual-ATE coincide). However, in the presence of informative cluster size, \citet{kahan2023demystifying} demonstrated that the treatment coefficient estimator under a linear mixed model (without covariates) targets neither cluster-ATE nor individual-ATE. These findings suggest that model specification in CRTs can inadvertently dictate the estimands being estimated. To that end, \emph{model-robust} methods for inferring cluster-ATE and individual-ATE are of growing interest. Several frequentist methods have been proposed \citep{benitez2023defining, balzer2019new, balzer2023two, nugent2024blurring, wang2026handling}, for example, based on linear regression \citep{su2021model} and the vehicle of efficient influence function \citep{wang2024model}. {\color{black}Relatedly, a strand of recent work has characterized how these estimands diverge under informative cluster sizes and clarifies which weighting schemes and working models yield consistent estimators, in parallel-arm CRTs \citep{wang2022two}, CRTs with a baseline period \citep{lee2025should}, and cluster-randomized crossover trials \citep{lee2025estimated}.}

The primary objective of this article is to develop \emph{model-robust} Bayesian analysis methods for CRTs, delivering estimand-aligned inference even when analyses are based on potentially misspecified Bayesian hierarchical models. First, starting with a fitted Bayesian hierarchical model, we consider a model-based g-computation estimator (a direct standardization approach), with and without covariate adjustment, for estimating both the cluster-ATE and the individual-ATE. We then apply the model-robust standardization technique in \citet{li2025model} to the fitted Bayesian hierarchical model to mitigate potential bias due to model misspecification. We establish that, under a Bayesian formulation and via empirical-process arguments, this model-robust point estimator is root-$M$ consistent under CRT randomization (with $M$ being the number of clusters), even when the working outcome model is misspecified. Second, we examine the frequentist property of Bayesian interval estimators derived from the posterior distributions, especially under violations of model assumptions. Previous research \citep{hong2020model} has shown that posterior distributions under model misspecification may exhibit bias, resulting in misleading inference. Building upon \citet{antonelli2022causal} (whose focus was on independent rather than correlated data), we propose a calibrated variance estimator that integrates posterior samples with cluster-level bootstrap. Finally, we explore how Bayesian nonparametric priors (specifically the BART priors) may contribute to the quality of inference by relaxing parametric assumptions commonly made in linear mixed models for CRT analyses. {\color{black} Our development can be best understood within the Calibrated Bayes framework of \citet{rubin1984bayesianly} and \citet{little2006calibrated}, which advocates Bayesian modeling and inference in principle, while calibrating the resulting inference to frequentist operating characteristics in practice. Concretely, we retain the Bayesian workflow, specifying priors on the working outcome model and summarizing the posterior samples for point estimation, but we additionally target the nominal frequentist coverage of the interval estimators under cluster randomization, even when the working outcome model is possibly misspecified. Thus, throughout this article, the Bayesian machinery serves as a flexible modeling and estimation device, whereas the frequentist coverage under cluster randomization serves as the external yardstick against which our interval estimators are calibrated and evaluated.}

Collectively, our study seeks to address the following three interconnected questions:
\begin{enumerate}[itemsep=0ex]
\item[Q1] How can we construct point estimators within Bayesian hierarchical models that remain robust to potential model misspecification when analyzing CRTs?
\item[Q2] What posterior inferential strategy can offer reliable frequentist coverage properties for interval estimation under conditions of model misspecification in CRTs?
\item[Q3] To what extent can Bayesian nonparametric methods, such as BART \citep{chipman_bart_2010}, improve the estimation of cluster- and individual-ATE in CRTs?
\end{enumerate}

The remainder of this article is organized as follows. In Section \ref{sec: Notation}, we introduce the notation, assumptions and the two estimands of interest: the cluster-ATE and individual-ATE. In Sections \ref{sec:Model-based g-computation} and \ref{sec:Model-robust formulation of the estimand}, we review two types of estimand formulation method that can be used to target cluster-ATE and individual-ATE, both with and without covariate adjustment. In Section \ref{sec: Bayesian inference}, we then detail how to combine each point estimation strategy with the posterior samples from a fitted Bayesian hierarchical model, and introduce a calibrated variance estimation approach that can restore the frequentist property. In Section \ref{sec: Simulation studies}, we report on a series of simulations that examine the finite-sample performance of point and variance estimation methods under data-generating processes of varying complexity. We provide an illustrative data example in Section \ref{sec: Case Study}, and offer concluding remarks in Section \ref{sec: Discussion}.

\section{Estimands and model-assisted estimation}
\subsection{Notation and setup} \label{sec: Notation}
We consider a parallel-arm CRT, where data are collected among $M$ clusters involving a total of $N$ individuals. Each cluster $i$ has $N_i$ individuals $(i=1, \ldots, M)$, and the total number of individuals is $N=\sum_{i=1}^M N_i$. We consider the observed cluster size as the true cluster size $N_i$ and do not consider within-cluster subsampling. We let $A_i \in \{0,1\}$ be cluster-level treatment indicator for cluster $i$, with $A_i = 1$ indicating the assignment to the treatment arm and $A_i = 0$ indicating the assignment to the control arm. In a CRT, individuals within a cluster receive the same treatment condition, and we consider intention-to-treat estimands throughout. We let subscript $i$ represent the cluster index $(i = 1, \ldots, M)$ and $j$ represent the individual index within cluster $i$ $(j = 1, \ldots, N_i)$. Define $\bm{X}_{ij}$ as the $q$-dimensional vector of baseline covariates for individual $j$ in cluster $i$, including both individual-level and cluster-level characteristics. We further let $\bm{X}_i = (\bm{X}_{i1}, \ldots, \bm{X}_{iN_i})^{\top}$ be the collection of all baseline covariates in cluster $i$, where $\bm{X}_i \in \mathbb{R}^{N_i \times q}$. 

For individual $j$ in cluster $i$, we define $Y_{ij}$ as the observed outcome. Let $\bm{Y}_i = \left(Y_{i1},\ldots,Y_{iN_i}\right)^{\top}$ be the vector of all observed outcomes in cluster $i$. Under the potential outcomes framework, the individual potential outcome is defined as $Y_{ij}(a)$ under treatment condition $a\in\{0,1\}$. Under the cluster-level Stable Unit Treatment Value Assumption (SUTVA), we have $Y_{ij}=A_iY_{ij}(1)+(1-A_i)Y_{ij}(0)$. Cluster-level SUTVA requires a well-defined intervention, and implies that individual potential outcomes in cluster $i$ depend only on their own cluster's treatment assignment, but not the assignments to any other clusters.
Let $\bm{Y}_i(a) = \left\{Y_{i1}(a), \ldots, Y_{iN_i}(a)\right\} \in \mathbb{R}^{N_i}$ be the collection of potential outcomes for cluster $i$ under condition $a$. Then, the collection of random variables in cluster $i$ is $\bm{O}_i^{\text{full}} = \left\{\bm{Y}_i(1), \bm{Y}_i(0), N_i, \bm{X}_i\right\}$. The complete data $\left\{\left(\bm{O}_1^{\text{full}}, A_1\right), \ldots, \left(\bm{O}_M^{\text{full}}, A_M\right)\right\}$ is only partially observed. We assume that the random variable vectors from different clusters are mutually independent, but within the same cluster, there can be arbitrary correlations among outcomes, covariates and cluster size.

We consider two typical types of marginal causal estimands described in \citet{kahan2023estimands}. These estimands are cluster-ATE and individual-ATE, considering the unit of inference at the cluster level and the individual level, respectively. The cluster-ATE addresses the expected change in outcomes due to treatment across the population of clusters. This estimand weights each cluster equally, irrespective of its size $N_i$. On the other hand, the individual-ATE addresses the expected change in outcomes due to treatment for the population of individuals, and mimics the conventional estimand in an individually-randomized trial. The individual-ATE weights each individual equally, regardless of the cluster to which they belong. We pursue a super population framework in defining estimands (a detailed comparison between finite-population and super-population estimands for CRTs is given in \citet{kahan2023demystifying}), such that the observed data are a random sample from an infinite hypothetical population of clusters. {\color{black}The cluster size $N_i$ itself is also a random variable drawn from an underlying distribution, rather than a quantity fixed by design. The cluster-ATE ($\Delta_C$) and individual-ATE ($\Delta_I$) are therefore defined by invoking expectations over the super-population of clusters. In particular, the individual-ATE, which weights clusters by their size, is equivalently interpreted as the average treatment effect over the population of individuals induced by this cluster size distribution.} Let function $f$ be a predetermined smooth function that defines the scale of the effect measure, the two types of estimands are defined as 
\begin{equation}
\Delta_C \equiv f\left(\mu_{C}(1),\mu_{C}(0)\right), \quad \Delta_I \equiv f\left(\mu_{I}(1),\mu_{I}(0)\right).
\label{eq:estimand}
\end{equation}
where for $a\in \{0,1\}$,
\begin{equation}
\mu_{C}(a) = \mathbb{E}\left(\frac{1}{N_i}\sum_{j=1}^{N_i} Y_{ij}(a)\right),
\label{eq:mu_C}
\end{equation}
\begin{equation}
\mu_{I}(a) = \frac{\mathbb{E}\left(\sum_{j=1}^{N_i} Y_{ij}(a)\right)}{\mathbb{E}\left(N_i\right)}.
\label{eq:mu_I}
\end{equation}
Here, setting $f(x, y) = x - y$ leads to the difference estimands; setting $f(x, y) = {x}/{y}$ and $f(x, y) = {x (1 - y)}/\{y (1 - x)\}$ lead to the risk ratio estimand and odds ratio estimands, respectively. 

To identify the cluster-ATE and individual-ATE, the core challenge lies in the fact that for each individual, we only observe the potential outcome for the actual treatment received, denoted as $Y_{ij}^{\mathrm{obs}} \equiv Y_{ij}=Y_{ij}\left(A_i\right)$, which is factual. The other potential outcome, represented by $Y_{ij}^{\mathrm{mis}}=Y_{ij}\left(1-A_i\right)$, remains counterfactual \citep{li2023bayesian}. In CRTs, the treatment indicator $A_i$ is randomized such that $A_i \perp \left\{\bm Y_i(0), \bm Y_i(1), \bm X_i, N_i\right\}$. For every cluster in the study, the probability of receiving the treatment, $\pi=\mathbb{P}\left(A_i=1\right)\in(0,1)$, is also often a constant. Due to cluster randomization, the marginal distribution of the potential outcomes can be identified from the observed data. To see this, 
using the Law of Iterated Expectation, the estimands of interest can be further expressed as functions of the observed data as
\begin{equation}
\mu_{C}(a) = \mathbb{E}\left\{\mathbb{E}\left(\overline{Y}_i \mid  A_i=a, \bm{X}_i, N_i\right)\right\},
\label{eq:expression for mu_C}
\end{equation}
\begin{equation}
\mu_{I}(a) = \frac{\mathbb{E}\left\{{\mathbb{E}\left(Y_{i+} \mid  A_i=a, \bm{X}_i, N_i\right)}\right\} }{\mathbb{E}\left(N_i\right)},
\label{eq:expression for mu_I}
\end{equation}
where $\overline{Y}_i=N_i^{-1}\sum_{j=1}^{N_i} Y_{ij}$ and $Y_{i+}=\sum_{j=1}^{N_i} Y_{ij}$ are the cluster-average and cluster-total outcomes. 

In practice, \eqref{eq:expression for mu_C} and \eqref{eq:expression for mu_I} imply that estimation of $\mu_{C}(a)$ and $\mu_{I}(a)$ reduces to estimating the conditional mean functions 
\[
m_C(a,\bm x,n)=\mathbb{E}\!\left(\overline{Y}_i \mid A_i=a,\bm{X}_i=\bm x,N_i=n\right),\qquad
m_I(a,\bm x,n)=\mathbb{E}\!\left(Y_{i+} \mid A_i=a,\bm{X}_i=\bm x,N_i=n\right).
\] 
These functions can be modeled using linear mixed models or flexible nonparametric regressions (e.g., spline-based or machine-learning methods). Given estimators $\widehat m_C$ and $\widehat m_I$, plug-in (g-computation) estimators are obtained by averaging over the empirical distribution of $\left(\bm{X}_i,N_i\right)$ across observed clusters. Equivalently, one may integrate $\widehat m_C$ and $\widehat m_I$ over a fitted distribution for $\left(\bm{X}_i,N_i\right)$. Under standard identification conditions (consistency, exchangeability, and positivity), these plug-in estimators target the desired estimands; see Section~\ref{sec:Model-based g-computation} for an implementation using posited linear mixed models.


\subsection{Model-based g-computation formula} \label{sec:Model-based g-computation}
Next, we review two distinct approaches to formulating the estimand. We begin with the model-based approach. In what follows, we focus on continuous outcomes and structure the discussion around the commonly used linear mixed model. In the context of parallel-arm CRTs, a linear mixed model that adjust for covariates with main effects and treatment-covariate interactions (referred to as the analysis of covariance II model \citep{wang2026mixed, yang2020sample}) is given by, for $i=1, \ldots, M$ and $j=1, \ldots, N_i$,
\begin{align}
Y_{i j}=\beta_0+\beta_1 A_i +\bm \beta_2^{\top} \bm X_{ij}+\beta_3 N_{i}+\bm \beta_4^{\top} A_i \bm X_{ij}+ \beta_5 A_iN_i+ \phi_{i} + \epsilon_{ij},
\label{eq:LMM}
\end{align}
where $\phi_i \sim \mathcal{N}\left(0, \sigma_{\phi}^2\right)$ is the random effect for cluster $i$ that measures the cluster-specific departure from the overall mean, $\epsilon_{i j} \sim \mathcal{N}\left(0, \sigma_{\epsilon}^2\right)$ is the residual error for individual $j$ in cluster $i$, and $\beta_0, \beta_1, \left(\bm \beta_2^{\top}, \beta_3\right)$, and $\left(\bm \beta_4^{\top}, \beta_5\right)$ stand for the intercept, the main effect of treatment, {\color{black}the main effects of covariates}, and the treatment-covariate interaction effects, respectively. Under this working model, the cluster random effects satisfy $\mathbb{E}\left(\phi_i\right)=0$ and are independent of $(A_i,\bm X_{ij},N_i)$; hence they integrate out of the conditional means used for standardization. In CRTs, it is commonly assumed within mixed model frameworks that random effects are mutually independent and are further independent of the treatment assignment and all covariates \citep{wang2026mixed}. Within this framework, the fraction of the entire variance in $Y_{ij}$ attributable to the between-group variation, represented by $\sigma_{\phi}^2/\left(\sigma_{\phi}^2 + \sigma_{\epsilon}^2\right)$, is known as the ICC \citep{murray1998design}. {\color{black}In applied analyses, the treatment coefficient is often reported directly as the treatment effect. However, there is no guarantee that the coefficient will always correspond to either the cluster-ATE or the individual-ATE. Under informative cluster size, the coefficient targets a size-dependent weighted average of cluster-specific effects that matches neither marginal estimand. Once treatment-covariate interactions or baseline covariates are included, it represents a covariate-conditional rather than a marginal effect. Under a nonlinear link, such as the logistic mixed model for binary outcomes, a conditional coefficient does not equal the marginal effect even when the model is correctly specified, owing to non-collapsibility. The issues are further compounded under model misspecification, where the coefficient need not correspond to any well-defined marginal causal quantity. These considerations motivate the subsequent estimators, which target the marginal cluster-ATE and individual-ATE directly.}

In particular, we can define the covariate adjusted model-based g-computation formulas as ${\Delta}_C^{\text{mb}} = f \left({\mu}_C^{\text{mb}}(1), {\mu}_C^{\text{mb}}(0)\right)$, and ${\Delta}_I^{\text{mb}} = f \left({\mu}_I^{\text{mb}}(1), {\mu}_I^{\text{mb}}(0)\right)$, where
\begin{align}
{\mu}_C^{\text{mb}}(a) &=\beta_0+\beta_1 a + \mathbb{E}\left\{{N_i}^{-1}\sum_{j=1}^{N_i} \left(\bm \beta_2^{\top} \bm X_{ij}+\beta_3 N_{i}+\bm \beta_4^{\top} a \bm X_{ij}+ \beta_5 aN_i\right)\right\}, \label{eq:model-based mu_C}\\
{\mu}_I^{\text{mb}}(a) &={\beta}_0+{\beta}_1 a +\frac{\mathbb{E}\left\{ \sum_{j=1}^{N_i} \left({\bm\beta}_2^{\top} \bm X_{ij}+{\beta}_3 N_{i}+{\bm\beta}_4^{\top} a \bm X_{ij}+ {\beta}_5 aN_i\right)\right\}}{\mathbb{E}\left(N_i\right)}. \label{eq:model-based mu_I}
\end{align}
In Equation~\eqref{eq:model-based mu_C}--\eqref{eq:model-based mu_I}, expectations are taken over the joint distribution of $(\bm X_i,N_i)$ across clusters. Intuitively, both formulas involve calculating the mean of model predictions for each cluster when the treatment is set to $a$ \citep{wang2024model}. This approach is standard in the causal inference literature for independent data, and is sometimes referred to as the population standardization approach to estimate marginal estimands in observational studies \citep{rosenbaum1987model}.  

A special case of the linear mixed model in Equation~\eqref{eq:LMM} is the unadjusted analysis, in which neither baseline covariates nor cluster-size adjustments are included. Under this unadjusted outcome model, the parameter vectors $\left(\bm \beta_2^{\top}, \beta_3\right)$ and $(\bm \beta_4^{\top}, \beta_5)$ in the above analysis of covariance model are set to zero, and the resulting model-based g-computation \citep{rosenbaum1987model} formulas reduce to 
${\Delta}_C^{\text{mb,unadj}} = f \left({\mu}_C^{\text{mb,unadj}}(1), {\mu}_C^{\text{mb,unadj}}(0)\right)$
, and ${\Delta}_I^{\text{mb,unadj}} = f \left({\mu}_I^{\text{mb,unadj}}(1), {\mu}_I^{\text{mb,unadj}}(0)\right)$
, where ${\mu}_C^{\text{mb,unadj}}(a) = {\beta}_0+{\beta}_1 a$
, and ${\mu}_I^{\text{mb,unadj}}(a)={\beta}_0+{\beta}_1 a$. 
Therefore, in their mathematical forms, the g-computation formulas without baseline covariates lead to identical expressions, and consequently the same estimated numerical values, using both the cluster-ATE and the individual-ATE g-formulas (even though the true cluster-ATE and individual-ATE can differ due to informative cluster size). This reflects an implicit limitation of the unadjusted model-based g-computation approach. For example, the difference estimands are equal under this approach, ${\Delta}_C^{\text{mb,unadj}} = \beta_0 + \beta_1 \times 1 - (\beta_0 + \beta_1 \times 0) = \beta_1 = {\Delta}_I^{\text{mb,unadj}}$. However, in the presence of informative cluster size, the true values of the cluster-ATE and individual-ATE are not necessarily equal \citep{kahan2023estimands}. Consequently, this unadjusted model-based g-computation approach is expected to be biased under informative cluster size.

\subsection{Model-robust standardization formula}\label{sec:Model-robust formulation of the estimand}

For model-based covariate adjustment to provide model-robust inference, \citet{wang2024model} demonstrated that it generally requires independence between the cluster size and cluster characteristics, an assumption often violated in the presence of informative cluster size. Cluster size can be influenced by geographical features or other contextual factors. Thus, aside from the g-computation formulation for cluster-ATE and individual-ATE, which may subject to outcome model misspecification bias, we further consider the model-robust standardization formula introduced in \citet{li2025model} to leverage the known randomization mechanism in CRTs and avoid reliance on correct outcome model specification. The estimators in \citet{li2025model} are frequentist in nature and serve as the foundation for our proposed Bayesian analogs, and they also demonstrate superior performance over conventional treatment effect coefficient estimators in the presence of informative cluster size. The model-robust standardization formula under covariate adjustment are given by ${\Delta}_C^{\text{mr}} = f \left( {\mu}_C^{\text{mr}}(1),  {\mu}_C^{\text{mr}}(0)\right)$ and ${\Delta}_I^{\text{mr}} = f \left( {\mu}_I^{\text{mr}}(1),  {\mu}_I^{\text{mr}}(0)\right)$,
where
\begin{align}
 {\mu}_C^{\text{mr}}(a) &=\mathbb{E}\left\{{\mathbb{E}}\left(\overline{Y}_i \mid A_i=a, \bm{X}_i, N_i\right) + \frac{\mathbb{I}\left(A_i=a\right)\left(\overline{Y}_i- {\mathbb{E}}\left(\overline{Y}_i \mid A_i=a, \bm{X}_i, N_i\right)\right)}{\pi^a\left(1-\pi\right)^{1-a}}\right\}, \label{eq:model-robust mu_C}\\
 {\mu}_I^{\text{mr}}(a) &= \frac{1}{\mathbb{E}\left(N_i\right)} \cdot \mathbb{E}\left\{\mathbb{E}\left(Y_{i+} \mid A_i=a, \bm{X}_i, N_i\right) + \frac{\mathbb{I}\left(A_i=a\right)\left(Y_{i+}- {\mathbb{E}}\left(Y_{i+} \mid A_i=a, \bm{X}_i, N_i\right)\right)}{\pi^a\left(1-\pi\right)^{1-a}}\right\},\label{eq:model-robust mu_I}
\end{align}
where ${\mathbb{E}}\left(\overline{Y}_i \mid A_i=a, \bm{X}_i, N_i\right)$ and ${\mathbb{E}}\left(Y_{i+} \mid A_i=a, \bm{X}_i, N_i\right)$ denote the conditional expectations of the average outcome and the sum outcome, respectively, within cluster $i$, given cluster-specific baseline covariates and cluster size, which can be estimated using any appropriate outcome regression model. 

Naturally, the conditional mean of the cluster-average outcome and the cluster-total outcome within cluster are induced from an individual-level outcome model, for which there are numerous options. For example, when the mean function is equal to zero, ${\mu}_C^{\text{mr}}(a)$ and ${\mu}_I^{\text{mr}}(a)$ become nonparametric two-sample mean estimators with different weights. If the mean function is induced by the unadjusted linear mixed model, the conditional expectation of the cluster-average outcome becomes ${\mathbb{E}}\left(\overline{Y}_i \mid A_i=a, \bm{X}_i, N_i\right)= {\mathbb{E}}\left(\overline{Y}_i \mid A_i=a\right)= {\beta}_0+ {\beta}_1 a$, and the conditional expectation of the cluster-total outcome becomes ${\mathbb{E}}\left(Y_{i+} \mid A_i=a, \bm{X}_i, N_i\right) = N_i \times \left({\beta}_0+ {\beta}_1 a \right)$. Under this setting, the first term in both ${\mu}_C^{\text{mr}}(a)$ and ${\mu}_I^{\text{mr}}(a)$ are equal, reflecting the unadjusted model's mean structure. However, the second term in each expression incorporates residuals that differ due to the use of cluster-level versus individual-level aggregation, thereby enabling adjustment for informative cluster size. Under model~\eqref{eq:LMM}, which is the mixed model with linear adjustment for covariates and treatment-covariate interactions, we obtain the following expression of the cluster-average outcome
${\mathbb{E}}\left(\overline{Y}_i \mid A_i=a, \bm{X}_i, N_i\right)= {\beta}_0+ {\beta}_1 a + \left(\frac{1}{N_i}\sum_{j=1}^{N_i} {\bm\beta}_2^{\top} \bm X_{ij}\right)+ {\beta}_3 N_{i}+\left(\frac{1}{N_i}\sum_{j=1}^{N_i} {\bm\beta}_4^{\top} a \bm X_{ij}\right)+  {\beta}_5 aN_i$, and the cluster-total outcome ${\mathbb{E}}\left({Y}_{i+} \mid A_i=a, \bm{X}_i, N_i\right)= {\beta}_0 N_i+ {\beta}_1 a N_i+ \left(\sum_{j=1}^{N_i} {\bm\beta}_2^{\top} \bm X_{ij}\right)+ {\beta}_3 N_{i}^2+\left(\sum_{j=1}^{N_i} {\bm\beta}_4^{\top} a \bm X_{ij}\right)+  {\beta}_5 aN_i^2$. Although this standardization procedure relies on a working outcome model for $\overline{Y}_i$ or ${Y}_{i+}$, these efficient influence function-based model-robust estimators can achieve consistency of the target estimands, namely the cluster-ATE and individual-ATE, even under outcome model misspecification, since the randomization probability $\pi_i$ in CRTs is often predetermined and controlled by the investigator. A proof of model robustness is provided in \citet{li2025model}. {\color{black} We emphasize that, throughout this article, the term \emph{model-robust} refers to this large-sample property of the \emph{point estimator}: its consistency for the target estimand under cluster randomization does not require the working outcome model to be correctly specified. This should be distinguished from the calibration introduced in Section~\ref{sec: Bayesian inference}, which concerns the variance estimation rather than the point estimator and ensures that the uncertainty of the estimator is correctly quantified. Neither property alone delivers valid interval coverage. Correct coverage requires both an approximately unbiased point estimator (a correctly centered interval) and correctly quantified uncertainty (a correctly sized interval), together with an asymptotic normal approximation for the point estimator.}

\section{Leveraging Bayesian hierarchical modeling for estimating cluster-average and individual-average treatment effects}\label{sec: Bayesian inference}
{\color{black}Before detailing the estimation procedure, we clarify how the Bayesian and frequentist perspectives are combined in our work, as the two roles are conceptually distinct. The Bayesian framework governs modeling and point estimation. We posit a working outcome model, place priors on its parameters, and obtain posterior draws that are then mapped, through either the g-computation formula or the model-robust standardization formula, into posterior draws of the target estimand. The frequentist component contributes to the evaluation and calibration of uncertainty. The key tension we resolve in the following is that a standard posterior quantile interval, which propagates only the uncertainty in the working model parameters, need not achieve this frequentist coverage once the working model is misspecified or once the estimand depends on the observed data beyond the model parameters (as is the case for the model-robust standardization estimator). The calibrated variance estimator reconciles the two perspectives by augmenting the posterior parameter uncertainty with a cluster-level resampling component that recovers the sampling variability of the estimator, thereby aiming to restore the nominal frequentist coverage while preserving the Bayesian inference workflow.}

\subsection{Point estimation}
There are a range of literature on Bayesian methods \citep{carlin1997bayes, gelman1995bayesian}, as well as review papers specifically addressing the application of Bayesian methods to randomized clinical trials \citep{spiegelhalter1994bayesian, kadane1995prime}. Assuming we have unknown quantities $\bm{\theta}$, standard frequentist methods make inferences based on a presumed likelihood $p(\bm Y \mid \bm{\theta})$. Bayesian approaches enhance this by incorporating a prior distribution $p(\bm{\theta})$, which represents our initial uncertainty regarding $\bm{\theta}$ prior to considering the data. In Bayesian analysis, the selection of an appropriate prior distribution remains an important consideration, aiming either to incorporate relevant external information or to provide a non-informative formulation in certain scenarios \citep{spiegelhalter2001bayesian}. 
In what follows, we anchor our discussion in the linear mixed model presented in Section \ref{sec:Model-based g-computation}. We assume the following weakly-informative priors for $\boldsymbol{\beta}$, $\sigma_{\phi}^2$ and $\sigma_{\epsilon}^2 $:
$$
\begin{aligned}
\boldsymbol{\beta} & \sim \mathcal{N}_p\left(\bm 0,\bm L_0\right)=\mathcal{N}_p\left(\bm 0,100 \boldsymbol{I}_p\right), \text { where } \boldsymbol{\beta}=\left(\beta_0, \beta_1, \bm \beta_2^{\top}, \beta_3,\bm \beta_4^{\top},\beta_5\right)^{\top}, \\
\sigma_{\epsilon}^2 & \sim \operatorname{IG}(a,b)=\operatorname{IG}(0.001,0.001),\\
\sigma_{\phi}^2 & \sim \operatorname{IG}(\alpha,c)=\operatorname{IG}(0.001,0.001).
\end{aligned}
$$
Given that these priors are conjugate priors for the linear mixed model, we can simplify the process of updating our beliefs using data because the posterior distribution belongs to the same family of distributions as the prior. Inverse-gamma priors are a common conjugate choice for variance components, although they may perform less well when the true variance is near zero. As a sensitivity analysis, one could instead adopt weakly-informative priors on the standard deviation scale, such as half-Cauchy or uniform distributions. When conjugate priors are used, the mathematical form of the full conditional distributions can often be explicitly determined, which simplifies calculations. This is particularly beneficial when working with large datasets, as it reduces computational costs. The derived full conditional distributions are presented as follows:
$$
\begin{aligned}
\boldsymbol{\beta}\mid \bm Y,\bm \phi,\sigma_{\phi}^2,\sigma_{\epsilon}^2 & \sim \mathcal{N}\left(\left(\frac{\bm H^\top\bm H}{\sigma_{\epsilon}^2}+\bm L_0^{-1}\right)^{-1} \cdot \frac{\bm H^\top \left(\bm Y - \bm Z \bm \phi\right)}{\sigma_{\epsilon}^2},\left(\frac{\bm H^\top\bm H}{\sigma_{\epsilon}^2}+\bm L_0^{-1}\right)^{-1}\right),\\
\phi_i \mid \bm Y , \phi_1, \ldots,\phi_{i-1},\phi_{i+1},\ldots,\phi_M,\boldsymbol{\beta},\sigma_{\phi}^2,\sigma_{\epsilon}^2 &\sim \mathcal{N}\left(\frac{\sigma_{\phi}^2 \cdot \sum I{(\bm Z_{ki} =1)}\left(\bm Y_{k}-\bm H_{k}^\top\boldsymbol{\beta}\right)}{\sigma_{\epsilon}^2+\sigma_{\phi}^2 \cdot N_i},\frac{\sigma_{\phi}^2 \cdot \sigma_{\epsilon}^2}{\sigma_{\epsilon}^2+\sigma_{\phi}^2 \cdot N_i}\right),\\
\sigma_{\phi}^2 \mid \bm Y,\boldsymbol{\beta},\bm \phi,\sigma_{\epsilon}^2 & \sim \operatorname{IG}\left(\alpha+\frac{M}{2},c+\frac{1}{2}\bm \phi^\top \bm \phi\right),\\
\sigma_{\epsilon}^2\mid \bm Y,\boldsymbol{\beta},\bm \phi,\sigma_{\phi}^2 & \sim \operatorname{IG}\left(a+\frac{N}{2},b+\frac{1}{2}\left(\bm Y - \bm H \bm \beta - \bm Z \bm \phi\right)^\top\left(\bm Y - \bm H \bm \beta - \bm Z \bm \phi\right)\right),
\end{aligned}
$$
where $\bm Y$ is the $N \times 1$ vector of all observed outcomes, $\bm H$ is the $N \times p$ design matrix of all fixed effect covariates with corresponding $p\times 1$ vector of fixed effect regression coefficients, $\bm Z$ is the $N \times M$ indicator matrix for cluster membership, $\bm \phi = \left(\phi_1, \ldots,\phi_M\right)^\top$, $I(\cdot)$ is the indicator function, which equals $1$ if the condition inside is true and $0$ otherwise. With the explicit forms of these full conditional distributions established, we can construct a Gibbs sampling algorithm to generate posterior samples for all model parameters. This allows us to estimate the g-computation estimands defined in Section \ref{sec:Model-based g-computation} and model-robust estimands defined in Section \ref{sec:Model-robust formulation of the estimand}, either with or without adjusting for covariates. Consider $B$ samples, $\{\boldsymbol{\beta}^{(b)}\}_{b=1}^B$, drawn from the posterior distribution. Let $\widehat{\Delta}$ serve as a general notation for the point estimator of either cluster-ATE or individual-ATE, derived using a model-based g-computation formula (e.g., Equations \eqref{eq:model-based mu_C} and \eqref{eq:model-based mu_I}) or a model-robust standardization formula (e.g., Equations \eqref{eq:model-robust mu_C} and \eqref{eq:model-robust mu_I}). We define the point estimator $\widehat{\Delta}$ as follows
$$
\widehat{\Delta} = E_{\boldsymbol{\beta} \mid \boldsymbol{D}}[\Delta(\boldsymbol{D}, \boldsymbol{\beta})] \approx \frac{1}{B} \sum_{b=1}^B \Delta(\boldsymbol{D}, \boldsymbol{\beta}^{(b)}),
$$
where $\Delta(\boldsymbol{D}, \boldsymbol{\beta}^{(b)})$ represents the estimated value computed using the observed dataset $\boldsymbol{D}$ and the parameter values from the $b$-th draw, $\boldsymbol{\beta}^{(b)}$. Note that while we specify conjugate priors for the linear mixed model in this example, the framework is flexible and can accommodate a variety of prior choices. Previous work on Bayesian methods for CRTs has explored diverse prior specifications, for instance, log-uniform priors for cluster-level random effect variances and residual error variances \citep{spiegelhalter2001bayesian}, fitted half-normal priors or informative beta priors for the ICC \citep{turner2001bayesian}. These alternative priors can also be incorporated within our framework. For the point estimator derived from the model-robust standardization, under CRT randomization (known treatment assignment), we prove in Web Appendix 1, via empirical-process arguments \citep{van2000asymptotic}, that this estimator is model-robust and root-$M$ consistent, even under Bayesian outcome model misspecification. We note that this result requires only Assumptions 1 and 5 together with the Donsker condition stated in the Preliminaries of the Web Appendix, and in particular does not invoke any posterior contraction toward the true conditional mean. This result is formalized in Proposition~\ref{prop:mr_consistency}. Let $\widehat{\mu}_{\omega}^{\text{mr}}(a)$ denote the model-robust estimator of the arm-specific mean for treatment arm $a$, where $\omega \in \{C, I\}$ indexes the cluster- and individual-level estimands, respectively. This estimator, $\widehat{\mu}_{\omega}^{\text{mr}}(a)$, uses the working models to estimate the conditional means $\mathbb{E}\left(\overline{Y}_i\mid A_i=a,\bm{X}_i,N_i\right)$ when $\omega=C$ and $\mathbb{E}\left(Y_{i+}\mid A_i=a,\bm{X}_i,N_i\right)$ when $\omega=I$, which enter the corresponding augmentation terms in Equations \eqref{eq:model-robust mu_C} and \eqref{eq:model-robust mu_I}.

\begin{proposition}\label{prop:mr_consistency}
\emph{Under the true law $P_0$, we assume that the fitted outcome regression functions contributing to the posterior point estimator are contained in a fixed pointwise measurable class $\mathcal F$ that is $P_0$-Donsker with a square-integrable envelope. Then, under mild regularity conditions on the data-generating process stated in the Preliminaries of the Web Appendix and under cluster randomization, the model-robust point estimator $\widehat{\mu}_\omega^{\text{mr}}(a)$ achieves root-$M$ consistency:
$$\widehat{\mu}_\omega^{\text{mr}}(a) - \mu_\omega(a) = O_p(M^{-1/2}),$$
even under outcome model misspecification.}
\end{proposition}

\subsection{Uncalibrated posterior inference}
A $95$\% credible interval for a parameter is defined as the range between a lower and an upper bound such that, given the observed data, the posterior probability that the parameter lies within this range is at least $95$\%, where there are many options for choosing the lower and upper bound. The most common method for determining credible intervals is to calculate the $2.5$-th quantile and the $97.5$-th quantile of the sorted posterior samples from MCMC methods. The empirical distribution function $F\left(\boldsymbol{\beta}\right)$ at a given $\boldsymbol{\beta}$ is the proportion of the posterior draws that are less than or equal to $\boldsymbol{\beta}$. It can be defined as $F\left(\boldsymbol{\beta}\right)=\frac{1}{B} \sum_{b=1}^B I\left(\boldsymbol{\beta}^{(b)} \leq \boldsymbol{\beta}\right)$.
The quantile is obtained by finding the value $\boldsymbol{\beta}$ among the posterior draws for which the empirical distribution function $F\left(\boldsymbol{\beta}\right)$ equals the desired quantile $v$. Mathematically, it can be expressed as 
$$Q_v=\inf \left\{b: F\left(\boldsymbol{\beta}^{(b)}\right) \geq v \right\}.$$ In words, $Q_v$ is the smallest value such that at least $v$ proportion of the posterior draws are less than or equal to $Q_v$. Then, uncalibrated posterior inference constructs the $95$\% credible interval as $\left(Q_{0.025},Q_{0.975}\right)$.

When the outcome model is misspecified, standard Bayesian posterior inference does not guarantee nominal frequentist coverage, because it accounts only for uncertainty in the posterior distribution of the misspecified outcome model parameters and not for additional sampling variability in the data \citep{antonelli2022causal}. The model-robust formulation in Equations \eqref{eq:model-robust mu_C} and \eqref{eq:model-robust mu_I} incorporates both the parameter vector $\bm{\beta}$ from the conditional expectation and aggregated observed outcomes in $\bm{D}$, motivating a correction to the posterior variance to better reflect sampling variability. Specifically, in Section \ref{sec:var_correct} we extend the variance correction approach of \citet{antonelli2022causal} to derive a formally justified, calibrated variance estimator. 

\subsection{Calibrated variance estimation}\label{sec:var_correct}

We quantify uncertainty in inferring causal estimands by combining the posterior distribution of parameters, conditional on the observed data, with a computationally efficient resampling procedure. Given the design of CRTs, we adopt a \emph{cluster-level} resampling approach, extending the individual-level resampling proposed by \citet{antonelli2022causal}. First, we generate $K$ new datasets, $\boldsymbol{D}^{(1)}, \ldots, \boldsymbol{D}^{(K)}$, by sampling clusters with replacement from the empirical distribution of the observed clusters. {\color{black}Here we resample clusters while fixing only the total number of clusters, without stratifying by treatment arm. An arm-stratified alternative that additionally preserves the observed treatment allocation is examined in Web Appendix 9, where it is shown to yield nearly identical results. There, we also provide discussions on extending the resampling scheme to stratified, pair-matched, and covariate-constrained designs.} For each bootstrapped datasets and posterior parameter samples, we compute $\Delta\left(\boldsymbol{D}^{(k)}, \boldsymbol{\beta}^{(b)}\right)$ for $k=1, \ldots, K$ and $b=1, \ldots, B$. Here, $\boldsymbol{\beta}^{(b)}$ represents a sample from the posterior distribution of $\boldsymbol{\beta}$ given the original observed data $\boldsymbol{D}$, and is fixed across different bootstrap samples. By calculating the posterior distribution of the estimator over multiple resampled datasets, we can estimate the potential variability from the data. Define $G_{\Delta} = \left[ \Delta(\bm{D}^{(k)}, \bm{\beta}^{(b)}) \right]_{k=1,\ldots,K;\,b=1,\ldots,B}$,
which is a matrix with $K$ rows and $B$ columns, where each entry represents the estimated value computed using the $k$-th resampled dataset and the $b$-th posterior draw of model parameters. For each $k = 1, \ldots, K$, we approximate $E_{\boldsymbol{\beta} \mid D}\left[\Delta\left(\boldsymbol{D}^{(k)}, \boldsymbol{\beta}\right)\right]$ by averaging over the $B$ columns in row $k$. This results in a mean vector of length $K$, given by $\frac{1}{B}G_{\Delta}\mathbf{1}$, where $\mathbf{1}$ denotes a vector of all ones of length $B$. We then estimate the variability using the sample variance of the $K$ means, denoted by $\operatorname{Var}_{D^{(k)}}\left\{E_{\boldsymbol{\beta} \mid D}\left[\Delta\left(\boldsymbol{D}^{(k)}, \boldsymbol{\beta}\right)\right]\right\}$.

However, this variance has not yet accounted for the uncertainty that arises from potential variability in the posterior distribution of model parameters. To address this issue, the final, calibrated variance formula is 
\begin{align}
\operatorname{Var}\left(\widehat{\Delta}\right)=\operatorname{Var}_{\boldsymbol{D}^{(k)}}\left\{E_{\boldsymbol{\beta} \mid \boldsymbol{D}}\left[\Delta\left(\boldsymbol{D}^{(k)}, \boldsymbol{\beta}\right)\right]\right\}+\operatorname{Var}_{\boldsymbol{\beta} \mid \boldsymbol{D}}\left[\Delta\left(\boldsymbol{D}, \boldsymbol{\beta}\right)\right].
\label{eq:calibrated variance}
\end{align}
In this variance formula, the first term quantifies variability arising from the data by evaluating the posterior distribution of the estimator across resampled datasets. The second term addresses the uncertainty arising from parameter variability, which is identical to the posterior variance per the uncalibrated posterior inference. {\color{black}Intuitively, the two terms capture two distinct sources of uncertainty. The first reflects the sampling variability of the clusters themselves, that is, how much the estimate would change had we observed a different sample of clusters from the population, while the second reflects our uncertainty about the parameters of the Bayesian working model given the observed data. Uncalibrated posterior inference, which constructs intervals directly from the quantiles of the posterior draws, reflects only this second source of uncertainty. This is particularly consequential for the model-robust estimator, whose form in Equations~\eqref{eq:model-robust mu_C} and \eqref{eq:model-robust mu_I} depends not only on the parameter vector $\bm{\beta}$ but also directly on the aggregated observed outcomes. Because the posterior distribution of $\bm{\beta}$ conditions on the observed data, it cannot convey how much this data-dependent component would itself vary across repeated samples of clusters. Standard posterior intervals therefore only quantify how uncertain we are about the working model, but not how much the estimate would vary across repeated samples, and can substantially undercover. By contrast, the g-computation estimator depends on the data only through $\bm{\beta}$, so when the outcome model is correctly specified the posterior variance alone already captures the relevant uncertainty, and the uncalibrated interval should perform adequately in large samples.} {\color{black}The interval obtained from this calibrated variance should be interpreted as a calibrated Bayesian summary in the sense of \citet{rubin1984bayesianly} and \citet{little2006calibrated}. We refer to it as a calibrated interval throughout, rather than a credible interval, since it incorporates a cluster-level resampling component beyond the posterior of the working model. The point estimate and posterior summaries are Bayesian, while the interval is constructed from a Bayesian working model and calibrated to attain frequentist coverage.}

If we use this calibrated posterior inference for the model-robust estimator for cluster-ATE using linear mixed model with covariate adjustment, we would execute the following steps.

\begin{enumerate}
    \item Draw $B$ posterior samples $\{\boldsymbol{\beta}^{(b)}\}_{b=1}^B \sim\Pi \left(\boldsymbol{\beta}\mid \boldsymbol{D} \right)$ from the fitted linear mixed model (Bayesian setting), where $\boldsymbol{D}=\left\{\bm{Y}_i, A_i, N_i, \bm{X}_i\right\}_{i=1}^M$.
    \item Compute the point estimate.
    $$\widehat{\Delta}_C^{\text{mr}} \approx \frac{1}{B} \sum_{b=1}^B \widehat{\Delta}_C^{\text{mr}}(\boldsymbol{D}, \boldsymbol{\beta}^{(b)})=\frac{1}{B} \sum_{b=1}^B f \left(\widehat{\mu}_C^{\text{mr}}\left(1,\boldsymbol{D}, \boldsymbol{\beta}^{(b)}\right), \widehat{\mu}_C^{\text{mr}}\left(0,\boldsymbol{D}, \boldsymbol{\beta}^{(b)}\right)\right),$$
    where $$\widehat{\mu}_C^{\text{mr}}\left(a,\boldsymbol{D}, \boldsymbol{\beta}^{(b)}\right) =\frac{1}{M} \sum_{i=1}^M\left\{\frac{I\left(A_i=a\right)\left(\overline{Y}_i-\widehat{\mathbb{E}}^{(b)}\left(\overline{Y}_i \mid A_i=a, \bm{X}_i, N_i\right)\right)}{\pi^a\left(1-\pi\right)^{1-a}}+\widehat{\mathbb{E}}^{(b)}\left(\overline{Y}_i \mid A_i=a, \bm{X}_i, N_i\right)\right\},$$
    $$\widehat{\mathbb{E}}^{(b)}\left(\overline{Y}_i \mid A_i=a, \bm{X}_i, N_i\right)={\beta}_0^{(b)}+{\beta}_1^{(b)} a + \left(\frac{1}{N_i}\sum_{j=1}^{N_i}{{\bm\beta}_2^{(b)}}^{\top} \bm X_{ij}\right)+{\beta}_3^{(b)} N_{i}+\left(\frac{1}{N_i}\sum_{j=1}^{N_i}{{\bm\beta}_4^{(b)}}^{\top} a \bm X_{ij}\right)+ {\beta}_5^{(b)} aN_i.$$

    \item Generate $K$ new datasets, $\boldsymbol{D}^{(1)}, \ldots, \boldsymbol{D}^{(K)}$, by resampling clusters with replacement from the empirical distribution of the data, where $\boldsymbol{D}^{(k)}=\left\{\bm{Y}_i^{(k)}, A_i^{(k)}, N_i^{(k)}, \bm{X}_i^{(k)}\right\}_{i=1}^M$.

    \item For each $\boldsymbol{D}^{(k)}$, compute the estimated cluster-ATE based on the model-robust estimator, $\widehat{\Delta}_C^{\text{mr}}(\boldsymbol{D}^{(k)}, \boldsymbol{\beta}^{(b)})$, for $b \in \{1,\ldots,B\}$.
    $$\widehat{\Delta}_C^{\text{mr}}(\boldsymbol{D}^{(k)}, \boldsymbol{\beta}^{(b)})=f\left(\widehat{\mu}_C^{\text{mr}}\left(1,\boldsymbol{D}^{(k)}, \boldsymbol{\beta}^{(b)}\right), \widehat{\mu}_C^{\text{mr}}\left(0,\boldsymbol{D}^{(k)}, \boldsymbol{\beta}^{(b)}\right)\right),$$
    where
    $$
    \begin{aligned}
    \widehat{\mu}_C^{\text{mr}}\left(a,\boldsymbol{D}^{(k)}, \boldsymbol{\beta}^{(b)}\right) =& \frac{1}{M} \sum_{i=1}^M\left\{\frac{I\left(A_i=a\right)\left(\overline{Y}_i^{(k)}-\widehat{\mathbb{E}}^{(b)}\left(\overline{Y}_i^{(k)} \mid A_i=a, \bm{X}_i^{(k)}, N_i^{(k)}\right)\right)}{\pi^a\left(1-\pi\right)^{1-a}}\right.\\
    &\left.+\widehat{\mathbb{E}}^{(b)}\left(\overline{Y}_i^{(k)} \mid A_i=a, \bm{X}_i^{(k)}, N_i^{(k)}\right)\right\},
    \end{aligned}
    $$
    $$
    \begin{aligned}
    \widehat{\mathbb{E}}^{(b)}\left(\overline{Y}_i^{(k)} \mid A_i=a, \bm{X}_i^{(k)}, N_i^{(k)}\right)=&{\beta}_0^{(b)}+{\beta}_1^{(b)} a + \left(\frac{1}{N_i^{(k)}}\sum_{j=1}^{N_i^{(k)}}{{\bm\beta}_2^{(b)}}^{\top} \bm X_{ij}^{(k)}\right)+{\beta}_3^{(b)} N_{i}^{(k)}\\
    &+\left(\frac{1}{N_i^{(k)}}\sum_{j=1}^{N_i^{(k)}}{{\bm\beta}_4^{(b)}}^{\top} a \bm X_{ij}^{(k)}\right)+ {\beta}_5^{(b)} aN_i^{(k)}.
    \end{aligned}$$
    \item The potential variability from the data is estimated by the variance of $\left(\widehat{\Delta}_C^{\text{mr}}\left(\boldsymbol{D}^{(1)}, \boldsymbol{\beta}\right),\ldots, \widehat{\Delta}_C^{\text{mr}}\left(\boldsymbol{D}^{(K)}, \boldsymbol{\beta}\right)\right)$, where $\widehat{\Delta}_C^{\text{mr}}\left(\boldsymbol{D}^{(k)}, \boldsymbol{\beta}\right) \approx \frac{1}{B}\sum_{b=1}^B \widehat{\Delta}_C^{\text{mr}}\left(\boldsymbol{D}^{(k)}, \boldsymbol{\beta}^{(b)}\right).$

    \item The uncertainty arising from parameter variability is then estimated by computing the variance of posterior samples of the estimated cluster-ATE given the original observed dataset, which is the variance of $\left(\widehat{\Delta}_C^{\text{mr}}(\boldsymbol{D}, \boldsymbol{\beta}^{(1)}), \ldots, \widehat{\Delta}_C^{\text{mr}}(\boldsymbol{D}, \boldsymbol{\beta}^{(B)})\right)$. This estimated variance is identical to the posterior variance in uncalibrated posterior inference.

    \item The calibrated variance estimation is obtained by summing the estimated data variability and parameter variability from steps 5 to 6. With this calibrated variance estimate, the calibrated $95$\% interval can be derived using an asymptotic normal approximation.
    
\end{enumerate}
This example illustrates the variance correction procedure for posterior inference of cluster-ATE using a covariate adjusted, model-robust estimator. This procedure can be easily adapted to both model-based and model-robust estimators, with or without leveraging covariates, for estimating cluster-ATE or individual-ATE. The following Proposition~\ref{calibrated_var_consistency} establishes the consistency of the calibrated variance estimator under correct specification of the outcome regression model, whose posterior contracts at a sufficiently fast rate. The detailed assumptions and proof are provided in the Preliminaries and Appendix 2 of the Web Appendix.

\begin{proposition}\label{calibrated_var_consistency}
\emph{Assuming cluster randomization, and that the outcome model contracts at a rate faster than $M^{-1/4}$. Then, under mild regularity conditions on the data-generating process and the posterior distribution, the variance estimator $\widehat{V}=\operatorname{Var}_{\boldsymbol{D}^{(k)}}\left\{E_{\boldsymbol{\beta}\mid\boldsymbol{D}}\left[\Delta(\boldsymbol{D}^{(k)},\boldsymbol{\beta})\right]\right\}+\operatorname{Var}_{\boldsymbol{\beta}\mid\boldsymbol{D}}\left[\Delta(\boldsymbol{D},\boldsymbol{\beta})\right]$ is consistent, i.e., $\widehat{V} - V = o_p(M^{-1})$, where $V = \operatorname{Var}_{\boldsymbol{D}}\left\{E_{\boldsymbol{\beta} \mid \boldsymbol{D}}\left[\Delta\left(\boldsymbol{D}, \boldsymbol{\beta}\right)\right]\right\}$.}
\end{proposition}

Under randomization and a well-behaved outcome model, the calibrated variance estimator converges to the true variance as the number of clusters increases. {\color{black}Beyond the standard identification assumption already introduced in Section \ref{sec: Notation}, several regularity conditions are required. First, the cluster sizes are assumed to be bounded, drawn from a distribution with finite support, ruling out outlying cluster sizes. Second, the Bayesian working model is required to learn the true outcome mean sufficiently well as the number of clusters grows. Its posterior should concentrate around the true conditional mean at a fast enough rate and should not place substantial mass on values far from the truth, which holds for standard Bayesian outcome models when they are correctly specified or when they are flexible enough to approximate the truth. Third, no single cluster dominates or is negligible in the analysis, in the sense that the cluster weights are bounded away from zero and infinity. This holds automatically for the cluster-ATE (where the weights are all equal to one) and, for the individual-ATE (where the weights are proportional to cluster size) setup.} {\color{black}We stress that the second condition, which requires the posterior to concentrate around the true conditional mean, is substantively stronger than what is needed for Proposition~\ref{prop:mr_consistency} and is not compatible with arbitrary model misspecification. Proposition~\ref{calibrated_var_consistency} should therefore be read as a result for a working model that is correctly specified or sufficiently flexible to approximate the truth, whereas the behavior of the calibrated variance estimator under misspecification is assessed empirically in Section~\ref{sec: Simulation studies}}.

Specifically, our proof in Web Appendix 2 shows that, when the outcome model is correctly specified, the contribution of posterior parameter uncertainty to the variability of the model-robust estimator becomes asymptotically negligible. {\color{black} Conversely, under model misspecification, the proposed variance estimator tends to overestimate the true variance, yielding conservative uncertainty quantification. However, that conservative variance estimation does not by itself guarantee valid coverage. The calibrated variance corrects the quantification of uncertainty, but it cannot correct bias in the point estimator. When the point estimator carries substantial bias, the interval is centered away from the truth and coverage can fall below nominal even though the variance is conservative.} To further evaluate the practical utility and robustness of this correction, we will empirically investigate the finite-sample performance of the calibrated variance estimator under several Bayesian working models. Specifically, we first consider the case where the working outcome model is specified as a linear mixed model, which, while computationally convenient, may be prone to misspecification in complex data-generating processes. Under such misspecification, for instance when the model omits key nonlinearity features, the calibrated variance for the model-robust estimator overestimates the variance, yielding wider intervals. By contrast, when BART's flexible model space can better approximate the true data-generating mechanism, the contraction condition becomes more plausible, although our proof relies on a smooth finite-dimensional expansion and does not directly cover BART; we therefore examine its performance empirically as the number of clusters increases. Accordingly, we examine mixed-effects BART, which automatically accommodates nonlinearities and higher-order interactions and can mitigate misspecification bias by exploring a richer model space. Together with the linear mixed model analysis, the subsequent numerical results provide empirical evidence on the reliability and generality of the calibrated variance estimator.

\section{Simulation Studies}\label{sec: Simulation studies}

\subsection{Simulation design}

We conduct three simulation studies to evaluate the empirical performance of model-based g-computation and model-robust standardization estimators under combinations of covariate adjustment (with or without), outcome models (linear mixed model or BART), and posterior variance estimation methods (uncalibrated or calibrated), as summarized in Table~\ref{tab:Estimator_Variance_Combos}. The first simulation study examines the performance of the model-based g-computation estimator combined with the uncalibrated posterior variance estimation for constructing credible intervals. We evaluate the impact of covariate adjustment on the efficiency and accuracy of estimating the cluster-ATE and the individual-ATE, and investigate the impact of outcome model misspecification on posterior inference. The second simulation study compares the model-based g-computation estimator and the model-robust standardization approach, both incorporating covariate adjustment via a linear mixed model, but under potential model misspecification. We also assess the frequentist coverage of uncalibrated and calibrated posterior variance estimators paired with each of the two point estimators. The third simulation study investigates the performance of Bayesian nonparametric methods under both model-based and model-robust frameworks, with and without posterior variance correction, across scenarios with model misspecification. We specifically focus on BART-based estimators to their linear mixed model counterparts, focusing on frequentist coverage, relative bias, and robustness to model misspecification in small-sample settings. Methods corresponding to empty cells in Table~\ref{tab:Estimator_Variance_Combos} are not the focus of our simulation studies, as they are not expected to yield additional insights beyond those already covered by the selected combinations. To maintain focus and avoid redundancy, we prioritize evaluating only the marked methods in Table \ref{tab:Estimator_Variance_Combos}. All twelve methods (full combination of all factors), nevertheless, are implemented and compared in the data application section.

\begin{table}[ht]
\centering
\caption{Categorization of methods. The table presents combinations of estimators (model-based and model-robust), posterior variance estimation methods (uncalibrated and calibrated), covariate adjustment (with or without), and outcome models (linear mixed model and BART). Cells marked with (1), (2), and (3) indicate the focus of the first, second, and third simulation studies, respectively.}
\label{tab:Estimator_Variance_Combos}
\renewcommand{\arraystretch}{1.3}
\setlength{\tabcolsep}{10pt}
\begin{tabular}{llcccc}
    \toprule
    \multirow{2}{*}{Covariate adjustment} & \multirow{2}{*}{Outcome model} & \multicolumn{2}{c}{Model-based g-computation} & \multicolumn{2}{c}{Model-robust standardization} \\
    \cmidrule(lr){3-4} \cmidrule(lr){5-6}
    & & Uncalibrated & Calibrated & Uncalibrated & Calibrated \\
    \midrule
    Unadjusted & Linear mixed model & (1) & & & \\
    \midrule
    \multirow{2}{*}{Adjusted} 
      & Linear mixed model & (1) \& (2) & (2) & (2) & (2) \\
      & BART               & (3)      &      &      & (3) \\
    \bottomrule
\end{tabular}
\end{table}

Our data-generating processes are as follows. The covariate-generating models are adapted from the simulation experiment in \citet{wang2024model}. Let $(N_i, C_{i1}, C_{i2})$, for $i=1, \ldots, M$, be independent draws from distribution $\mathcal{P}^N \times \mathcal{P}^{C_1 | N} \times \mathcal{P}^{C_2 | N, C_1}$, where $\mathcal{P}^N$ is uniform over support $(10,50)$, $\mathcal{P}^{C_1 | N} = \mathcal{N}(N / 10,4)$, $\mathcal{P}^{C_2 | N, C_1} = \mathcal{B}\left[\operatorname{expit}\left\{\log (N / 10) C_1\right\}\right]$ is a Bernoulli distribution with $\operatorname{expit}(x) = \left(1+e^{-x}\right)^{-1}$. The coefficient of variation of cluster size is approximately $0.394$. Next, for each individual, we generate the two individual-level covariates from $X_{ij1} \sim \mathcal{B}\left(N_i / 50\right)$, and $X_{ij2} \sim \mathcal{N}\left\{\frac{\left(\sum_{j=1}^{N_i} X_{ij1}\right) \times \left(2 C_{i2}-1\right)}{N_i}, 9\right\}$. Then, we draw a cluster random intercept
$\phi_i\sim \mathcal N\left(0,\sigma_\phi^2\right)$ with $\sigma_\phi^2=0.25$ and generate potential outcomes
$Y_{ij}(A_i) \sim \mathcal N\left(\eta_{ij}(A_i)+\phi_i,\ 1\right)$,
so that $\mathrm{ICC}=\sigma_\phi^2/(\sigma_\phi^2+1)=0.2$. Here, $\eta_{ij}(A_i)$ denotes the fixed effect component of the outcome-generating model, which varies across the designed scenarios. We consider three data-generating processes of increasing complexity:
\begin{enumerate}
    \item[(i)] $\eta_{ij}(A_i)= 0.3A_i+ \frac{N_i}{200} + \frac{C_{i1}}{3} + \frac{2 C_{i2}-1}{5} + \frac{A_iC_{i2}}{10} +\frac{1}{4} \left(X_{ij1}+ X_{ij2}\right) + \frac{A_i N_i}{200} + \frac{A_i X_{ij1}}{10}$,

    \item[(ii)] $\eta_{ij}(A_i)= -0.1 - \frac{0.3}{1 + \exp(-6(N_i + C_{i1} + C_{i2} + X_{ij1} + X_{ij2}))}+ 0.2 X_{ij1} X_{ij2} + 0.3 (\frac{N_i}{100} + C_{i1} + C_{i2}) A_i - \frac{1}{1 + \exp(-4(X_{ij1} + X_{ij2}))} A_i$,

    \item[(iii)] $\eta_{ij}(A_i)=\frac{N_iA_i}{60}+\frac{1}{2}N_i\sin(C_{i1}) \cdot (2C_{i2} - 1) + \frac{1}{2} e^{X_{ij1}} \cdot \left|X_{ij2}\right| + \frac{1}{4}A_i \cdot X_{ij2} \cdot  log(\left|C_{i1}\right|)$.
\end{enumerate}
The observed outcomes are generated by independently sampling $A_i \sim \mathcal{B}(0.5)$ and defining $Y_{ij} = A_i Y_{ij}(1) + (1 - A_i) Y_{ij}(0)$. Scenario (iii) is adapted from the continuous potential outcomes model in \citet{wang2024model}. {\color{black}
Scenario (iii) is intentionally designed to be the most challenging setting. First, the outcome surface is strongly nonlinear, involving trigonometric, exponential, absolute-value, and logarithmic transformations together with nonlinear treatment-covariate interactions. Second, the covariates follow a nested, cross-level dependence structure: the cluster size $N_i$ enters the generation of both the cluster-level covariates ($C_{i1}$, $C_{i2}$) and the individual-level covariates ($X_{ij1}$, $X_{ij2}$), and the cluster-level covariates also shape the individual-level covariates. As a result, much of the signal driving the outcome is carried at the cluster level and is propagated downward through this hierarchy, rather than varying freely across individuals. This adds additional challenges for estimation. Although each cluster contributes many individual observations, the independent replication that a flexible model can exploit to learn the cluster-level and cross-level structure is governed by the number of clusters $M$, not the number of individuals $N$. When $M$ is small, the effective information available to disentangle these nested, cluster-size-dependent relationships is limited, and a highly adaptive working model such as BART is prone to overfitting the few clusters observed. Scenario (iii) thus serves as a stress test for the estimators when flexible methods are vulnerable.}

\subsection{Implementation and performance metrics} \label{sec: Implementation and performance metrics}

In each simulation study scenario, we consider a relatively small ($M = 30$) or moderate ($M = 60$ or $M = 90$) number of clusters. In the first simulation study, each scenario is independently replicated $R = 500$ times. We generate $3,000$ posterior draws using MCMC, discarding the first $1,000$ as burn-in. In the second study, which involves a more complex data-generating process, each scenario is replicated $R = 1,000$ times, with $2,000$ MCMC draws and the first $1,000$ discarded as burn-in. For the third simulation study, BART results are based on $R = 250$ replications. We generate $10,000$ posterior draws using MCMC, discarding the first $5,000$ as burn-in. BART requires longer chains than linear mixed models to ensure convergence, due to its nonparametric and data-adaptive structure, which introduces greater model flexibility. Convergence is assessed using traceplots of the cluster-ATE and individual-ATE as well as the Geweke diagnostic. All simulations were carried out using R version  4.3.1.

We evaluate the overall accuracy in the estimation of cluster-ATE and individual-ATE for each method using relative bias. There are $B$ posterior samples of the model parameter generated for each simulated dataset $r$, and $\left(\widehat{\Delta}^{(r,1)}, \cdots, \widehat{\Delta}^{(r,B)}\right)$ is the sequence of estimated posterior cluster-ATE or individual-ATE based on the simulated dataset $r$, $r \in \{1,\ldots,R\}$. The estimated value of cluster-ATE or individual-ATE of dataset $r$ is $\widehat{\Delta}^{(r)} = \frac{1}{B}\left(\widehat{\Delta}^{(r,1)}+ \cdots + \widehat{\Delta}^{(r,B)}\right)$. The final estimate is the average of the posterior mean across all simulations, $\widehat{\Delta}=\frac{1}{R} \sum_{r=1}^{R} \widehat{\Delta}^{(r)}$. Then, relative bias is $\left(\widehat{\Delta}-\Delta\right)/\Delta$,
where the true value of the estimand $\Delta$, for each scenario is calculated from a large sample, a simulated population with $100,000$ clusters \citep{zhu2024leveraging}. Relative bias evaluates the accuracy of the estimation in comparison to the actual value and provides a sense of how large the error is in relation to the size of the true ATE. It is particularly useful in our scenario since the magnitude of the true value is important to consider. A smaller value of absolute relative bias indicates higher accuracy in estimation. The Monte Carlo Standard Deviation (MCSD) is then calculated for each method to measure the spread of the estimated values across different simulation runs, i.e., different Monte Carlo replications. A smaller MCSD indicates that the method is more efficient and is given by
$\text{MCSD} = \sqrt{\frac{1}{R-1} \sum_{r=1}^{R} \left(\widehat{\Delta}^{(r)} - \widehat{\Delta}\right)^2}$. The average estimated standard error (AESE) is defined as the average, across simulation replicates, of the posterior standard error estimates of $\widehat{\Delta}^{(r)}$, that is, the posterior standard deviation of the draws around their posterior mean within each replicate:
$$
\text{AESE} = \frac{1}{R}\sum_{r=1}^{R}\sqrt{\frac{1}{B-1} \sum_{b=1}^{B} \left(\widehat{\Delta}^{(r,b)} - \widehat{\Delta}^{(r)}\right)^2}.
$$ The match between the MCSD and AESE suggests that the posterior standard error estimates are well-calibrated and appropriately reflect the sampling variability across replicates. In addition, we calculate the $95$\% interval coverage, using both calibrated and uncalibrated variance estimates, for each method to assess whether it achieves nominal frequentist coverage for the cluster-ATE and individual-ATE, thereby supporting valid estimand-aligned inference. Empirical coverage is computed as the proportion of $95$\% intervals that contain the true value of the estimand across all $R$ replicated datasets. To compare the efficiency of estimators and the impact of leveraging or not leveraging covariates, we calculate the relative efficiency, which is dividing the estimated variance of the nonparametric estimator across all simulated datasets without covariate adjustment by the estimated variance of the estimator with covariate adjustment. The nonparametric unadjusted estimator is obtained by $\widehat{\mu}_C^{\text{np}}(a) =\frac{1}{M} \sum_{i=1}^M\left\{\frac{I\left(A_i=a\right)\overline{Y}_i}{\pi^a\left(1-\pi\right)^{1-a}}\right\}$, and $\widehat{\mu}_I^{\text{np}}(a) = \frac{1}{\sum_{i=1}^M N_i}\sum_{i=1}^M N_i\left\{\frac{I\left(A_i=a\right)\overline{Y}_i}{\pi^a\left(1-\pi\right)^{1-a}}\right\}$. This metric quantifies the percentage improvement in efficiency gained through covariate adjustment using either the g-computation or model-robust estimator. This metric also facilitates efficiency comparison between the g-computation and model-robust estimators. 

\subsection{Simulation results}

\subsubsection{Covariate adjustment in g-computation under potential model misspecification}

\begin{table}[ht]
\centering
\caption{Results in the first simulation experiment from $500$ simulations. $\Delta_C$ denotes true value of cluster-average treatment effect. $\Delta_I$ denotes true value of individual-average treatment effect. ``Adjusted'' indicates that the fitted model is a linear mixed model that includes covariate main effects and treatment–covariate interactions. ``Unadjusted'' indicates that the linear mixed model includes only a treatment indicator. $M$ denotes the number of clusters. ``Relative Bias'' refers to the relative bias of the estimated value expressed as a percentage, with standard errors shown in parentheses. ``Coverage'' refers to the $95$\% credible interval coverage across all simulated datasets expressed as a percentage. ``MCSD'' denotes Monte Carlo standard deviation. ``AESE'' denotes average of estimated standard error.}
\begin{tabular}{llrrrrrr}
\toprule
{Data-generating Model} & {Fitted Model} & {M} & {Estimand} & {Relative Bias (SE)} & Coverage & {MCSD} & {AESE} \\
\midrule
\multirow{12}{*}{\shortstack{Scenario (i)\\ $\Delta_C=0.590$\\ $\Delta_I=0.625$}} & \multirow{6}{*}{Adjusted} 
& \multirow{2}{*}{30} & cluster-ATE & -2.1 (1.7) & 95.0 & 0.226 & 0.247 \\
&&& individual-ATE & -1.5 (1.7) & 95.4 & 0.244 & 0.255 \\
& & \multirow{2}{*}{60} & cluster-ATE & -2.8 (1.1) & 95.0 & 0.145 & 0.146 \\
&&& individual-ATE & -3.1 (1.1) & 94.4 & 0.151 & 0.154 \\
& & \multirow{2}{*}{90} & cluster-ATE & -1.9 (0.8) & 95.0 & 0.112 & 0.116 \\
&&& individual-ATE & -1.5 (0.9) & 94.0 & 0.121 & 0.122 \\
\cmidrule{2-8}
& \multirow{6}{*}{Unadjusted} 
& \multirow{2}{*}{30} & cluster-ATE & -1.9 (4.9) & 95.6 & 0.652 & 0.681 \\
&&& individual-ATE & -7.5 (4.7) & 95.6 & 0.652 & 0.681 \\
& & \multirow{2}{*}{60} & cluster-ATE & 0.9 (3.4) & 95.0 & 0.449 & 0.471 \\
&&& individual-ATE & -4.8 (3.2) & 95.2 & 0.449 & 0.471 \\
& & \multirow{2}{*}{90} & cluster-ATE & 0.9 (2.9) & 93.6 & 0.388 & 0.383 \\
&&& individual-ATE & -4.8 (2.8) & 93.6 & 0.388 & 0.383 \\
\midrule
\multirow{12}{*}{\shortstack{Scenario (ii)\\ $\Delta_C=0.679$\\ $\Delta_I=0.842$}} & \multirow{6}{*}{Adjusted} 
& \multirow{2}{*}{30} & cluster-ATE & -3.8 (2.3) & 82.0 & 0.354 & 0.260 \\
&&& individual-ATE & -3.8 (2.0) & 82.8 & 0.372 & 0.266 \\
& & \multirow{2}{*}{60} & cluster-ATE & -2.6 (1.5) & 83.2 & 0.222 & 0.156 \\
&&& individual-ATE & -3.6 (1.2) & 83.6 & 0.227 & 0.162 \\
& & \multirow{2}{*}{90} & cluster-ATE & -1.8 (1.2) & 83.0 & 0.181 & 0.124 \\
&&& individual-ATE & -1.9 (1.0) & 80.2 & 0.193 & 0.129 \\
\cmidrule{2-8}
& \multirow{6}{*}{Unadjusted} 
& \multirow{2}{*}{30} & cluster-ATE & -2.6 (2.9) & 93.4 & 0.435 & 0.428 \\
&&& individual-ATE & -21.5 (2.3) & 91.2 & 0.435 & 0.428 \\
& & \multirow{2}{*}{60} & cluster-ATE & 0.7 (1.9) & 94.0 & 0.281 & 0.294 \\
&&& individual-ATE & -18.8 (1.5) & 92.2 & 0.281 & 0.294 \\
& & \multirow{2}{*}{90} & cluster-ATE & 1.3 (1.6) & 93.2 & 0.250 & 0.240 \\
&&& individual-ATE & -18.3 (1.3) & 87.2 & 0.250 & 0.240 \\
\bottomrule
\end{tabular}
\label{tab:combined_first_experiment}
\end{table}

The first simulation experiment focuses on the model-based g-computation estimator and the uncalibrated method for generating credible intervals, which use quantiles of the posterior samples. For the fitted model, we compare the performance of a linear mixed model that adjusts for covariates with main effects and treatment-covariate interactions, as specified in Equation~\eqref{eq:LMM}, to the performance of an unadjusted linear mixed model. Table~\ref{tab:combined_first_experiment} summarizes the results for the first simulation experiment. Under data-generating process (i), the cluster-ATE $\Delta_C=0.590$ and the individual-ATE $\Delta_I=0.625$, whereas under data-generating process (ii), the cluster-ATE $\Delta_C=0.679$ and the individual-ATE $\Delta_I=0.842$. 

Specifically, across both scenarios, the proposed g-computation estimator with covariate adjustment consistently outperforms its unadjusted counterpart in terms of efficiency. This improvement highlights how leveraging covariate information can refine outcome prediction, leading to tighter posterior samples and more efficient estimation of the marginal causal estimands. In scenario (i), where the linear mixed model is the true underlying outcome model, both g-computation estimators, with and without covariate adjustment, show negligible bias and achieve nominal coverage. However, in scenario (ii), the fitted linear mixed models fail to account for the non-linear transformations and interactions inherent in the true outcome-generating model, leading to a misspecified parametric working model. In this case, the unadjusted linear mixed model yields a biased point estimator but occasionally achieves coverage closer to the nominal level. In contrast, the covariate-adjusted approach produces a less biased point estimator, yet its credible intervals consistently fall short of nominal coverage. It suggests that in the presence of informative cluster size, adjusting for covariates, even with a misspecified model, can improve efficiency but may compromise nominal coverage. Conversely, not adjusting for covariates may achieve nominal coverage in some cases by chance, at the cost of reduced efficiency. Therefore, without further modification of the estimators and the inferential approach, these commonly used methods fail to fulfill the promise of estimand-aligned inference under model misspecification in CRTs.

\subsubsection{Improving estimation and inference under model misspecification: comparison of point estimator and posterior variance correction combinations} 

\begin{figure}[htbp]
    \centering
    \includegraphics[width=1\textwidth]{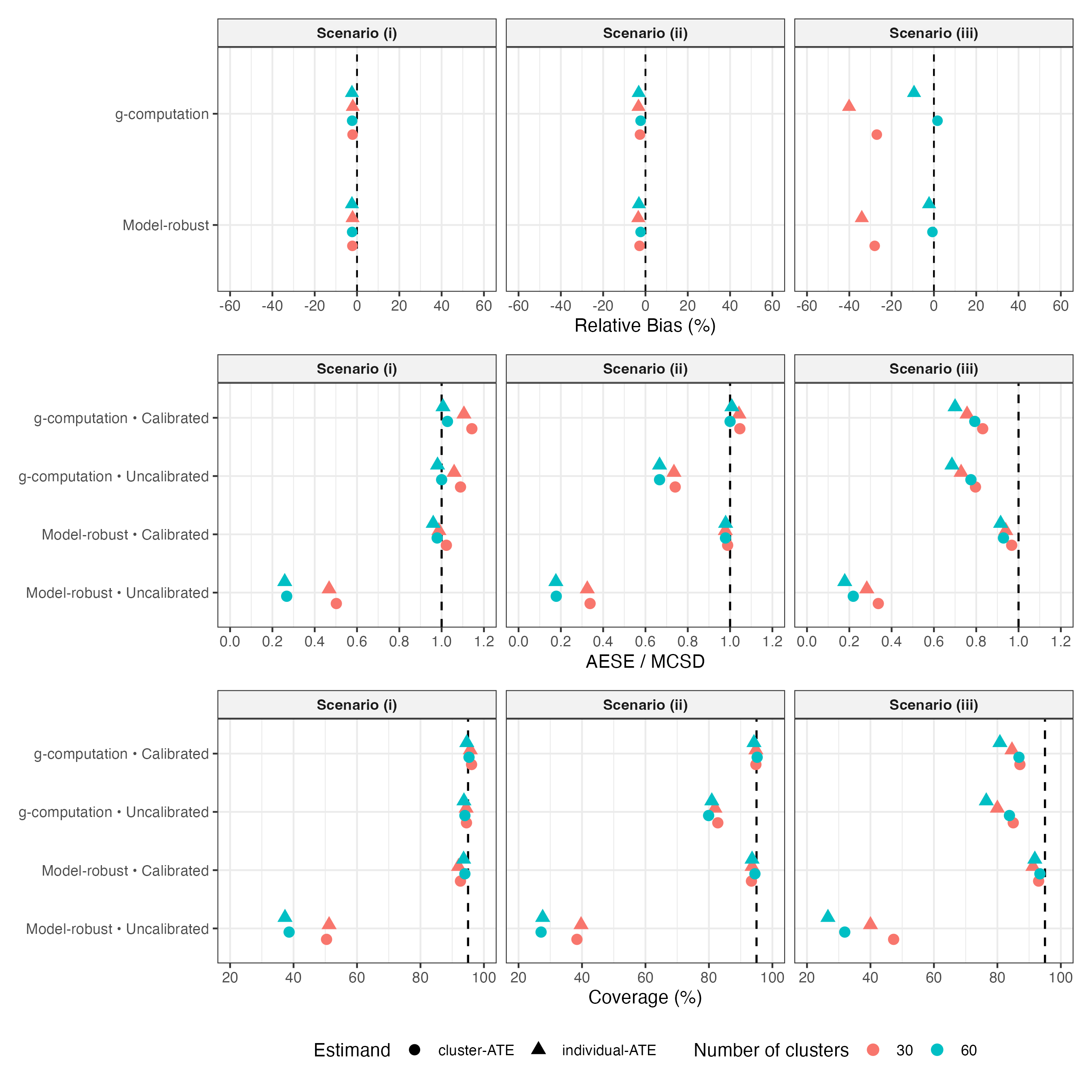}
    \caption{Comparison of relative bias, AESE-to-MCSD ratio, and coverage of $95$\% intervals from the second simulation experiment based on $1,000$ replications for scenario (i), (ii), and (iii), estimating the cluster-average treatment effect (cluster-ATE) and individual-average treatment effect (individual-ATE). A linear mixed model, which adjusts for covariates with main effects and treatment–covariate interactions, is used as the fitted model. The estimation methods are categorized into g-computation and model-robust approaches, each with calibrated and uncalibrated posterior inference. For the calibrated variance estimation, $100$ resampled datasets are used within each simulation to compute posterior variance. Shapes distinguish estimands (circles for cluster-ATE and triangles for individual-ATE), and colors indicate the number of clusters ($M=30$ in red, $M=60$ in blue). Reference lines are included for each metric: dashed at $0$ for relative bias (no systematic over or underestimation), dashed at $1$ for AESE-to-MCSD ratio (perfect calibration where AESE equals MCSD), and dashed at $95$\% for coverage (indicating the nominal $95$\% level).}
    
    \label{fig:second simulation experiment combined}
\end{figure}

The second simulation experiment compares four combinations of point estimators and posterior variance estimation methods: the model-based g-computation and model-robust estimators, each paired with either the uncalibrated or the calibrated variance estimation method. A linear mixed model with covariate adjustment is used as the fitted model throughout this experiment. Our goal is to resolve the issue shown in the first experiment by combining a model-robust estimator and a calibrated posterior variance estimation method. To summarized the simulation results, Table A in Web Appendix 3 reports numerical results under data-generating process (i), under which the linear mixed model is the true underlying data-generating model. Tables B and C in Web Appendix 3 present the numerical results under data-generating processes (ii) and (iii), which involve non-linear relationships between the covariates and the outcomes, and potentially more complicated interactions. To facilitate comparison, Figure \ref{fig:second simulation experiment combined} visualizes simulation results across all three scenarios. It includes three panels summarizing relative bias, the ratio of AESE to MCSD, and empirical coverage of $95$\% intervals. Results are shown for four combinations of estimation approaches. The relative bias panel includes results only for the two point estimators, as bias is not affected by the posterior variance estimation. In all panels, results are presented separately for the cluster-ATE and individual-ATE estimands, distinguished by shape, and for two total numbers of clusters ($M=30$ and $M=60$), distinguished by color. Web Appendix 6 provides the relative efficiency comparisons for the proposed point estimators across all three scenarios, evaluating the efficiency gains of covariate adjusted g-computation and model-robust estimators relative to the unadjusted moment estimator.

Under scenario (i), Figure \ref{fig:second simulation experiment combined} shows that both the g-computation and model-robust estimators yield negligible bias and achieve nominal coverage for both the cluster-ATE and individual-ATE when paired with calibrated variance estimation. In contrast, the uncalibrated variance estimation method for constructing credible intervals, based on posterior quantiles, fails to achieve nominal frequentist coverage for the model-robust estimator, although it performs adequately when paired with the g-computation estimator. This difference arises because the g-computation estimator depends solely on the outcome model parameters. When the outcome model is correctly specified, the posterior distribution from Bayesian estimation appropriately reflects the uncertainty. However, the model-robust estimator involves standardization that depends on both model parameters and the observed data, introducing additional variability that the uncalibrated posterior variance fails to capture. Regarding the AESE-to-MCSD ratio, we also observe that all combinations yield values close to one under moderate sample sizes, except for the model-robust estimator paired with uncalibrated posterior inference. Notably, the model-robust estimator with calibrated posterior inference achieves AESE-to-MCSD ratios near one even in small-sample settings, demonstrating accurate calibration of uncertainty despite limited sample size.

Under scenarios (ii) and (iii), where the working model is incorrectly specified, the model-robust estimator shows smaller, or at least comparable, relative bias in estimating both the individual-ATE and cluster-ATE compared to the g-computation estimator. Furthermore, only the model-robust estimator, when paired with the calibrated variance estimation method, consistently achieves nominal coverage for both cluster-ATE and individual-ATE. This is also reflected in the AESE-to-MCSD ratios in Figure \ref{fig:second simulation experiment combined}, where only the model-robust estimator with calibrated inference consistently yields values near one under scenarios (ii) and (iii), indicating well-calibrated uncertainty estimates that align with the empirical variability observed across replications. The model-based g-computation estimator depends solely on model parameters and not on the observed outcome data. Therefore, when the outcome model is correctly specified, the uncalibrated posterior inference can still provide valid uncertainty quantification. However, the g-computation estimator becomes invalid when the outcome model is misspecified. Under more complex data-generating processes, this misspecification leads to under coverage of the intervals. To maintain small bias and achieve nominal coverage under complex data-generating processes, the model-robust estimator paired with calibrated posterior variance estimation is recommended. {\color{black}We further emphasize that calibration and the choice of point estimator play distinct, non-interchangeable roles. Calibration restores valid uncertainty quantification, but it operates only on the variance and leaves the point estimate unchanged. It therefore cannot compensate for a biased point estimator. The g-computation estimator can carry substantial relative bias, and applying the calibrated variance does not remove this bias, so its intervals still fail to attain nominal coverage, as seen from the calibrated g-computation interval in scenario (iii). Calibration is thus not a substitute for an estimand-aligned point estimator. It must be paired with the model-robust standardization estimator, which remains consistent under misspecification, rather than relied upon to repair the shortcomings of the model-based g-computation.}

{\color{black} The severity with which uncalibrated posterior inference underestimates uncertainty for the model-robust estimator is worth explaining. Across the three scenarios, the uncalibrated intervals attain empirical coverage of only roughly 27\% to 51\%, whereas the corresponding calibrated intervals recover coverage close to the nominal level, approximately 91\% to 95\%. The precise numerical results are reported in Web Appendix 3. This severity follows from the relative magnitudes of the two variance components. As shown in the proof of Proposition~\ref{calibrated_var_consistency}, the posterior parameter variance retained by uncalibrated inference is $o_p(M^{-1})$, whereas the cluster-level sampling variability that it discards is of order $M^{-1}$. Uncalibrated intervals for the model-robust estimator therefore do not merely omit a small correction, but discard the dominant source of variability. A further consequence of the differing rates at which the two components shrink is that the underestimation of uncertainty becomes more severe as the number of clusters increases. In all three scenarios, the coverage of the uncalibrated model-robust intervals drops further when moving from $M=30$ to $M=60$. The posterior contracts faster than the sampling variability, so the gap between the uncalibrated variance and the true variance widens with $M$. Calibration eliminates this problem by restoring the omitted component, and the calibrated model-robust intervals attain near-nominal coverage across all scenarios and cluster sizes.
}

Finally, as shown in Web Appendix 6, the relative efficiency of both the model-based and model-robust estimators relative to the unadjusted nonparametric estimator is always greater than $1$ across all data-generating processes, further confirming that incorporating covariate information improves estimation efficiency. Under scenario (i), Web Appendix 6 demonstrates that both the g-computation and model-robust estimators achieve substantial efficiency gains relative to the unadjusted moment estimator, with comparable levels of improvement. When the working model is incorrectly specified, i.e, scenario (ii) and (iii), the relative efficiency of the model-robust estimator consistently outperforms or is at least comparable to that of the g-computation estimator. {\color{black}To benchmark the proposed calibrated Bayesian approach against an established frequentist alternative, we additionally compare with the model-robust standardization estimator of \citet{li2025model}, implemented with both a GEE working model with an exchangeable working correlation and a linear mixed working model, each using the cluster jackknife variance estimator. The results are reported in Web Appendix 7. The two working models under frequentist paradigm yield nearly identical results. Focusing on the model-robust estimator with calibrated inference, the frequentist and Bayesian approaches reveal a trade-off. In terms of coverage, the frequentist jackknife intervals are slightly better, staying at or just above the nominal level, whereas the calibrated Bayesian intervals may dip slightly below nominal in the small samples, though they remain close to it throughout. In terms of the point estimator, the calibrated Bayesian estimator exhibits a smaller MCSD in the small-sample comparisons. The frequentist jackknife intervals attain nominal coverage at the cost of a larger AESE, and hence wider intervals. Under the most complex data-generating process, scenario (iii), both the frequentist and Bayesian model-robust estimators exhibit non-negligible relative bias when the number of clusters is small, reflecting the intrinsic challenge of point estimation under this limited sample size setting.}

\subsubsection{Leveraging BART for estimating cluster-ATE and individual-ATE}

\begin{figure}[htbp]
    \centering
    \includegraphics[width=\textwidth]{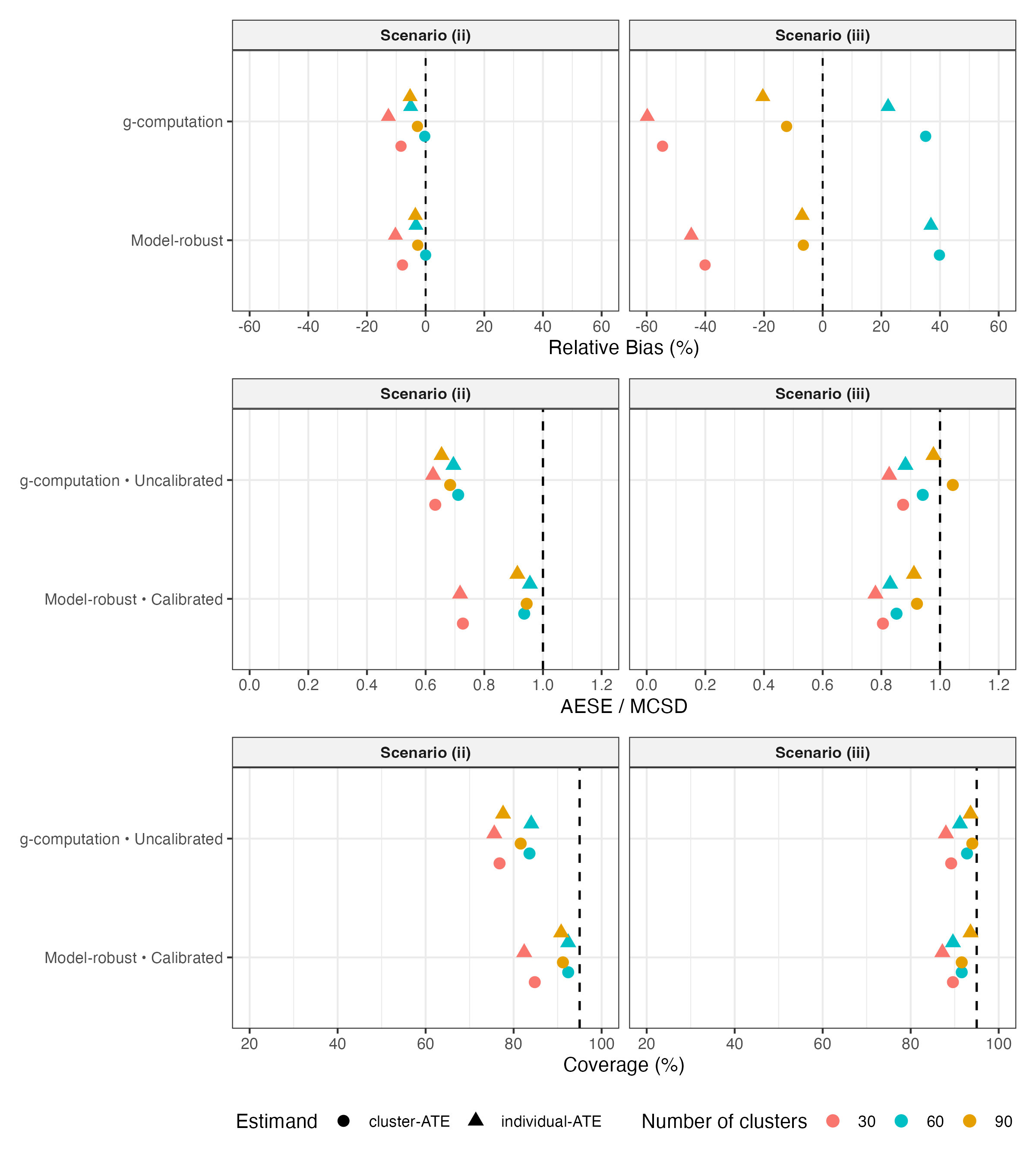}
    \caption{Comparison of relative bias, AESE-to-MCSD ratio, and coverage of $95$\% intervals from the third simulation experiment based on $250$ replications for scenario (ii), and (iii), estimating the cluster-average treatment effect (cluster-ATE) and individual-average treatment effect (individual-ATE). A mixed-effects BART is used as the fitted model. The estimation methods are categorized into g-computation with uncalibrated posterior inference and model-robust standardization with calibrated posterior inference. For the calibrated posterior inference, $100$ resampled datasets are used within each simulation to compute variance. Shapes distinguish estimands (circles for cluster-ATE and triangles for individual-ATE), and colors indicate the number of clusters ($M=30$ in red, $M=60$ in blue, $M=90$ in orange). Reference lines are included for each metric: dashed at $0$ for relative bias (no systematic over or underestimation), dashed at $1$ for AESE-to-MCSD ratio (perfect calibration where AESE equals MCSD), and dashed at $95$\% for coverage (indicating the nominal $95$\% level).}
    
    \label{fig:third simulation experiment combined}
\end{figure}

While linear mixed models provide a widely used parametric framework for estimating treatment effects in CRTs, their validity can be sensitive to model misspecification. In complex settings involving non-linear relationships or high-dimensional covariates, misspecification of the outcome model can lead to biased or inefficient inference. To address these challenges, Bayesian methods offer a flexible alternative by incorporating nonparametric and machine learning techniques that relax strong parametric assumptions. BART \citep{chipman_bart_2010}, a nonparametric Bayesian ensemble method that constructs a sum-of-trees model, can be used to mitigate concerns about parametric misspecification. It consists of two key components: an ensemble of regression trees summed to model outcome mean, and a regularization prior that constrains the tree parameters to prevent overfitting \citep{chipman_bart_2010}. Recent extensions such as mixed-effects BART \citep{spanbauer2021nonparametric} incorporate cluster-specific random intercepts into the BART framework to account for within-cluster correlation in clustered data. It combines the flexibility of BART to model the fixed effect portion with hierarchical modeling, allowing it to accommodate both non-linearity and within-cluster dependence in CRTs. Suppose that the final outcome $Y_{ij}$ is related to generic covariates $A_i$, $\bm{X}_{ij}$, and $N_i$, the mixed-effects BART is given by
\begin{align}
Y_{ij} =\sum_{l=1}^L h\left(A_i, \bm{X}_{ij}, N_i; T_l, \Lambda_l\right) + \phi_i + \epsilon_{ij},
\label{eq:mixed BART}
\end{align}
where $\phi_i \sim \mathcal{N}\left(0, \sigma_{\phi}^2\right)$ represents the cluster-level random effect, and $\epsilon_{i j} \sim \mathcal{N}\left(0, \sigma_{\epsilon}^2\right)$ denotes the individual-level residual error. The mean function is modeled as a sum of $L$ regression trees, expressed as $\sum_{l=1}^L h\left(A_i, \bm{X}_{ij}, N_i; T_l, \Lambda_l\right)$, where each $T_l$ represents the structure of the $l$-th binary tree, and $\Lambda_l$ denotes a set of parameters values associated with each of the terminal nodes in tree $T_l$ \citep{chipman_bart_2010}. The function $h\left(\cdot\right)$ assigns a terminal node value in $\Lambda_l$ to $\{A_i, \bm{X}_{ij}, N_i\}$ according to binary tree $T_l$. Increasing $L$ allows for greater model flexibility, though it also imposes higher computational costs. A common default of $L=50$ trees has been found to provide adequate performance in practice, as suggested by \citet{bleich2014variable}. This formulation mirrors the linear mixed model structure while replacing the fixed component with a data-adaptive ensemble model, allowing for flexible modeling of fixed effects within the hierarchical framework. In the following, we implement this mixed-effects BART model using the \verb|mxbart| function from the R package \verb|mxBART| \citep{mxBART}.

Using mixed-effects BART with covariate adjustment as the working outcome models, Figure \ref{fig:third simulation experiment combined} visualizes results from BART-based methods under scenarios (ii) and (iii), characterized by non-linear outcome models and informative cluster size. Results are shown for two estimation approaches, which are g-computation with uncalibrated posterior inference and model-robust standardization with calibrated posterior inference. The organization of the figure follows the same structure as Figure \ref{fig:second simulation experiment combined}, with three panels displaying relative bias, the AESE-to-MCSD ratio, and coverage of $95$\% intervals. However, we now consider three total numbers of clusters ($M=30$, $60$, and $90$), which are distinguished by color. We additionally include a larger number of clusters ($M=90$) in this simulation, as tree-based ensemble methods like BART usually require larger sample sizes to stabilize posterior inference and reduce variability in predictions. Web Appendix 4 presents the numerical results corresponding to these BART-based simulations. {\color{black}The Bayesian procedures incur a non-trivial computational cost. For the most demanding configuration, mixed-effects BART with the model-robust estimator and calibrated variance estimation ($10,000$ posterior draws with $5,000$ discarded as burn-in and $100$ bootstrap resamples), the full simulation of $250$ replications under the most complex scenario required approximately $2.68$, $3.52$, and $4.13$ hours for $M=30$, $60$, and $90$ clusters, respectively, when run on high-performance computing facilities with $35$ cores. The calibrated variance estimation adds only modest overhead, since the bootstrap resamples reuse the original posterior draws without additional model refitting. As the replications, bootstrap resamples, and MCMC chains can all be computed in parallel, the runtime decreases approximately linearly with the number of available cores.}

We first compare the performance of the linear mixed model and BART, each combined with the model-based g-computation estimator and uncalibrated posterior inference. While BART paired with g-computation and uncalibrated intervals occasionally attains close to nominal coverage, this is not consistently guaranteed. In contrast, the linear mixed model under the same estimation strategy consistently fails to achieve nominal coverage under model misspecification. Broadly speaking, when combined with g-computation and uncalibrated intervals, neither the linear mixed model nor BART supports nominal coverage under non-linear data-generating processes, under informative cluster size. Moreover, in small-sample settings, $M = 30$ or $M = 60$, BART shows larger relative bias compared to the linear mixed model, and does not consistently demonstrate efficiency gains as measured by MCSD shown in Web Appendix 4. Thus, while BART combined with g-computation and uncalibrated inference may offer some improvement in coverage over its linear mixed model counterpart, it still falls short when compared to the linear mixed model paired with the model-robust estimator and calibrated variance estimation, which ultimately delivers more reliable coverage performance.

We next compare BART and the linear mixed model when both are used in conjunction with model-robust standardization and calibrated variance estimation. In the presence of informative cluster size and non-linear data-generating processes, BART combined with model-robust standardization and calibrated intervals can approach nominal coverage when the number of clusters is at least moderate. However, the linear mixed model paired with this combination more consistently achieves nominal coverage, even in small-sample settings. Although both approaches are embedded under the model-robust standardization formula, the linear mixed model's parametric structure can avoid over-fitting and helps stabilize variance estimation. In contrast, BART's flexible and data-adaptive nature may lead to overfitting in small-sample settings, resulting in underestimated uncertainty. This issue is addressed as the number of clusters (i.e., independent units) increases to $M=90$.

Focusing specifically on scenarios (ii) and (iii), which involve informative cluster size, comparing BART combined with g-computation and uncalibrated inference to BART combined with model-robust approach and calibrated inference, the latter usually achieves coverage close to the nominal level in cases where the former does not. We then explore whether informative cluster size (effect modification at the cluster level) plays a key role in this simulation comparison. Hence, we design and evaluate a scenario without an informative cluster size, with details of the data-generating process and corresponding mixed-effects BART results provided in Web Appendix 5. Interestingly, when the cluster size is non-informative, both approaches yield small bias and coverage close to the nominal level, even with a small number of clusters.

{\color{black} The results under scenario (iii) clearly illustrate that calibration can only correct the quantification of uncertainty, but not the bias in the point estimator. With a small number of clusters, the BART-based model-robust point estimator may carry substantial relative bias, and although the calibrated variance is applied, the empirical coverage falls below the nominal level. Calibration is therefore necessary but not sufficient for valid inference in a finite sample. It must be paired with a point estimator that is itself approximately unbiased in this scenario to achieve nominal coverage.} {\color{black}To better understand the weaker small-sample performance of BART under scenario (iii), we conducted a detailed sensitivity analysis reported in Web Appendix 8. In this analysis, we retain the nonlinear data-generating model of scenario (iii) exactly, including its dependence on cluster size (so that informative cluster size is preserved and the cluster-ATE and individual-ATE remain distinct), but we generate the cluster-level and individual-level covariates independently of the cluster size, thereby alleviating the nested, cross-level covariate dependence. Interestingly, under this simplification, the relative bias of the BART-based estimators decreases substantially across all cluster sizes. This experiment isolates the source of the difficulty: the large small-sample bias of BART under scenario (iii) is also driven primarily by the nested, cluster-size-dependent covariate structure rather than by informative cluster size alone or by an inherent limitation of BART. It also suggests that BART remains a competitive working model in complex but less hierarchically entangled settings, while its flexibility should be exercised with caution when the covariate structure is deeply nested and the number of clusters is small.}

\section{Illustrative data application}
\label{sec: Case Study}
\begin{table}[ht]
\centering
\begin{threeparttable}
\caption{Analysis of the Pain Program for Active Coping and Training (PPACT) data. Each analysis using the linear mixed model (LMM) as the fitted model was performed with a total of $2,000$ MCMC iterations. The first $1,000$ iterations were discarded as burn-in. Each analysis using BART as the fitted model was performed with a total of $10,000$ MCMC iterations, with the first $5,000$ discarded as burn-in. Convergence was assessed using traceplots of the cluster-ATE and individual-ATE as well as the Geweke diagnostic. {\color{black}The unadjusted LMM rows correspond to the working model $Y_{ij}=\beta_0+\beta_1 A_i+\phi_i+\epsilon_{ij}$. For the g-computation estimator, this unadjusted specification does not target either the cluster-ATE or the individual-ATE: as shown in Section~\ref{sec:Model-based g-computation}, both estimators reduce to the treatment coefficient $\beta_1$, so the two estimand columns report the same point estimate and interval.}}
\label{tab:PPACT analysis results}
    \begin{tabular}{clllrrrr}
        \toprule
        \multirow{2}{*}{\textbf{Fitted Model}} &\multirow{2}{*}{\textbf{Covariate}}& \multirow{2}{*}{\textbf{Estimator}} & \multirow{2}{*}{\makecell{\textbf{Posterior} \\ \textbf{Inference}}} & \multicolumn{2}{c}{\textbf{cluster-ATE}} & \multicolumn{2}{c}{\textbf{individual-ATE}} \\
        \cmidrule(lr){5-6} \cmidrule(lr){7-8}
        &&& & \textbf{Estimate} & \textbf{95\% C.I.} & \textbf{Estimate} & \textbf{95\% C.I.} \\
        \midrule
        \multirow{8}{*}{LMM} & \multirow{4}{*}{Unadjusted} &\multirow{2}{*}{g-computation}& Uncalibrated& -0.627 & (-0.977, -0.274) & -0.627 & (-0.977, -0.274) \\
        &&& Calibrated& -0.627 & (-0.984, -0.269) & -0.627 & (-0.984, -0.269) \\
        \cmidrule(r){3-8} 
        &&\multirow{2}{*}{Model-robust}& Uncalibrated *& -0.661 & (-0.661, -0.661) & -0.616 & (-0.622, -0.610) \\
        &&& Calibrated& -0.661 & (-1.082, -0.241) & -0.616 & (-0.990, -0.242) \\
        \cmidrule(r){2-8} 
        & \multirow{4}{*}{Adjusted} &\multirow{2}{*}{g-computation}& Uncalibrated& -0.612 & (-0.980, -0.254) & -0.530 & (-0.878, -0.188) \\
        &&& Calibrated& -0.612 & (-0.991, -0.233) & -0.530 & (-0.888, -0.171) \\
        \cmidrule(r){3-8} 
        &&\multirow{2}{*}{Model-robust}& Uncalibrated& -0.594 & (-0.658, -0.530) & -0.520 & (-0.566, -0.472) \\
        &&& Calibrated& -0.594 & (-0.954, -0.235) & -0.520 & (-0.849, -0.191) \\
        \midrule
        \multirow{4}{*}{BART}& \multirow{4}{*}{Adjusted} &\multirow{2}{*}{g-computation}& Uncalibrated& -0.516 & (-0.850, -0.188) & -0.514 & (-0.842, -0.187) \\
        &&& Calibrated& -0.516 & (-0.846, -0.186) & -0.514 & (-0.843, -0.185) \\
        \cmidrule(r){3-8} 
        &&\multirow{2}{*}{Model-robust}& Uncalibrated& -0.588 & (-0.674, -0.505) & -0.526 & (-0.594, -0.457) \\
        &&& Calibrated& -0.588 & (-0.926, -0.250) & -0.526 & (-0.837, -0.215) \\
        \bottomrule
    \end{tabular}

\begin{tablenotes}
      \item[*] When using the model-robust estimator without covariate adjustment and without posterior variance correction, the estimated cluster-ATE remains identical across all posterior draws. {\color{black}This occurs because, given that the observed data have equal numbers of clusters in each intervention arm, the treatment coefficient cancels in the model-robust standardization formula, and the estimator reduces to the nonparametric unadjusted estimator specified in Section~\ref{sec: Implementation and performance metrics}, which depends only on the observed data and not on the model parameters. As a result, the estimator takes the same value across all posterior draws, and the uncalibrated posterior interval has zero width; the calibrated interval, which additionally accounts for the sampling variability of the data, has positive width. This degeneracy is specific to the cluster-ATE under the unadjusted LMM model, as the covariate-adjusted estimator and the individual-ATE retain their dependence on the model parameters. This explains the row marked with an asterisk, where the $95$\% credible interval has zero width, with upper and lower bounds coinciding with the point estimate.}
\end{tablenotes}
\end{threeparttable}
\end{table}

We demonstrate our methods to analyze data from the Pain Program for Active Coping and Training (PPACT) \citep{debar2022primary}. PPACT is a pragmatic CRT evaluating the effect of a cognitive behavioral therapy intervention embedded in primary care versus usual care for treating long-term opioid users with chronic pain. The study included adult health plan members from participating primary care providers with at least six months of continuous enrollment, a pain diagnosis within the past year, and long-term opioid use, defined as a supply of short-acting opioids for more than $90$ days over a $120$-day period, or at least two dispensations of long-acting opioids in the past $180$ days. Patients were excluded if they had severe cognitive impairment, a current malignant cancer diagnosis, hospice or end-of-life care within past year, were undergoing intensive addiction treatment, or were unwilling to participate in the skills training or lifestyle change components of the intervention. The study involves a total of $106$ clinic-based clusters, with $53$ randomized to the intervention and $53$ to usual care in a 1:1 ratio. The primary hypothesis is that patients receiving the intervention would have a greater reduction in pain impact compared to those receiving usual care. The primary outcome is the self-reported PEGS score at $12$ months, a continuous measure that assesses pain impact as a composite of pain intensity and interference with enjoyment of life, general activity, and sleep. Patient-level clinical covariates are extracted from the electronic health record and other existing clinical information systems, including both demographic and diagnostic variables. The baseline covariates we consider include cluster size, patient age at randomization, gender, disability, current smoking status, body mass index, alcohol and drug abuse diagnosis in $6$ months prior to randomization, comorbidity defined as having two or more specified chronic medical conditions, depression diagnosis in $6$ months prior to randomization, number of different pain types, average morphine milligram equivalents dose per day in $6$ months prior to randomization, and an indicator for whether this average daily dose exceeded $90$. Our analysis includes a total of $705$ individuals from the $106$ clusters, selected for having complete data on all relevant baseline covariates and outcomes. 

We apply twelve combinations of Bayesian methods to analyze the PPACT trial, including g-computation and model-robust estimators, with and without posterior variance correction, under both linear mixed models and BART. 
For linear mixed models, we consider both covariate-adjusted and unadjusted specifications, with each analysis run for $2,000$ MCMC iterations and a $1,000$-iteration burn-in. {\color{black}The covariate-adjusted specification is model~\eqref{eq:LMM}, which includes the covariate main effects, cluster size, and their interactions with treatment. The unadjusted specification is $Y_{ij}=\beta_0+\beta_1 A_i+\phi_i+\epsilon_{ij}$, containing only a treatment indicator and a cluster random intercept. As shown in Section~\ref{sec:Model-based g-computation}, the g-computation estimator under this unadjusted model reduces to the treatment coefficient $\beta_1$ itself, so the corresponding rows of Table~\ref{tab:PPACT analysis results} report exactly what a standard unadjusted linear mixed model analysis would report.} For BART, only the covariate adjusted specification is used, as omitting covariates would limit the model's strength to capture non-linear and interactive effects that tree-based methods are designed to handle. Each BART analysis is run for $10,000$ iterations with a $5,000$-iteration burn-in. Convergence is evaluated using traceplots of the cluster-ATE and individual-ATE and the Geweke diagnostic. The results from all twelve methods are summarized in Table~\ref{tab:PPACT analysis results}. Across all twelve methods, spanning g-computation and model-robust estimators, with and without posterior variance correction, under both linear mixed models and BART, we observe that the $95$\% intervals for both the cluster-ATE and individual-ATE consistently exclude zero. It provides strong evidence that the cognitive behavioral therapy is effective in reducing pain impact among long-term opioid users with chronic pain, compared to usual care.

Building on insights from the simulation studies, we interpret the findings in Table~\ref{tab:PPACT analysis results} with several key observations. The unadjusted g-computation approach, identified in our simulations as among the least reliable, gives the same point estimates for the cluster-ATE and individual-ATE. {\color{black}As shown in Section~\ref{sec:Model-based g-computation}, this is because the unadjusted LMM g-computation estimators for the cluster-ATE and individual-ATE both reduce to the coefficient of the treatment indicator, so their estimates necessarily coincide. This masks the distinction between the definitions of the two estimands, indicating that the method is not estimand-aligned under informative cluster size.} {\color{black}In other words, these two columns should not be read as separate estimates of two distinct estimands. The reported value is simply the treatment coefficient of the unadjusted linear mixed model, which targets neither the cluster-ATE nor the individual-ATE under informative cluster size.} {\color{black}Note that a degeneracy affects the unadjusted model-robust estimator for the cluster-ATE, whose uncalibrated interval has zero width. Web Appendix 10 shows that this occurs exactly when the realized share of weight in each arm equals the randomization probability, in which case the estimator no longer depends on the model parameters.} In contrast, the most trustworthy strategy supported by our simulation results, is the model-robust standardization estimator with covariate adjustment and calibrated posterior variance. Both linear mixed models and BART can be appropriate outcome models, especially given the relatively large number of clusters, which supports stable estimation even with BART. Under these preferred methods, the difference between the estimated cluster-ATE and individual-ATE becomes more apparent. 

Focusing on the best-performing strategy, model-robust estimator with covariate adjustment and calibrated posterior inference, we find that BART and linear mixed model yield comparable results. Specifically, while the $95$\% calibrated intervals are slightly wider under the linear mixed model, the point estimates and interval bounds are quite close to those obtained using BART, suggesting that a linear specification may adequately approximate the true underlying data-generating process. Using a linear mixed model, the estimated cluster-ATE is $-0.594$ with a $95$\% calibrated interval of $(-0.954, -0.235)$, and the individual-ATE is $-0.520$ with a $95$\% calibrated interval of $(-0.849, -0.191)$. With BART, the estimated cluster-ATE is $-0.588$ with a $95$\% calibrated interval of $(-0.926, -0.250)$, and the individual-ATE is $-0.526$ with a $95$\% calibrated interval of $(-0.837, -0.215)$. These results indicate that, when accounting for the appropriate unit of inference, the expected reduction in pain impact, measured by the PEGS score at 12 months due to the cognitive behavioral therapy intervention, is $0.588$ at the cluster level and $0.526$ at the individual level. Both reductions are statistically significant at the 5\% level and demonstrate the effectiveness of the cognitive behavioral therapy among long-term opioid users with chronic pain.

\section{Discussion} \label{sec: Discussion}
Despite the growing importance of CRTs in health research, Bayesian analyses of such trials have often focused on model-based parameters of hierarchical models rather than on explicitly defined causal estimands. This disconnect can lead to ambiguity in interpretation, as posterior inferences may implicitly target quantities that depend on model specification rather than on the intended unit of inference. To address this methodological gap, we develop estimand-aligned Bayesian inferential strategies for CRTs that preserve the causal interpretation of treatment effects. Building on the distinctions proposed by \citet{kahan2023estimands}, we formalize two target estimands, the cluster-ATE and the individual-ATE, which diverge in the presence of informative cluster size. We then survey different Bayesian estimators that integrates the g-computation formula and model-robust standardization to estimate these estimands, thereby improving the frequentist coverage properties of Bayesian inference. Through extensive simulation studies, we evaluate the finite-sample performance of the proposed estimators under varying outcome models, covariate adjustment strategies, and posterior variance estimation methods. Our findings highlight the appeal of combining Bayesian hierarchical models with model-robust standardization formula to target model-free estimands, particularly in the presence of informative cluster size. By leveraging this formulation, we extended the approach in \citet{antonelli2022causal} to generate a calibrated variance estimator that jointly accounts for parameter uncertainty and sampling variability, thereby improving the frequentist calibration of the resulting Bayesian intervals. Simulation results demonstrate that, even under misspecified parametric working models, the model-robust estimator derived from a Bayesian linear mixed model achieves valid estimand-aligned inference for both the cluster-ATE and individual-ATE when combined with the calibrated variance approach. For applied analysts, these findings suggest that standard Bayesian hierarchical models can be extended to deliver causally interpretable and well-calibrated inference simply by post-processing posterior samples through model-robust standardization and variance correction, offering a practical route toward principled estimand-aligned Bayesian analysis of CRTs.

In our numerical studies, we also consider Bayesian nonparametric models to enhance flexibility in modeling outcomes and dependencies among observations. Specifically, mixed-effects BART \citep{spanbauer2021nonparametric} combines the standard BART framework for fixed effects with hierarchical random effects, providing a natural extension for CRTs. Our simulation studies show that mixed-effects BART, when combined with model-robust standardization and calibrated variance estimation, tends to achieve small bias and near-nominal coverage for both cluster-ATE and individual-ATE, under model misspecification and informative cluster size, {\color{black}particularly when the number of clusters is large. A limitation concerns the finite-sample behavior of mixed-effects BART. Under the most complex data-generating process (scenario (iii)), BART-based estimators can exhibit substantial relative bias when the number of clusters is limited, and this bias may persist to a non-negligible degree even at $M=60$. A sensitivity simulation analysis indicates that this behavior is largely attributable to the difficulty of learning a nested, cluster-size-dependent covariate structure from few clusters, compounded by a highly nonlinear outcome surface. When the covariates are generated independently of cluster size, the bias is substantially reduced. In practice, therefore, when the number of clusters is small and the covariate structure is potentially complex, the more parsimonious Bayesian linear mixed model, paired with model-robust standardization and calibrated variance estimation, is expected to provide more stable inference. Mixed-effects BART becomes more advantageous as the number of clusters grows, where its flexibility can be exploited without incurring excessive overfitting.} When g-computation is used with uncalibrated posterior inference, neither the linear mixed model nor BART attains nominal coverage under model misspecification. These findings further underscore the importance of coupling model-robust standardization with calibrated variance estimation to ensure estimand-aligned inference under informative cluster size, even when leveraging flexible Bayesian nonparametric models. Future work should further investigate the strengths of mixed-effects BART for CRTs with larger numbers of clusters or high-dimensional covariates, and explore alternative posterior calibration methods, such as the Bayesian bootstrap approach \citep{yiu2025semiparametric}, that provide valid inference for low-dimensional functionals while retaining the adaptivity of nonparametric Bayesian models.

The broader motivation of this work resonates with the Calibrated Bayes framework articulated by \citet{rubin1984bayesianly} and \citet{little2006calibrated}, which advocates Bayesian reasoning in principle yet calibration to real-world frequencies in practice. As Rubin discussed in his 1984 article \citep{rubin1984bayesianly}, ``\emph{frequency calculations are useful for making Bayesian statements scientific, scientific in the sense of capable of being shown wrong by empirical test; here the technique is the calibration of Bayesian probabilities to the frequencies of actual events\ldots, The applied statistician should be Bayesian in principle and calibrated to the real world in practice---appropriate frequency calculations help to define such a tie}''. \citet{little2006calibrated} later formalized this synthesis as a statistical philosophy that ``\emph{uses frequentist methods for model assessment and Bayesian methods for inference under the model}''. In this spirit, our proposed estimand-aligned Bayesian approach offers a calibrated option that reconciles Bayesian hierarchical inference with frequentist performance guarantees, rather than prescribing calibration as a universal tool, in the context of CRTs. By explicitly aligning posterior inference with model-free estimands and empirical coverage properties, we achieve model-robust analysis under cluster randomization within a Bayesian framework. We believe this work enriches the Bayesian presence in the realm of model-assisted analysis of randomized experiments---an area where Bayesian methods are sparsely discussed but can be valued for their flexibility, coherence, and ability to integrate prior and other design-based information.

\section*{Acknowledgements}
Research in this article was supported by a Patient-Centered Outcomes Research Institute Award\textsuperscript{\textregistered} (PCORI\textsuperscript{\textregistered} Award ME-2022C2-27676). The statements presented in this article are solely the responsibility of the authors and do not necessarily represent the official views of PCORI\textsuperscript{\textregistered}, its Board of Governors, or the Methodology Committee.

\section*{Data Availability Statement}
The sample R code to implement the proposed methods can be found at the GitHub website at \url{https://github.com/Ruyi-Liu/Model-robust-Bayesian-inference-in-cluster-randomized-trials}. The PPACT data that were used to illustrate the proposed approach in Section \ref{sec: Case Study} are publicly available at \url{https://github.com/PainResearch/PPACT}.


\bibliographystyle{apalike}
\bibliography{references}

\clearpage
\newgeometry{left=0.7in, right=0.7in, top=1in, bottom=1in} 

\appendix
\setcounter{figure}{0}
\renewcommand{\thefigure}{A\arabic{figure}}

\setcounter{table}{0}
\renewcommand{\thetable}{\Alph{table}}

\renewcommand{\theequation}{A\arabic{equation}}
\setcounter{equation}{0}
  
\title{Supplementary Materials}

\section*{Preliminaries: notation, decomposition, and assumptions}

This section collects the notation, decomposition, and regularity conditions used throughout Web Appendices 1 and 2. We follow the decomposition and approach of \citet{antonelli2022causal} and adapt the proof for CRT settings. Throughout, we assume the treatment assignment model is correctly specified and known due to randomization. Under a more general representation, the model-robust estimator for $\Delta_\omega$, $\omega \in \{C,I\}$, takes the form of
\[
\widehat{\Delta}_{\omega}^{\text{mr}} = f(\widehat{\mu}_{\omega}^{\text{mr}}(1), \widehat{\mu}_{\omega}^{\text{mr}}(0)),
\]
with
\[
\widehat{\mu}_{\omega}^{\text{mr}}(a) = \left( \sum_{i=1}^M \frac{{\omega}_i}{\overline{{\omega}}} \right)^{-1} \sum_{i=1}^M \frac{{\omega}_i}{\overline{{\omega}}} \left\{\frac{\mathbb{I}\left(A_i=a\right)\left(\overline{Y}_i-\widehat{\mathbb{E}}\left(\overline{Y}_i \mid A_i=a, \bm{X}_i, N_i\right)\right)}{\pi^a\left(1-\pi\right)^{1-a}}+\widehat{\mathbb{E}}\left(\overline{Y}_i \mid A_i=a, \bm{X}_i, N_i\right)\right\},
\]
where $\overline{{\omega}} = M^{-1} \sum_{i=1}^M {\omega}_i$ is the average weight across clusters, $M$ is the number of clusters, $A_i$ is the treatment assignment for cluster $i$, $\overline{Y}_i = N_i^{-1} \sum_{j=1}^{N_i} Y_{ij}$ is the observed mean outcome in cluster $i$, $N_i$ denotes the size of cluster $i$, and $\bm{X}_i$ represents the collection of all covariates within cluster $i$. Specifically, setting ${\omega}_i = 1$ yields the cluster-ATE, while ${\omega}_i = N_i$ corresponds to the individual-ATE. We will focus on the proof for $\mu_{\omega}(a)$ and it is trivial to then extend to $\Delta_\omega$. Then, 
$$
\begin{aligned}
\widehat{\mu}_{\omega}^{\text{mr}}(a) -{\mu}_{\omega}(a) 
&= \left( \sum_{i=1}^M \frac{{\omega}_i}{\overline{{\omega}}} \right)^{-1} \sum_{i=1}^M \frac{{\omega}_i}{\overline{{\omega}}}  \left[ \frac{\mathbb{I}(A_i = a)}{\pi_a} (\overline{Y}_i - \widehat{m}_{ai}) + \widehat{m}_{ai} - {\mu}_{\omega}(a) \right]\\
&= \left( \sum_{i=1}^M{\omega}_i \right)^{-1} \sum_{i=1}^M {\omega}_i  \left[ \frac{\mathbb{I}(A_i = a)}{\pi_a} (\overline{Y}_i - \widehat{m}_{ai}) + \widehat{m}_{ai} - {\mu}_{\omega}(a) \right],
\end{aligned}
$$ 
where $\pi_a \equiv \mathbb{P}(A_i = a) = \pi^a\left(1-\pi\right)^{1-a}$ for simplicity, $\widehat{m}_{ai} = \widehat{\mathbb{E}}\left(\overline{Y}_i \mid A_i=a, \bm{X}_i, N_i\right)$ is the estimated conditional mean (from the outcome model), and $m_{ai}$ is the true conditional mean. We assume a parametric working model for the outcome regression and apply Taylor series expansions for $\widehat{m}_{ai}$ in the proof. 

We use the following decomposition
\[
\widehat{\mu}_\omega^{\text{mr}}(a) - {\mu}_\omega(a) = A + B
\]
where
\begin{align*}
A &= \left( \sum_{i=1}^M \omega_i \right)^{-1} \sum_{i=1}^M \omega_i\, (\widehat{m}_{ai} - m_{ai})\left(1 - \frac{\mathbb{I}(A_i = a)}{\pi_a}\right), \\
B &= \left( \sum_{i=1}^M \omega_i \right)^{-1} \sum_{i=1}^M \omega_i \left[ \frac{\mathbb{I}(A_i = a)}{\pi_a}(\overline{Y}_i - m_{ai}) + m_{ai} - {\mu}_\omega(a) \right].
\end{align*}

\paragraph{Assumptions}
Let $P_0 \in \mathcal{P}$ denote the true (unknown) probability distribution that generates the observed data. In our setting, this corresponds to the distribution over all cluster-level data tuples: $(A_i, \bm{X}_i, N_i, \bm{Y}_i)$, where $A_i$ is the treatment assignment, $\bm{X}_i$ denotes the matrix of covariates for cluster $i$, $N_i$ is the cluster size, and $\bm{Y}_i$ is the vector of observed outcomes for individuals in cluster $i$. The collection of random variables in cluster $i$ is $\bm{ {O}}_i^{full} = \left\{\bm{Y}_i(1), \bm{Y}_i(0), N_i, \bm{X}_i\right\}$. The complete data $\left\{\left(\bm{ {O}}_1^{full}, A_1\right), \ldots, \left(\bm{ {O}}_M^{full}, A_M\right)\right\}$ is only partially observed.

\noindent \textbf{Assumption 1 (Super-population)}:
\begin{itemize}
    \item[(a)] Random variables $ \bm{ {O}}_1^{full}, \ldots, \bm{ {O}}_M^{full} $ are mutually independent.
    \item[(b)] The total number of individuals in cluster $i$, $N_i$, follows an unknown distribution $\mathcal{P}^N$ over a finite support on $\mathbb{N}^{+}$.
    \item[(c)] Given $N_i$, $\left\{\bm{Y}_i(1), \bm{Y}_i(0), \bm{X}_i\right\}$ follows an unknown distribution $\mathcal{P}^{\left\{\bm{Y}(1), \bm{Y}(0), \bm{X}\right\}\mid N}$ with finite second moments.
    \item [(d)] $\displaystyle \sup_{P_0} \mathbb{E}_{P_0} \left[ \left( \overline{Y}_i - m_{ai} \right)^2 \right] \leq K_y < \infty$.
\end{itemize}

According this Assumption, the random variable vectors from different clusters are independent of each other, but within the same cluster, there can be correlation among outcomes, covariates and cluster size. Meanwhile, the collection of random variables in each cluster, $\{\bm{{O}}_1^{full}, \ldots, \bm{ {O}}_M^{full} \}$, are marginally identically distributed according to $\mathcal{P}^N \times \mathcal{P}^{\left\{\bm{Y}(1), \bm{Y}(0), \bm{X}\right\}\mid N}$. This technical condition is useful for managing the differences in the dimensions of $\bm{{O}}_i^{full}$ among various clusters.

\noindent \textbf{Assumption 2 (Bounds on the error of posterior distributions)}: $\displaystyle \sup_{P_0} \mathbb{E}_{P_0} \operatorname{Var}_M \left( \widehat{m}_{ai} - m_{ai} \mid \boldsymbol{D}_i \right) \leq K_m < \infty$.

It states that the posterior distribution of the conditional mean of the cluster-level outcome has bounded variance around its true value. This assumption guarantees that the posterior does not assign excessive mass far from the truth and is typically mild in applications, especially when the outcome is categorical or bounded.

\noindent \textbf{Assumption 3 (Posterior contraction of the outcome model)}:
There exists a sequence $\epsilon_{M} \to 0$ and a constant $Q > 0$, independent of $\epsilon_{M}$, such that
\[
\sup_{P_0} \mathbb{E}_{P_0} \mathbb{P}_M \left( \frac{1}{\sqrt{M}} \| \widehat{\bm m}_a - \bm m_a \| > Q \epsilon_{M} \middle| \boldsymbol{D} \right) \to 0,
\]
where $\| \cdot \|$ denotes the Euclidean norm taken over the $M$ clusters. This assumption states that the posterior distribution of the outcome model contracts to the true conditional mean function at rate $\epsilon_{M}$. We will use a Taylor series expansion for $\widehat{m}_{ai}$ of the following form
\[
\widehat{m}_{ai} = {m}_{ai} + d_1(\bm{X}_i,N_i, \bm{\beta}^*)^\top (\bm{\beta} - \bm{\beta}^*) + (\bm{\beta} - \bm{\beta}^*)^\top H_1(\bm{X}_i,N_i, \tilde{\bm{\beta}})(\bm{\beta} - \bm{\beta}^*),
\]
where $\bm{\beta}^*$ is the true value of $\bm{\beta}$, $\tilde{\bm{\beta}}$ is a point between $\bm{\beta}$ and $\bm{\beta}^*$, and $m(\cdot)$ has bounded first and second derivatives.

\noindent\textbf{Assumption 4 (Moments of the posterior distribution)}:
    \begin{itemize}
        \item [(a)] $\mathbb{E}_{\bm{\beta}|\bm D}[\|\bm \beta - \bm \beta^*\|_2^4] = o_p(M^{-1})$
        \item [(b)] $\mathbb{E}_{\bm D}\left[\mathbb{E}_{\bm{\beta}|\bm D}[\|\bm \beta - \bm \beta^*\|_2^8]\right]^{1/2} = o_p(M^{-1})$
    \end{itemize}
These assumptions assume the convergence rates of the outcome regression models are slightly faster than $M^{-1/4}$.

\noindent\textbf{Assumption 5 (Regularity conditions on cluster weights).}  
There exist constants \( 0 < b_{\omega} < B_{\omega} < \infty \), independent of \( M \), such that for all clusters \( i = 1, \ldots, M \),
\[
b_{\omega} \leq \omega_i \leq B_{\omega}.
\]
This ensures that no individual cluster receives vanishingly small or disproportionately large weight.

\newpage
\clearpage

\section*{Appendix 1. Consistency of the model-robust point estimator}

Utilizing the decomposition and the assumptions listed above in the Preliminaries, we have $\widehat{\mu}_\omega^{\text{mr}}(a) - {\mu}_\omega(a) = A + B$, where
\begin{align*}
A &= \left( \sum_{i=1}^M \omega_i \right)^{-1} \sum_{i=1}^M \omega_i\, (\widehat{m}_{ai} - m_{ai})\left(1 - \frac{\mathbb{I}(A_i = a)}{\pi_a}\right), \\
B &= \left( \sum_{i=1}^M \omega_i \right)^{-1} \sum_{i=1}^M \omega_i \left[ \frac{\mathbb{I}(A_i = a)}{\pi_a}(\overline{Y}_i - m_{ai}) + m_{ai} - {\mu}_\omega(a) \right].
\end{align*}
We examine the contraction rates for each component of $\widehat{\mu}_\omega^{\text{mr}}(a) - {\mu}_\omega(a) = A + B$, and show that, even under outcome model misspecification, the estimator remains root-$M$ consistent, that is, $\widehat{\mu}_\omega^{\text{mr}}(a) - {\mu}_\omega(a)$ contracts at rate $M^{-1/2}$.

\paragraph{Contraction rate of component $B$}
We begin with component $B$, which does not depend on the posterior distribution of the outcome model. Define
\[
\zeta_i \;\equiv\; \omega_i\left[\frac{\mathbb{I}(A_i=a)}{\pi_a}\left(\overline{Y}_i - m_{ai}\right) + m_{ai} - \mu_\omega(a)\right],
\qquad\text{so that}\qquad
B = \left(\sum_{i=1}^M \omega_i\right)^{-1}\sum_{i=1}^M \zeta_i .
\]
Under cluster randomization, $A_i$ is independent of $(\bm{X}_i, N_i, \bm{Y}_i(a))$, so that $\mathbb{E}_{P_0}\left[\omega_i \frac{\mathbb{I}(A_i=a)}{\pi_a}\left(\overline{Y}_i - m_{ai}\right)\right] = 0$; moreover, by the definition $\mu_\omega(a)=\mathbb{E}_{P_0}(\omega_i m_{ai})/\mathbb{E}_{P_0}(\omega_i)$, we have $\mathbb{E}_{P_0}\left[\omega_i\left(m_{ai} - \mu_\omega(a)\right)\right] = 0$ for both $\omega_i = 1$ (cluster-ATE) and $\omega_i = N_i$ (individual-ATE). Hence $\mathbb{E}_{P_0}(\zeta_i) = 0$. Under Assumption 1 and independent cluster randomization, the $\zeta_i$ are mutually independent across clusters. In addition, Assumptions 1(c)--(d) and 5 imply, for the fixed true law $P_0$, that $\mathbb{E}_{P_0}(\zeta_i^2) \leq K_{\zeta,P_0} < \infty$. Therefore,
\begin{align*}
\mathbb{E}_{P_0} \mathbb{P}_M \left(|B| > C\epsilon_M \mid \boldsymbol{D}\right)
&= \mathbb{P}_{P_0}\left(\left|\left(\sum_{i=1}^M \omega_i\right)^{-1}\sum_{i=1}^M \zeta_i\right| > C\epsilon_M\right)\\
&\leq \mathbb{P}_{P_0}\left(\left|\sum_{i=1}^M \zeta_i\right| > C b_\omega M\epsilon_M\right)\\
&\leq \frac{\mathrm{Var}_{P_0}\left(\sum_{i=1}^M \zeta_i\right)}{C^2 b_\omega^2 M^2\epsilon_M^2}\\
&= \frac{\sum_{i=1}^M\mathrm{Var}_{P_0}(\zeta_i)}{C^2 b_\omega^2 M^2\epsilon_M^2}\\
&\leq \frac{K_{\zeta,P_0}}{C^2 b_\omega^2}\cdot\frac{1}{M\epsilon_M^2}.
\end{align*}
The equality holds because $B$ is a function of the observed data $\boldsymbol{D}$ and is therefore constant when conditioning on $\boldsymbol{D}$. The first inequality is an event inclusion following from $\sum_{i=1}^M\omega_i\geq b_\omega M$, and the second follows from Chebyshev's inequality applied to the centered numerator $\sum_i\zeta_i$. In particular, taking $\epsilon_M=M^{-1/2}$ gives
\[
\mathbb{P}_{P_0}\left(\sqrt{M}|B|>C\right)\leq \frac{K_{\zeta,P_0}}{C^2b_\omega^2},
\]
which can be made arbitrarily small by taking $C$ sufficiently large. Thus $B=O_p(M^{-1/2})$. The same bound also implies $B=o_p(\epsilon_M)$ for every sequence satisfying $\sqrt{M}\epsilon_M\to\infty$.

\paragraph{Contraction rate of component $A$}
Recall $A = \left( \sum_{i=1}^M \omega_i \right)^{-1} \sum_{i=1}^M \omega_i\, (\widehat{m}_{ai} - m_{ai})\left(1 - \frac{\mathbb{I}(A_i = a)}{\pi_a}\right)$, where the CRT randomization makes $\pi_a\in(0,1)$ known and the cluster weights satisfy $0<b_\omega\le \omega_i\le B_\omega<\infty$. Let $\mathcal{F}$ be a fixed pointwise measurable $P_0$-Donsker class with square-integrable envelope $F$ that contains the fitted outcome regression functions contributing to the posterior point estimator. Define the translated class $\mathcal{G} := \{g=f-m_a:\ f\in \mathcal F\}$, where $m_a(\bm{x},n):=\mathbb{E}\left(\overline{Y}_i \mid A_i=a, \bm{X}_i=\bm{x}, N_i=n\right)$. Since $m_a\in L_2(P_0)$ under Assumption 1(c), translation by the fixed function $m_a$ preserves the Donsker property; hence $\mathcal{G}$ is $P_0$-Donsker with a square-integrable envelope. Define the empirical and true measures
\begin{align*}
    P_M f := \frac{1}{M}\sum_{i=1}^M f(W_i), ~~~ P_0f := \E[f(W)], ~~~ W_i = (A_i,\bm{X}_i, N_i, \omega_i),
\end{align*}
and the empirical process $\G_M f := \sqrt{M}\{(P_M - P_0)f\}$. Under Assumption 1 and independent cluster randomization, the $W_i$ are independent and identically distributed. The weight $\omega_i$ is a function of $(\bm{X}_i, N_i)$, equal to $1$ for the cluster-ATE and to $N_i$ for the individual-ATE. Let $Z_i := 1 - \frac{\mathbb{I}(A_i=a)}{\pi_a}$, so that under cluster randomization $\E[Z_i \mid \bm{X}_i, N_i] = 0$ and $|Z_i| \leq \pi_a^{-1}$. For a fitted regression $\widehat m_a\in\mathcal F$, write $g_{\widehat m}:=\widehat m_a-m_a\in\mathcal G$. Dividing both the numerator and the denominator of $A$ by $M$, the component $A$ can be expressed as a ratio,
\begin{align*}
    A = \frac{P_M\left\{\omega\, Z\, g_{\widehat m}\right\}}{P_M \omega},
\end{align*}
and we treat the numerator and the denominator separately. For the numerator, define the index class
\begin{align*}
    \mathcal{H} := \{h_g(w) := \omega\, z\, g(\bm{x},n) : g \in \mathcal{G}\}.
\end{align*}
Since $\mathcal{G}$ is $P_0$-Donsker and the fixed multiplier $\omega Z$ is uniformly bounded, $|\omega Z|\le B_\omega/\pi_a$ by Assumption 5, the product class $\mathcal{H}$ is $P_0$-Donsker with a square-integrable envelope. Moreover, because $\E[Z \mid \bm{X}, N] = 0$ and both $\omega$ and $g$ are functions of $(\bm{X}, N)$, we have $P_0h_g = \E\{ \omega\,\E[Z\mid \bm{X}, N]\,g(\bm{X},N) \}=0$ for all $g \in \mathcal{G}$. Thus, for every $g\in\mathcal G$,
\begin{align*}
    P_M\left\{\omega Z g\right\} = (P_M - P_0) h_g = M^{-1/2} \G_M h_g.
\end{align*}
Because $\mathcal H$ is $P_0$-Donsker, $\G_M$ converges weakly in $\ell^\infty(\mathcal H)$ to a tight Gaussian process. Consequently,
\begin{align*}
    \sup_{h\in\mathcal H}|\G_M h|=O_p(1),
\end{align*}
and the bound is uniform over the possibly data-dependent index $g_{\widehat m}\in\mathcal G$. It follows that
\begin{align*}
    \left|P_M\left\{\omega Zg_{\widehat m}\right\}\right|
    \le M^{-1/2}\sup_{h\in\mathcal H}|\G_Mh|
    =O_p(M^{-1/2}).
\end{align*}
This uniform inequality also justifies posterior averaging. Specifically, if $g_\beta=\widehat m_a(\cdot;\beta)-m_a\in\mathcal G$ for the fitted regressions indexed by the posterior draws, then
\begin{align*}
    \left|\mathbb E_{\bm\beta\mid\bm D}P_M\left\{\omega Zg_\beta\right\}\right|
    &\le \mathbb E_{\bm\beta\mid\bm D}\left|P_M\left\{\omega Zg_\beta\right\}\right|\\
    &\le M^{-1/2}\sup_{h\in\mathcal H}|\G_Mh|
    =O_p(M^{-1/2}).
\end{align*}

For the denominator, Assumption 5 gives $P_M\omega\ge b_\omega$ for every $M$; equivalently, the law of large numbers gives $P_M\omega\to\E(\omega)\in[b_\omega,B_\omega]$ almost surely. Thus $(P_M\omega)^{-1}\le b_\omega^{-1}=O_p(1)$. Combining the numerator and denominator bounds yields
\begin{align*}
    A = O_p(M^{-1/2})(P_M\omega)^{-1} = O_p(M^{-1/2}).
\end{align*}
Combining the results for components $A$ and $B$, we conclude that $\widehat{\mu}_\omega^{\mathrm{mr}}(a) - {\mu}_\omega(a) = A + B = O_p(M^{-1/2})$, which establishes the overall root-$M$ consistency of the model-robust estimator.

\newpage
\clearpage

\section*{Appendix 2. Consistency of the calibrated variance estimator}

Throughout this appendix we adopt the notation, the decomposition $\widehat{\mu}_\omega^{\text{mr}}(a) - \mu_\omega(a) = A + B$, and Assumptions 1-5 introduced in the Preliminaries. Let $\Delta(\boldsymbol{D}, \bm{\beta})$ denote the model-robust estimator as a function of data and model parameters. The variance estimator is
\[
\widehat{V} = \operatorname{Var}_{\boldsymbol{D}^{(k)}} \left\{ \mathbb{E}_{\bm{\beta} \mid \boldsymbol{D}}[\Delta(\boldsymbol{D}^{(k)}, \bm{\beta})] \right\} + \operatorname{Var}_{\bm{\beta} \mid \boldsymbol{D}}[\Delta(\boldsymbol{D}, \bm{\beta})],
\]
and our target is
\[
V = \operatorname{Var}_{\boldsymbol{D}} \left\{ \mathbb{E}_{\bm{\beta} \mid \boldsymbol{D}}[\Delta(\boldsymbol{D}, \bm{\beta})] \right\}.
\]
Our goal is to show that $\widehat{V} - V = o_p(M^{-1})$ in a CRT, provided that the outcome model contracts at a sufficiently fast rate and the cluster weights are bounded away from zero and infinity. To achieve this, we analyze the asymptotic behavior of two main components separately. First, we show that $\operatorname{Var}_{\bm{\beta} \mid \bm{D}}[\Delta(\bm{D}, \bm{\beta})] = o_p(M^{-1})$. Next, we show that $\operatorname{Var}_{\bm{D}^{(k)}}\left\{\mathbb{E}_{\bm{\beta} \mid \bm{D}}[\Delta(\bm{D}^{(k)}, \bm{\beta})]\right\} = \operatorname{Var}_{\bm{D}}(B) + o_p(M^{-1})$, and that $\operatorname{Var}_{\bm{D}}\left\{\mathbb{E}_{\bm{\beta} \mid \bm{D}}[\Delta(\bm{D}, \bm{\beta})]\right\} = \operatorname{Var}_{\bm{D}}(B) + o_p(M^{-1})$. Together, these results imply that the proposed variance estimator is consistent.

\paragraph{Asymptotic Behavior of $\operatorname{Var}_{\bm{\beta} \mid \bm{D}}[\Delta(\bm{D}, \bm{\beta})]$}

For the CRT setting with known treatment mechanism, we have the decomposition
\[
\widehat{\mu}_\omega^{\text{mr}}(a) - {\mu}_\omega(a) = A + B.
\]
We will analyze $\operatorname{Var}_{\bm{\beta} \mid \bm{D}}[A]$ and $\operatorname{Var}_{\bm{\beta} \mid \bm{D}}[B]$. If each of these two components is $o_p(M^{-1})$, then the entire variance is of the same order, since the sum of variances and covariances remains within this rate. First, note that $B$ contains no unknown parameters (i.e., it depends only on observed outcomes and the true conditional mean $m_{ai}$), so $\operatorname{Var}_{\bm{\beta} \mid \bm{D}}[B] = 0$. Next, we analyze $\operatorname{Var}_{\bm{\beta} \mid \bm{D}}[A]$, which takes the form
\[
\operatorname{Var}_{\bm{\beta} \mid \bm{D}}[A] = \operatorname{Var}_{\bm{\beta} \mid \bm{D}} \left[ \left( \sum_{i=1}^M \omega_i \right)^{-1} \sum_{i=1}^M \omega_i \left(1 - \frac{\mathbb{I}(A_i = a)}{\pi_a} \right)(\widehat{m}_{ai} - m_{ai}) \right].
\]
Using the Taylor expansion of $\widehat{m}_{ai}$ around $\bm{\beta}^*$ (from Assumption 3), we get:
\[
\widehat{m}_{ai} - m_{ai} 
= d_1(\bm{X}_i, N_i, \bm{\beta}^*)^\top (\bm{\beta} - \bm{\beta}^*) + (\bm{\beta} - \bm{\beta}^*)^\top H_1(\bm{X}_i, N_i, \tilde{\bm{\beta}})(\bm{\beta} - \bm{\beta}^*).
\]
Substituting this into the variance and using the arithmetic mean and geometric mean inequality :
\begin{align*}
\operatorname{Var}_{\bm{\beta} \mid \bm{D}}[A] 
&= \operatorname{Var}_{\bm{\beta} \mid \bm{D}} \left[ \left( \sum_{i=1}^M \omega_i \right)^{-1} \sum_{i=1}^M \omega_i \left(1 - \frac{\mathbb{I}(A_i = a)}{\pi_a} \right) \times \left( d_1(\bm{X}_i, N_i, \bm{\beta}^*)^\top (\bm{\beta} - \bm{\beta}^*) + (\bm{\beta} - \bm{\beta}^*)^\top H_1(\bm{X}_i, N_i, \tilde{\bm{\beta}})(\bm{\beta} - \bm{\beta}^*) \right) \right] \\
&\leq 2 \operatorname{Var}_{\bm{\beta} \mid \bm{D}} \left[ \left( \sum_{i=1}^M \omega_i \right)^{-1} \sum_{i=1}^M \omega_i \left(1 - \frac{\mathbb{I}(A_i = a)}{\pi_a} \right) d_1(\bm{X}_i, N_i, \bm{\beta}^*)^\top (\bm{\beta} - \bm{\beta}^*) \right] \\
&\quad + 2 \operatorname{Var}_{\bm{\beta} \mid \bm{D}} \left[ \left( \sum_{i=1}^M \omega_i \right)^{-1} \sum_{i=1}^M \omega_i \left(1 - \frac{\mathbb{I}(A_i = a)}{\pi_a} \right) (\bm{\beta} - \bm{\beta}^*)^\top H_1(\bm{X}_i, N_i, \tilde{\bm{\beta}})(\bm{\beta} - \bm{\beta}^*) \right].
\end{align*}
We analyze the first variance term, $\operatorname{Var}_{\bm{\beta}\mid \bm{D}}\left[
\left( \sum_{i=1}^M \omega_i \right)^{-1}
\sum_{i=1}^M \omega_i \left(1 - \frac{\mathbb{I}(A_i=a)}{\pi_a}\right)\,d_1(\bm{X}_i,N_i,\bm{\beta}^*)^\top(\bm{\beta}-\bm{\beta}^*)
\right]$. By Assumption 5, the cluster weights satisfy \( b_\omega \leq \omega_i \leq B_\omega \) for all \( i \), where \( 0 < b_\omega < B_\omega < \infty \) are constants independent of \(M\). Therefore,
\[
b_\omega M \leq \sum_{i=1}^M \omega_i \leq B_\omega M,
\quad \text{which implies} \quad
\frac{1}{(B_\omega M)^2} \leq \left( \sum_{i=1}^M \omega_i \right)^{-2} \leq \frac{1}{(b_\omega M)^2}.
\]
Conditioning on the observed data \(\bm{D}\), the terms \(\left(1 - \frac{\mathbb{I}(A_i=a)}{\pi_a}\right)\) and \(d_1(\bm{X}_i,N_i,\bm{\beta}^*)\) are nonrandom. Applying Assumption~4, we can factor out \(\operatorname{Var}_{\bm{\beta}\mid \bm{D}}(\bm{\beta}-\bm{\beta}^*)\) and bound the variance by
\[
\frac{\lambda}{(b_\omega M)^2} \cdot
\left\| \sum_{i=1}^M \omega_i \left(1 - \frac{\mathbb{I}(A_i=a)}{\pi_a} \right) d_1(\bm{X}_i,N_i,\bm{\beta}^*) \right\|^2,
\]
where \(\lambda\) denotes the largest eigenvalue of 
\(\operatorname{Var}_{\bm{\beta}\mid \bm{D}}(\bm{\beta}-\bm{\beta}^*)\). Since the weighted sum inside the norm is \(O_p(\sqrt{M})\) by the multivariate central limit theorem, the entire expression is
\[
\operatorname{Var}_{\bm{\beta}\mid \bm{D}}\left[
\left( \sum_{i=1}^M \omega_i \right)^{-1}
\sum_{i=1}^M \omega_i \left(1 - \frac{\mathbb{I}(A_i=a)}{\pi_a} \right)\,
d_1(\bm{X}_i,N_i,\bm{\beta}^*)^\top(\bm{\beta}-\bm{\beta}^*)
\right] 
= o_p(M^{-1}).
\]
We now handle the second term.
$$
\begin{aligned}
&\operatorname{Var}_{\bm{\beta}\mid\bm{D}}\left[
\left( \sum_{i=1}^M \omega_i \right)^{-1} \sum_{i=1}^M  \omega_i  
\left(1 - \frac{\mathbb{I}(A_i=a)}{\pi_a}\right)
(\bm{\beta}-\bm{\beta}^*)^\top
H_1(\bm{X}_i,N_i,\widetilde{\bm{\beta}})
(\bm{\beta}-\bm{\beta}^*)
\right]\\
& \le 
\mathbb{E}_{\bm{\beta}\mid\bm{D}}\left[
\left(
\left( \sum_{i=1}^M \omega_i \right)^{-1} \sum_{i=1}^M  \omega_i  
\left(1 - \frac{\mathbb{I}(A_i=a)}{\pi_a}\right)
(\bm{\beta}-\bm{\beta}^*)^\top
H_1(\bm{X}_i,N_i,\widetilde{\bm{\beta}})
(\bm{\beta}-\bm{\beta}^*)
\right)^2
\right].
\end{aligned}
$$
Since $H_1(\cdot)$ has bounded operator norm and 
$\left(1 - \frac{\mathbb{I}(A_i=a)}{\pi_a}\right)$ is bounded by a positive constant, the term is 
bounded by a constant times $\|\bm{\beta}-\bm{\beta}^*\|_2^2$. Therefore, $\left((\bm{\beta}-\bm{\beta}^*)^\top
H_1(\bm{X}_i,N_i,\widetilde{\bm{\beta}})
(\bm{\beta}-\bm{\beta}^*)\right)^2
\le 
C\,\|\bm{\beta}-\bm{\beta}^*\|_2^4$ for some constant $C > 0$. Substituting and applying Cauchy-Schwarz inequality,
$$
\begin{aligned}
&\mathbb{E}_{\bm{\beta}\mid\bm{D}}\left[
\left(
\left( \sum_{i=1}^M \omega_i \right)^{-1} \sum_{i=1}^M  \omega_i  
\left(1 - \frac{\mathbb{I}(A_i=a)}{\pi_a}\right)
(\bm{\beta}-\bm{\beta}^*)^\top
H_1(\bm{X}_i,N_i,\widetilde{\bm{\beta}})
(\bm{\beta}-\bm{\beta}^*)
\right)^2
\right]\\
&\le \frac{1}{\left(b_{\omega}M\right)^2}\mathbb{E}_{\bm{\beta}\mid\bm{D}}\left[
\left(\sum_{i=1}^M \omega_i^2
\left(1 - \frac{\mathbb{I}(A_i=a)}{\pi_a}\right)^2\right)
\sum_{i=1}^M\left((\bm{\beta}-\bm{\beta}^*)^\top
H_1(\bm{X}_i,N_i,\widetilde{\bm{\beta}})
(\bm{\beta}-\bm{\beta}^*)
\right)^2
\right]\\
&\le 
\frac{C}{\left(b_{\omega}M\right)^2}\left(\sum_{i=1}^M \omega_i^2
\left(1 - \frac{\mathbb{I}(A_i=a)}{\pi_a}\right)^2\right) \sum_{i=1}^M \mathbb{E}_{\bm{\beta}\mid\bm{D}}\left[\|\bm{\beta}-\bm{\beta}^*\|_2^4\right]\\
& \le \frac{C^{\prime} }{M} \sum_{i=1}^M \mathbb{E}_{\bm{\beta}\mid\bm{D}}\left[\|\bm{\beta}-\bm{\beta}^*\|_2^4\right]\\
&=C^{\prime} \mathbb{E}_{\bm{\beta}\mid\bm{D}}\left[\|\bm{\beta}-\bm{\beta}^*\|_2^4\right].
\end{aligned}
$$
By Assumption 4, $\mathbb{E}_{\bm{\beta}\mid\bm{D}}\left[\|\bm{\beta}-\bm{\beta}^*\|_2^4\right] = o_p(M^{-1})$, so the entire expression is $o_p(M^{-1})$. Therefore,
$$\operatorname{Var}_{\bm{\beta}\mid\bm{D}}\left[
\left( \sum_{i=1}^M \omega_i \right)^{-1} \sum_{i=1}^M  \omega_i  
\left(1 - \frac{\mathbb{I}(A_i=a)}{\pi_a}\right)
(\bm{\beta}-\bm{\beta}^*)^\top
H_1(\bm{X}_i,N_i,\widetilde{\bm{\beta}})
(\bm{\beta}-\bm{\beta}^*)
\right]= o_p(M^{-1}).
$$
Putting these two terms together shows each variance term is $o_p(M^{-1})$, so $\operatorname{Var}_{\bm{\beta}\mid \bm{D}}[A] = o_p(M^{-1})$.
Since $\operatorname{Var}_{\bm{\beta}\mid \bm{D}}[B] = 0$, we conclude that
$$\operatorname{Var}_{\bm{\beta}\mid \bm{D}}[\Delta(\bm{D},\bm{\beta})] = o_p(M^{-1}).$$

\paragraph{Asymptotic Behavior of $\operatorname{Var}_{\boldsymbol{D}^{(k)}} \left\{ \mathbb{E}_{\bm{\beta} \mid \boldsymbol{D}}[\Delta(\boldsymbol{D}^{(k)}, \bm{\beta})] \right\}$}

Recall that we have
\[
\widehat{\mu}_\omega^{\text{mr}}(a) - {\mu}_\omega(a) = A + B.
\]
Our aim is to show that 
$$
\operatorname{Var}_{\boldsymbol{D}^{(k)}}\left\{\mathbb{E}_{\bm{\beta} \mid \boldsymbol{D}}[B]\right\}
\approx \operatorname{Var}_{\boldsymbol{D}}[B],
\quad \text{and} \quad \operatorname{Var}_{\boldsymbol{D}^{(k)}}\left\{\mathbb{E}_{\bm{\beta} \mid \boldsymbol{D}}[A]\right\} = o_p(M^{-1}).
$$
Together, these can imply that the total variance is $o_p(M^{-1})$. For $\operatorname{Var}_{\boldsymbol{D}^{(k)}}\left\{\mathbb{E}_{\bm{\beta} \mid \boldsymbol{D}}[B]\right\}$, we have 
$$\operatorname{Var}_{\boldsymbol{D}^{(k)}}\left\{\mathbb{E}_{\bm{\beta} \mid \boldsymbol{D}}[B]\right\}
= \operatorname{Var}_{\boldsymbol{D}^{(k)}}[B] 
= \operatorname{Var}_{\boldsymbol{D}^{(k)}}\left[
\left( \sum_{i=1}^M \omega_i^{(k)} \right)^{-1} \sum_{i=1}^M \omega_i^{(k)} \left[\frac{\mathbb{I}(A_i^{(k)} = a)}{\pi_a}
(\overline{Y}_i^{(k)} - m_{ai}^{(k)}) + m_{ai}^{(k)} - \mu_{\omega}(a)
\right]\right],$$
where $(\omega_i^{(k)}, A_i^{(k)}, \overline{Y}_i^{(k)}, m_{ai}^{(k)})$ come from the cluster‐level bootstrap sample $\boldsymbol{D}^{(k)}$. As $M \to \infty$, resampling from $\boldsymbol{D}^{(k)}$ converges to sampling from $\boldsymbol{D}$, so
\[
\operatorname{Var}_{\boldsymbol{D}^{(k)}}[B]
\approx
\operatorname{Var}_{\boldsymbol{D}}\left[
\left( \sum_{i=1}^M \omega_i \right)^{-1} \sum_{i=1}^M \omega_i \left[ \frac{\mathbb{I}(A_i = a)}{\pi_a}
(\overline{Y}_i - m_{ai}) + m_{ai} - \mu_{\omega}(a)
\right]\right]
= \operatorname{Var}_{\boldsymbol{D}}[B].
\]
Hence,
\[
\operatorname{Var}_{\boldsymbol{D}^{(k)}}\left\{\mathbb{E}_{\bm{\beta} \mid \boldsymbol{D}}[B]\right\}
\approx \operatorname{Var}_{\boldsymbol{D}}[B].
\]
This approximation follows from the cluster bootstrap’s consistency. For $\operatorname{Var}_{\boldsymbol{D}^{(k)}}\left\{
\mathbb{E}_{\bm{\beta} \mid \boldsymbol{D}}[A]
\right\}
=
\operatorname{Var}_{\boldsymbol{D}^{(k)}}\left\{
\mathbb{E}_{\bm{\beta} \mid \boldsymbol{D}} \left[
\left( \sum_{i=1}^M \omega_i^{(k)} \right)^{-1} \sum_{i=1}^M \omega_i^{(k)}  \left(1 - \frac{\mathbb{I}(A_i^{(k)} = a)}{\pi_a}\right)
(\widehat{m}_{ai}^{(k)} - m_{ai}^{(k)})
\right]
\right\}$, using a Taylor expansion of $\widehat{m}_{ai}^{(k)}$ around $m_{ai}^{(k)}$, together with Assumptions~3 (posterior contraction) and~4 (bounded posterior moments), we can obtain that $\operatorname{Var}_{\boldsymbol{D}^{(k)}}\left\{\mathbb{E}_{\bm{\beta} \mid \boldsymbol{D}}[A]\right\} = o_p(M^{-1})$ as demonstrated in the steps below.

\begin{align*}
\operatorname{Var}_{\boldsymbol{D}^{(k)}}\!\Bigl\{\mathbb{E}_{\bm{\beta} \mid \boldsymbol{D}}[A]\Bigr\}
&= \operatorname{Var}_{\boldsymbol{D}^{(k)}}\!\Biggl\{
     \mathbb{E}_{\bm{\beta} \mid \boldsymbol{D}}\Biggl[
       \left( \sum_{i=1}^M \omega_i^{(k)} \right)^{-1} \sum_{i=1}^M \omega_i^{(k)} 
       \Bigl(1 - \frac{\mathbb{I}(A_i^{(k)}=a)}{\pi_a}\Bigr)
       \bigl(\widehat{m}_{ai}^{(k)} - m_{ai}^{(k)}\bigr)
     \Biggr]
   \Biggr\} \\
&= \operatorname{Var}_{\boldsymbol{D}^{(k)}}\!\Biggl\{
     \mathbb{E}_{\bm{\beta} \mid \boldsymbol{D}}\Biggl[
       \left( \sum_{i=1}^M \omega_i^{(k)} \right)^{-1} \sum_{i=1}^M \omega_i^{(k)} 
       \Bigl(1 - \frac{\mathbb{I}(A_i^{(k)}=a)}{\pi_a}\Bigr) \\
     &\quad \times \Bigl(
         d_1\bigl(\bm{X}_i^{(k)}, N_i^{(k)}, \bm{\beta}^*\bigr)^\top(\bm{\beta} - \bm{\beta}^*) + (\bm{\beta}-\bm{\beta}^*)^\top H_1\bigl(\bm{X}_i^{(k)}, N_i^{(k)}, \tilde{\bm{\beta}}\bigr)(\bm{\beta} - \bm{\beta}^*)
       \Bigr)
     \Biggr]
   \Biggr\} \\
&\leq 2\operatorname{Var}_{\boldsymbol{D}^{(k)}}\!\Biggl\{
     \mathbb{E}_{\bm{\beta} \mid \boldsymbol{D}}\Biggl[
       \left( \sum_{i=1}^M \omega_i^{(k)} \right)^{-1} \sum_{i=1}^M \omega_i^{(k)} 
       \Bigl(1 - \frac{\mathbb{I}(A_i^{(k)}=a)}{\pi_a}\Bigr) \times d_1\bigl(\bm{X}_i^{(k)}, N_i^{(k)}, \bm{\beta}^*\bigr)^\top(\bm{\beta}-\bm{\beta}^*)
     \Biggr]
   \Biggr\} \\
&\quad + 2\operatorname{Var}_{\boldsymbol{D}^{(k)}}\!\Biggl\{
     \mathbb{E}_{\bm{\beta} \mid \boldsymbol{D}}\Biggl[
       \left( \sum_{i=1}^M \omega_i^{(k)} \right)^{-1} \sum_{i=1}^M \omega_i^{(k)} 
       \Bigl(1 - \frac{\mathbb{I}(A_i^{(k)}=a)}{\pi_a}\Bigr) \times (\bm{\beta}-\bm{\beta}^*)^\top H_1\bigl(\bm{X}_i^{(k)}, N_i^{(k)}, \tilde{\bm{\beta}}\bigr)(\bm{\beta}-\bm{\beta}^*)
     \Biggr]
   \Biggr\}.
\end{align*}
Then,
{\small
\setlength{\abovedisplayskip}{6pt}
\setlength{\belowdisplayskip}{6pt}
\begin{align*}
&\operatorname{Var}_{\boldsymbol{D}^{(k)}}\!\Biggl\{
     \mathbb{E}_{\bm{\beta} \mid \boldsymbol{D}}\Biggl[
       \left( \sum_{i=1}^M \omega_i^{(k)} \right)^{-1} \sum_{i=1}^M \omega_i^{(k)} 
       \Bigl(1 - \frac{\mathbb{I}(A_i^{(k)}=a)}{\pi_a}\Bigr) \times d_1\bigl(\bm{X}_i^{(k)}, N_i^{(k)}, \bm{\beta}^*\bigr)^\top(\bm{\beta}-\bm{\beta}^*)
     \Biggr]
   \Biggr\}\\
&=
\mathbb{E}_{\bm{\beta} \mid \boldsymbol{D}} \bigl[(\bm{\beta}-\bm{\beta}^*)^\top\bigr]\,
\operatorname{Var}_{\boldsymbol{D}^{(k)}}\!\Biggl\{
       \left( \sum_{i=1}^M \omega_i^{(k)} \right)^{-1} \sum_{i=1}^M \omega_i^{(k)} 
       \Bigl(1 - \frac{\mathbb{I}(A_i^{(k)}=a)}{\pi_a}\Bigr) \times d_1\bigl(\bm{X}_i^{(k)}, N_i^{(k)}, \bm{\beta}^*\bigr)^\top
\Biggr\}\,
\mathbb{E}_{\bm{\beta} \mid \boldsymbol{D}} [\bm{\beta}-\bm{\beta}^*]
\\
&=
\mathbb{E}_{\bm{\beta} \mid \boldsymbol{D}} \bigl[(\bm{\beta}-\bm{\beta}^*)^\top\bigr]\,
\left\{\mathbb{E}_{\boldsymbol{D}^{(k)}}\!\left[\left( \sum_{i=1}^M \omega_i^{(k)}\right)^{-2}\right]
\operatorname{Var}_{\boldsymbol{D}^{(k)}}\!\left\{
       \sum_{i=1}^M \omega_i^{(k)} \Bigl(1 - \frac{\mathbb{I}\bigl(A_i^{(k)}=a\bigr)}{\pi_a}\Bigr)
       d_1\Bigl(\bm{X}_i^{(k)},N_i^{(k)},\bm{\beta}^*\Bigr)^\top 
\right\} \right.\\
&\quad\quad + \left.\; \operatorname{Var}_{\boldsymbol{D}^{(k)}}\!\left[\left( \sum_{i=1}^M \omega_i^{(k)} \right)^{-1}\right]
\left\|\mathbb{E}_{\boldsymbol{D}^{(k)}}\!\left\{
       \sum_{i=1}^M \omega_i^{(k)} \Bigl(1 - \frac{\mathbb{I}\bigl(A_i^{(k)}=a\bigr)}{\pi_a}\Bigr)
       d_1\Bigl(\bm{X}_i^{(k)},N_i^{(k)},\bm{\beta}^*\Bigr)^\top
\right\}\right\|^2\right\}
\mathbb{E}_{\bm{\beta} \mid \boldsymbol{D}} [\bm{\beta}-\bm{\beta}^*]
\\ 
&=
M \cdot \mathbb{E}_{\bm{\beta} \mid \boldsymbol{D}} \bigl[(\bm{\beta}-\bm{\beta}^*)^\top\bigr]\,
\mathbb{E}_{\boldsymbol{D}^{(k)}}\!\left[\left( \sum_{i=1}^M \omega_i^{(k)}\right)^{-2}\right]
\underbrace{\operatorname{Var}_{\boldsymbol{D}^{(k)}}\!\left\{
       \frac{1}{\sqrt{M}}\sum_{i=1}^M \omega_i^{(k)} \Bigl(1 - \frac{\mathbb{I}\bigl(A_i^{(k)}=a\bigr)}{\pi_a}\Bigr)
       d_1\Bigl(\bm{X}_i^{(k)},N_i^{(k)},\bm{\beta}^*\Bigr)^\top 
\right\}}_{\textstyle\Sigma^{(k)}}\,\mathbb{E}_{\bm{\beta} \mid \boldsymbol{D}} [\bm{\beta}-\bm{\beta}^*]\\
&\quad\quad + M^2 \cdot \mathbb{E}_{\bm{\beta} \mid \boldsymbol{D}} \bigl[(\bm{\beta}-\bm{\beta}^*)^\top\bigr]\,\; \operatorname{Var}_{\boldsymbol{D}^{(k)}}\!\left[\left( \sum_{i=1}^M \omega_i^{(k)} \right)^{-1}\right]
\left\|\mathbb{E}_{\boldsymbol{D}^{(k)}}\!\left\{\frac{1}{M}
       \sum_{i=1}^M \omega_i^{(k)} \Bigl(1 - \frac{\mathbb{I}\bigl(A_i^{(k)}=a\bigr)}{\pi_a}\Bigr)
       d_1\Bigl(\bm{X}_i^{(k)},N_i^{(k)},\bm{\beta}^*\Bigr)^\top
\right\}\right\|^2
\mathbb{E}_{\bm{\beta} \mid \boldsymbol{D}} [\bm{\beta}-\bm{\beta}^*]
\\ 
& \le M \cdot 
\mathbb{E}_{\bm{\beta} \mid \boldsymbol{D}} \bigl[(\bm{\beta}-\bm{\beta}^*)^\top\bigr]\,
\frac{1}{\left(b_{\omega}M\right)^2} \lambda^{(k)}\mathbb{E}_{\bm{\beta} \mid \boldsymbol{D}} [\bm{\beta}-\bm{\beta}^*]+ M^2 \cdot O(M^{-3}) \cdot
O(1) \cdot \mathbb{E}_{\bm{\beta} \mid \boldsymbol{D}} \bigl[(\bm{\beta}-\bm{\beta}^*)^\top\bigr]\,\; \mathbb{E}_{\bm{\beta} \mid \boldsymbol{D}} [\bm{\beta}-\bm{\beta}^*]
\\ 
&\le \left(\frac{\lambda^{(k)}}{b_{\omega}^2M} +O(M^{-1})\right)
\mathbb{E}_{\bm{\beta} \mid \boldsymbol{D}} [(\bm{\beta}-\bm{\beta}^*)^\top]\,
\mathbb{E}_{\bm{\beta} \mid \boldsymbol{D}}[\bm{\beta}-\bm{\beta}^*]
\\
& =o_p\bigl(M^{-1}\bigr).
\end{align*}
\par}
$\lambda^{(k)}$ is the maximum eigenvalue of $\Sigma^{(k)}$. The third line follows from the law of total variance. The fifth line follows from the Delta method, Mean Value Theorem, and the boundedness of the relevant terms. The last line is implied by Assumption 4. Using the fact that the Hessian matrix is bounded, and Jensen’s inequality, we have the following
\begin{align*}
&\operatorname{Var}_{\boldsymbol{D}^{(k)}} \left\{
  \mathbb{E}_{\bm{\beta} \mid \boldsymbol{D}} \left[
    \left( \sum_{i=1}^M \omega_i^{(k)} \right)^{-1} \sum_{i=1}^M \omega_i^{(k)} 
    \Bigl(1 - \tfrac{\mathbb{I}(A_i^{(k)} = a)}{\pi_a}\Bigr)\,
    (\bm{\beta} - \bm{\beta}^*)^\top 
    H_1\bigl(\bm{X}_i^{(k)}, N_i^{(k)}, \tilde{\bm{\beta}}\bigr)
    (\bm{\beta} - \bm{\beta}^*)
  \right]
\right\}
\\
&=
\operatorname{Var}_{\boldsymbol{D}^{(k)}} \left\{
  \left( \sum_{i=1}^M \omega_i^{(k)} \right)^{-1} \sum_{i=1}^M \omega_i^{(k)} 
  \Bigl(1 - \tfrac{\mathbb{I}(A_i^{(k)} = a)}{\pi_a}\Bigr)\,
  \mathbb{E}_{\bm{\beta} \mid \boldsymbol{D}} \Bigl(
    (\bm{\beta} - \bm{\beta}^*)^\top 
    H_1(\cdot)\,
    (\bm{\beta} - \bm{\beta}^*)
  \Bigr)
\right\}
\\
&\le
\mathbb{E}_{\boldsymbol{D}^{(k)}} \left[
  \left( \sum_{i=1}^M \omega_i^{(k)} \right)^{-2} \left(\sum_{i=1}^M \omega_i^{(k)}
  \Bigl(1 - \tfrac{\mathbb{I}(A_i^{(k)} = a)}{\pi_a}\Bigr)\,
  \mathbb{E}_{\bm{\beta} \mid \boldsymbol{D}} \Bigl(
    (\bm{\beta} - \bm{\beta}^*)^\top 
    H_1(\cdot)\,
    (\bm{\beta} - \bm{\beta}^*)
  \Bigr)\right)^2
\right]
\\
&\le \tfrac{1}{\left(b_{\omega}M\right)^2}
\mathbb{E}_{\boldsymbol{D}^{(k)}} \left[ \left(\sum_{i=1}^M \omega_i^{(k)}
  \Bigl(1 - \tfrac{\mathbb{I}(A_i^{(k)} = a)}{\pi_a}\Bigr)\,
  \mathbb{E}_{\bm{\beta} \mid \boldsymbol{D}} \Bigl(
    (\bm{\beta} - \bm{\beta}^*)^\top 
    H_1(\cdot)\,
    (\bm{\beta} - \bm{\beta}^*)
  \Bigr)\right)^2
\right]
\\
&\le \tfrac{1}{\left(b_{\omega}M\right)^2}
\mathbb{E}_{\boldsymbol{D}^{(k)}} \left[
\sum_{i=1}^M \left(\omega_i^{(k)}\right)^2
  \Bigl(1 - \tfrac{\mathbb{I}(A_i^{(k)} = a)}{\pi_a}\Bigr)^2\, \sum_{i=1}^M  \Bigl[
  \mathbb{E}_{\bm{\beta} \mid \boldsymbol{D}}
    \bigl( 
      (\bm{\beta} - \bm{\beta}^*)^\top 
      H_1(\cdot)\, 
      (\bm{\beta} - \bm{\beta}^*)
    \bigr)\Bigr]^2\right]
\\
&\le \tfrac{1}{\left(b_{\omega}M\right)^2}
\mathbb{E}_{\boldsymbol{D}^{(k)}} \left[
\sum_{i=1}^M \left(\omega_i^{(k)}\right)^2
  \Bigl(1 - \tfrac{\mathbb{I}(A_i^{(k)} = a)}{\pi_a}\Bigr)^2\, \sum_{i=1}^M  \Bigl[
  \mathbb{E}_{\bm{\beta} \mid \boldsymbol{D}}
    \bigl( \lambda
      (\bm{\beta} - \bm{\beta}^*)^\top 
      (\bm{\beta} - \bm{\beta}^*)
    \bigr)\Bigr]^2\right]
\\
&\le
\frac{C}{M^2}
\left(\sum_{i=1}^M\mathbb{E}_{\bm{\beta} \mid \boldsymbol{D}} \bigl[ \| \bm{\beta} - \bm{\beta}^* \|_2^2 \bigr]^2\right)
\mathbb{E}_{\boldsymbol{D}^{(k)}} \left[
  \sum_{i=1}^M \left(\omega_i^{(k)}\right)^2
  \left(1 - \tfrac{\mathbb{I}(A_i^{(k)} = a)}{\pi_a}\right)^2
\right]
\\
&= C^{\prime} \mathbb{E}_{\bm{\beta} \mid \boldsymbol{D}} \bigl[ \| \bm{\beta} - \bm{\beta}^* \|_2^2 \bigr]^2
\\
&= o_p\bigl(M^{-1}\bigr).
\end{align*}
\(\lambda\) denotes the largest eigenvalue of the Hessian matrix, and \(C\) and \(C'\) are positive constants. Combining this result with our preceding bounds yields
\[
\operatorname{Var}_{\boldsymbol{D}^{(k)}}\!\Bigl\{\mathbb{E}_{\bm{\beta}\mid \boldsymbol{D}}\bigl[\Delta\bigl(\boldsymbol{D}^{(k)}, \bm{\beta}\bigr)\bigr]\Bigr\}
=\operatorname{Var}_{\boldsymbol{D}}[B] + o_p\bigl(M^{-1}\bigr).
\]

\paragraph{Asymptotic behavior of $\operatorname{Var}_{\boldsymbol{D}} \left\{ \mathbb{E}_{\bm{\beta} \mid \boldsymbol{D}}[\Delta(\boldsymbol{D}, \bm{\beta})] \right\}$}

Reuse the expansion, we have
\begin{align*}
&\operatorname{Var}_{\boldsymbol{D}} \left\{ \mathbb{E}_{\bm{\beta} \mid \boldsymbol{D}}[A] \right\}\\
&= \operatorname{Var}_{\boldsymbol{D}} \left\{ \mathbb{E}_{\bm{\beta} \mid \boldsymbol{D}} \left[ 
\left( \sum_{i=1}^M \omega_i \right)^{-1} \sum_{i=1}^M \omega_i \left( 1 - \frac{\mathbb{I}(A_i = a)}{\pi_a} \right)(\widehat{m}_{ai} - m_{ai})
\right] \right\} \\
&= \operatorname{Var}_{\boldsymbol{D}} \left\{ \mathbb{E}_{\bm{\beta} \mid \boldsymbol{D}} \left[
\left( \sum_{i=1}^M \omega_i \right)^{-1} \sum_{i=1}^M \omega_i \left( 1 - \frac{\mathbb{I}(A_i = a)}{\pi_a} \right)
\left( d_1(\bm{X}_i, N_i, \bm{\beta}^*)^\top (\bm{\beta} - \bm{\beta}^*)
+ (\bm{\beta} - \bm{\beta}^*)^\top H_1(\bm{X}_i, N_i, \tilde{\bm{\beta}})(\bm{\beta} - \bm{\beta}^*)
\right)
\right] \right\} \\
&\leq 2 \operatorname{Var}_{\boldsymbol{D}} \left\{ \mathbb{E}_{\bm{\beta} \mid \boldsymbol{D}} \left[
\left( \sum_{i=1}^M \omega_i \right)^{-1} \sum_{i=1}^M \omega_i \left( 1 - \frac{\mathbb{I}(A_i = a)}{\pi_a} \right)
d_1(\bm{X}_i, N_i, \bm{\beta}^*)^\top (\bm{\beta} - \bm{\beta}^*)
\right] \right\} \\
&\quad + 2 \operatorname{Var}_{\boldsymbol{D}} \left\{ \mathbb{E}_{\bm{\beta} \mid \boldsymbol{D}} \left[
\left( \sum_{i=1}^M \omega_i \right)^{-1} \sum_{i=1}^M \omega_i \left( 1 - \frac{\mathbb{I}(A_i = a)}{\pi_a} \right)
(\bm{\beta} - \bm{\beta}^*)^\top H_1(\bm{X}_i, N_i, \tilde{\bm{\beta}})(\bm{\beta} - \bm{\beta}^*)
\right] \right\}
\end{align*}
We now handle the two variance terms in the upper bound of $\operatorname{Var}_{\boldsymbol{D}}\left\{ \mathbb{E}_{\bm{\beta} \mid \boldsymbol{D}}[A] \right\}$ separately. For the first term, we have
\begin{align*}
&\operatorname{Var}_{\boldsymbol{D}} \left\{ \mathbb{E}_{\bm{\beta} \mid \boldsymbol{D}} \left[ 
\left( \sum_{i=1}^M \omega_i \right)^{-1} \sum_{i=1}^M \omega_i \left(1 - \frac{\mathbb{I}(A_i = a)}{\pi_a} \right) 
d_1(\bm{X}_i, N_i, \bm{\beta}^*)^\top (\bm{\beta} - \bm{\beta}^*) \right] \right\} \\
&\le \mathbb{E}_{\boldsymbol{D}} \left\{ \left\| \left( \sum_{i=1}^M \omega_i \right)^{-1} \sum_{i=1}^M \omega_i \left(1 - \frac{\mathbb{I}(A_i = a)}{\pi_a} \right) 
d_1(\bm{X}_i, N_i, \bm{\beta}^*)^\top \mathbb{E}_{\bm{\beta} \mid \boldsymbol{D}}(\bm{\beta} - \bm{\beta}^*) \right\|^2 \right\} \\
&\le \frac{1}{\left(b_{\omega}M\right)^2} \mathbb{E}_{\boldsymbol{D}} \left\{ \left\| \sum_{i=1}^M \omega_i\left(1 - \frac{\mathbb{I}(A_i = a)}{\pi_a} \right) 
d_1(\bm{X}_i, N_i, \bm{\beta}^*)^\top \mathbb{E}_{\bm{\beta} \mid \boldsymbol{D}}(\bm{\beta} - \bm{\beta}^*) \right\|^2 \right\} \\
&= \frac{1}{\left(b_{\omega}M\right)^2} \mathbb{E}_{\boldsymbol{D}} \left\{ \left( \sum_{i=1}^M \omega_i \left(1 - \frac{\mathbb{I}(A_i = a)}{\pi_a} \right) 
d_1(\bm{X}_i, N_i, \bm{\beta}^*)^\top \right) \mathbb{E}_{\bm{\beta} \mid \boldsymbol{D}}(\bm{\beta} - \bm{\beta}^*) \right. \\
&\qquad \left. \times \mathbb{E}_{\bm{\beta} \mid \boldsymbol{D}}(\bm{\beta} - \bm{\beta}^*)^\top 
\left( \sum_{i=1}^M \omega_i \left(1 - \frac{\mathbb{I}(A_i = a)}{\pi_a} \right) 
d_1(\bm{X}_i, N_i, \bm{\beta}^*) \right) \right\} \\
&\le \frac{\lambda}{b_{\omega}^2M} \mathbb{E}_{\boldsymbol{D}} \left\{ \left( \frac{1}{\sqrt{M}}\sum_{i=1}^M \omega_i \left(1 - \frac{\mathbb{I}(A_i = a)}{\pi_a} \right) 
d_1(\bm{X}_i, N_i, \bm{\beta}^*)^\top \right) \left( \frac{1}{\sqrt{M}} \sum_{i=1}^M \omega_i \left(1 - \frac{\mathbb{I}(A_i = a)}{\pi_a} \right) 
d_1(\bm{X}_i, N_i, \bm{\beta}^*) \right) \right\} \\
&= \frac{\lambda}{b_{\omega}^2M}\mathbb{E}_{\boldsymbol{D}} \left\| \frac{1}{\sqrt{M}}\sum_{i=1}^M \omega_i\left(1 - \frac{\mathbb{I}(A_i = a)}{\pi_a} \right) 
d_1(\bm{X}_i, N_i, \bm{\beta}^*)^\top \right\|^2 \\
&\le \frac{\lambda}{b_{\omega}^2M} o_p(1) \\
&= o_p(M^{-1}),
\end{align*}
where $\lambda$ denotes the maximum eigenvalue of 
$\mathbb{E}_{\bm{\beta} \mid \boldsymbol{D}}(\bm{\beta} - \bm{\beta}^*)\mathbb{E}_{\bm{\beta} \mid \boldsymbol{D}}(\bm{\beta} - \bm{\beta}^*)^\top$.
The final result follows by applying Assumption 4 and the multivariate Central Limit Theorem. Now we focus on the second of the two terms.

\begin{align*}
&\operatorname{Var}_{\boldsymbol{D}} \left\{ \mathbb{E}_{\bm{\beta} \mid \boldsymbol{D}} \left[ \left( \sum_{i=1}^M \omega_i \right)^{-1} \sum_{i=1}^M \omega_i \left(1 - \frac{\mathbb{I}(A_i = a)}{\pi_a} \right) 
(\bm{\beta} - \bm{\beta}^*)^\top H_1(\bm{X}_i, N_i, \tilde{\bm{\beta}})(\bm{\beta} - \bm{\beta}^*) \right] \right\} \\
&= \operatorname{Var}_{\boldsymbol{D}} \left\{ \left( \sum_{i=1}^M \omega_i \right)^{-1} \sum_{i=1}^M \omega_i \left(1 - \frac{\mathbb{I}(A_i = a)}{\pi_a} \right) 
\mathbb{E}_{\bm{\beta} \mid \boldsymbol{D}} \left( (\bm{\beta} - \bm{\beta}^*)^\top H_1(\bm{X}_i, N_i, \tilde{\bm{\beta}})(\bm{\beta} - \bm{\beta}^*) \right) \right\} \\
&\le \mathbb{E}_{\boldsymbol{D}} \left\{ \left[ \left( \sum_{i=1}^M \omega_i \right)^{-1} \sum_{i=1}^M \omega_i \left(1 - \frac{\mathbb{I}(A_i = a)}{\pi_a} \right) 
\mathbb{E}_{\bm{\beta} \mid \boldsymbol{D}} \left( (\bm{\beta} - \bm{\beta}^*)^\top H_1(\bm{X}_i, N_i, \tilde{\bm{\beta}})(\bm{\beta} - \bm{\beta}^*) \right) \right]^2 \right\} \\
&\le \frac{1}{b_{\omega}^2M^2} \mathbb{E}_{\boldsymbol{D}} \left\{ \left( \sum_{i=1}^M \omega_i^2 \left(1 - \frac{\mathbb{I}(A_i = a)}{\pi_a} \right)^2 \right)
\left( \sum_{i=1}^M \mathbb{E}_{\bm{\beta} \mid \boldsymbol{D}}^2 \left[ (\bm{\beta} - \bm{\beta}^*)^\top H_1(\bm{X}_i, N_i, \tilde{\bm{\beta}})(\bm{\beta} - \bm{\beta}^*) \right] \right) \right\} \\
&\le \frac{1}{b_{\omega}^2M^2} \mathbb{E}_{\boldsymbol{D}} \left\{\left[\sum_{i=1}^M \omega_i^2\left(1 - \frac{\mathbb{I}(A_i = a)}{\pi_a} \right)^2 \right]^2\right\}^{1/2}
\mathbb{E}_{\boldsymbol{D}} \left[ \left( \sum_{i=1}^M \mathbb{E}_{\bm{\beta} \mid \boldsymbol{D}}^2 \left[ (\bm{\beta} - \bm{\beta}^*)^\top H_1(\bm{X}_i, N_i, \tilde{\bm{\beta}})(\bm{\beta} - \bm{\beta}^*) \right] \right)^2 \right]^{1/2} \\
&\le \frac{C}{M} \mathbb{E}_{\boldsymbol{D}} \left[ \left( \sum_{i=1}^M \mathbb{E}_{\bm{\beta} \mid \boldsymbol{D}}^2 \left[ (\bm{\beta} - \bm{\beta}^*)^\top H_1(\bm{X}_i, N_i, \tilde{\bm{\beta}})(\bm{\beta} - \bm{\beta}^*) \right] \right)^2 \right]^{1/2} \\
&\le \frac{C}{M}  \mathbb{E}_{\boldsymbol{D}} \left[ \left(  \sum_{i=1}^M \mathbb{E}_{\bm{\beta} \mid \boldsymbol{D}}^2 \left[ \lambda \| \bm{\beta} - \bm{\beta}^* \|_2^2 \right] \right)^2 \right]^{1/2} \\
&= \frac{C^{\prime}}{M} \mathbb{E}_{\boldsymbol{D}} \left[ \left(  \sum_{i=1}^M \mathbb{E}_{\bm{\beta} \mid \boldsymbol{D}}^2 \left[ \| \bm{\beta} - \bm{\beta}^* \|_2^2 \right]  \right)^2 \right]^{1/2} \\
&= C^{\prime} \mathbb{E}_{\boldsymbol{D}} \left[ \mathbb{E}_{\bm{\beta} \mid \boldsymbol{D}}^4 \left[ \| \bm{\beta} - \bm{\beta}^* \|_2^2 \right] \right]^{1/2}\\
&\le C^{\prime} \mathbb{E}_{\boldsymbol{D}} \left[ \mathbb{E}_{\bm{\beta} \mid \boldsymbol{D}} \left[ \| \bm{\beta} - \bm{\beta}^* \|_2^8 \right] \right]^{1/2}\\
&= o_p(M^{-1})
\end{align*}
Therefore, we have established that $\operatorname{Var}_{\boldsymbol{D}}\{\mathbb{E}_{\bm{\beta} \mid \boldsymbol{D}}[A]\}$ is $o_p(M^{-1})$, while $\operatorname{Var}_{\boldsymbol{D}}\{\mathbb{E}_{\bm{\beta} \mid \boldsymbol{D}}[B]\} = \operatorname{Var}_{\boldsymbol{D}}[B]$. Therefore, 
\[
\operatorname{Var}_{\boldsymbol{D}}\left\{\mathbb{E}_{\bm{\beta} \mid \boldsymbol{D}}[\Delta(D, \bm{\beta})]\right\} = \operatorname{Var}_{\boldsymbol{D}}[B] + o_p(M^{-1}).
\]
Combining this with the results from previous sections, we conclude that $$\widehat{V}- V = \operatorname{Var}_{\boldsymbol{D}^{(k)}} \left\{ \mathbb{E}_{\bm{\beta} \mid \boldsymbol{D}}[\Delta(\boldsymbol{D}^{(k)}, \bm{\beta})] \right\} + \operatorname{Var}_{\bm{\beta} \mid \boldsymbol{D}}[\Delta(D, \bm{\beta})] - \operatorname{Var}_{\boldsymbol{D}} \left\{ \mathbb{E}_{\bm{\beta} \mid \boldsymbol{D}}[\Delta(\boldsymbol{D}, \bm{\beta})] \right\} = o_p(M^{-1}),$$ which completes the proof.

\paragraph{Remark: the order of the leading variance component}

The proof above does not require the order of $\operatorname{Var}_{\boldsymbol{D}}[B]$ itself, since this term appears in both $\widehat{V}$ and $V$ and cancels in the difference. Nevertheless, it is useful to record its order, as it quantifies the dominant source of variability in the model-robust estimator. Define
\[
\zeta_i \equiv \omega_i\left[\frac{\mathbb{I}(A_i=a)}{\pi_a}\left(\overline{Y}_i - m_{ai}\right) + m_{ai} - \mu_\omega(a)\right],
\qquad\text{so that}\qquad
B = \left(\sum_{i=1}^M \omega_i\right)^{-1}\sum_{i=1}^M \zeta_i.
\]
Under cluster randomization, $A_i$ is independent of $(\bm{X}_i, N_i, \bm{Y}_i(a))$, so that
$\mathbb{E}_{P_0}\left[\omega_i \frac{\mathbb{I}(A_i=a)}{\pi_a}(\overline{Y}_i - m_{ai})\right] = 0$. Moreover, by the definition of $\mu_\omega(a)$, we have $\mathbb{E}_{P_0}\left[\omega_i\left(m_{ai} - \mu_\omega(a)\right)\right] = 0$ for both $\omega_i = 1$ (cluster-ATE) and $\omega_i = N_i$ (individual-ATE). Hence $\mathbb{E}_{P_0}(\zeta_i) = 0$. By Assumption 1, the $\zeta_i$ are mutually independent across clusters, and by Assumptions 1(c)--(d) and 5, they have uniformly bounded second moments, $\mathbb{E}_{P_0}(\zeta_i^2) \leq K_\zeta < \infty$. Since Assumption 5 further gives $\sum_{i=1}^M \omega_i \geq b_\omega M$, it follows that
\[
\operatorname{Var}_{\boldsymbol{D}}[B]
\;\leq\; \mathbb{E}_{P_0}\left[\left(\left(\sum_{i=1}^M \omega_i\right)^{-1}\sum_{i=1}^M \zeta_i\right)^{2}\right]
\;\leq\; \frac{1}{(b_\omega M)^2}\,\mathbb{E}_{P_0}\left[\left(\sum_{i=1}^M \zeta_i\right)^{2}\right]
\;=\; \frac{1}{(b_\omega M)^2}\sum_{i=1}^M \mathbb{E}_{P_0}(\zeta_i^2)
\;\leq\; \frac{K_\zeta}{b_\omega^{2}}\cdot\frac{1}{M}
\;=\; O(M^{-1}),
\]
where the equality uses the independence and mean-zero property of the $\zeta_i$. Therefore, the cluster-level sampling variability is of order $M^{-1}$, whereas the posterior parameter variance $\operatorname{Var}_{\bm{\beta}\mid\boldsymbol{D}}[\Delta(\boldsymbol{D},\bm{\beta})]$ was shown above to be $o_p(M^{-1})$. The latter is asymptotically negligible relative to the former, so uncalibrated posterior inference, which retains only the latter, discards the dominant component of the variance.

\newpage
\clearpage

\renewcommand{\thetable}{\Alph{table}}
\setcounter{table}{0}

\section*{Appendix 3. Simulation Results for the Second Experiment Across Three Scenarios}
\begin{table}[ht]
  \centering
    \caption{Results from the second simulation experiment based on $1,000$ replications for scenario (i), estimating the cluster-average treatment effect, $\Delta_C=0.590$, and the individual-average treatment effect, $\Delta_I=0.625$. A linear mixed model, which adjusts for covariates with main effects and treatment-covariate interactions, is used as the fitted model. For the calibrated variance estimation method, $100$ resampled datasets are used within each simulation to compute posterior variance. $M$ denotes the number of clusters. ``Relative Bias'' refers to the relative bias of the estimated value expressed as a percentage, with standard errors shown in parentheses. ``Coverage'' refers to the $95$\% interval coverage across all simulated datasets expressed as a percentage. ``MCSD'' denotes Monte Carlo standard deviation. ``AESE'' denotes average of estimated standard error. ``RE'' refers to the relative efficiency of the proposed estimator (either the g-computation estimator or the model-robust estimator) with covariate adjustment, compared to the unadjusted moment estimator.
    }
{
\begin{tabular}{ccrrrrrrr}
    \toprule
    Estimator & Posterior Inference & M & Estimand & Relative Bias (SE) & Coverage & MCSD & AESE & RE\\
    \midrule
    
\multirow{4}{*}{g-computation} & \multirow{4}{*}{Uncalibrated} 
& \multirow{2}{*}{30} & cluster-ATE & -2.1 (1.2) & 94.5 & 0.224 & 0.244 & 15.666\\
&&& individual-ATE & -2.0 (1.2) & 94.4 & 0.237 & 0.251 & 20.035\\
&& \multirow{2}{*}{60} & cluster-ATE & -2.3 (0.8) & 94.0 & 0.145 & 0.145 & 19.643\\
&&& individual-ATE & -2.4 (0.8) & 93.7 & 0.155 & 0.152 & 24.731\\
\midrule
\multirow{4}{*}{g-computation} & \multirow{4}{*}{Calibrated} 
& \multirow{2}{*}{30} & cluster-ATE & -2.1 (1.2) & 96.1 & 0.224 & 0.256 & 15.666\\
&&& individual-ATE & -2.0 (1.2) & 95.7 & 0.237 & 0.262 & 20.035\\
&& \multirow{2}{*}{60} & cluster-ATE & -2.3 (0.8) & 95.3 & 0.145 & 0.149 & 19.643\\
&&& individual-ATE & -2.4 (0.8) & 94.6 & 0.155 & 0.156 & 24.731\\
\midrule
\multirow{4}{*}{Model-robust} & \multirow{4}{*}{Uncalibrated} 
& \multirow{2}{*}{30} & cluster-ATE & -2.2 (1.2) & 50.4 & 0.223 & 0.112 & 15.806\\
&&& individual-ATE & -2.1 (1.2) & 51.2 & 0.235 & 0.110 & 20.378\\
&& \multirow{2}{*}{60} & cluster-ATE & -2.3 (0.8) & 38.6 & 0.142 & 0.038 & 20.482\\
&&& individual-ATE & -2.4 (0.8) & 37.3 & 0.151 & 0.039 & 26.059\\
\midrule
\multirow{4}{*}{Model-robust} & \multirow{4}{*}{Calibrated} 
& \multirow{2}{*}{30} & cluster-ATE & -2.2 (1.2) & 92.6 & 0.223 & 0.228 & 15.806\\
&&& individual-ATE & -2.1 (1.2) & 92.0 & 0.235 & 0.232 & 20.378\\
&& \multirow{2}{*}{60} & cluster-ATE & -2.3 (0.8) & 94.0 & 0.142 & 0.139 & 20.482\\
&&& individual-ATE & -2.4 (0.8) & 93.6 & 0.151 & 0.145 & 26.059\\
\bottomrule
\end{tabular}
}
  \label{tab:Antonelli_tsc1_nsim_1000}
\end{table}

\begin{table}[ht]
  \centering
    \caption{  
    Results from the second simulation experiment based on $1,000$ replications for scenario (ii), estimating the cluster-average treatment effect, $\Delta_C=0.679$, and the individual-average treatment effect, $\Delta_I=0.842$. A linear mixed model, which adjusts for covariates with main effects and treatment-covariate interactions, is used as the fitted model. For the calibrated variance estimation method, $100$ resampled datasets are used within each simulation to compute posterior variance. $M$ denotes the number of clusters. ``Relative Bias'' refers to the relative bias of the estimated value expressed as a percentage, with standard errors shown in parentheses. ``Coverage'' refers to the $95$\% interval coverage across all simulated datasets expressed as a percentage. ``MCSD'' denotes Monte Carlo standard deviation. ``AESE'' denotes average of estimated standard error. ``RE'' refers to the relative efficiency of the proposed estimator (either the g-computation estimator or the model-robust estimator) with covariate adjustment, compared to the unadjusted moment estimator.
    }
    {
    \begin{tabular}{ccrrrrrrr}
    \toprule
    Estimator & Posterior Inference & M & Estimand & Relative Bias (SE) & Coverage & MCSD & AESE & RE \\
    \midrule
    
\multirow{4}{*}{g-computation} & \multirow{4}{*}{Uncalibrated} 
& \multirow{2}{*}{30} & cluster-ATE & -2.6 (1.6) & 82.8 & 0.347 & 0.257 & 1.404 \\
&&& individual-ATE & -3.3 (1.3) & 81.9 & 0.358 & 0.263 & 1.517 \\
&& \multirow{2}{*}{60} & cluster-ATE & -2.3 (1.1) & 79.9 & 0.231 & 0.154 & 1.629 \\
&&& individual-ATE & -3.2 (0.9) & 80.9 & 0.240 & 0.160 & 1.738 \\

\midrule
\multirow{4}{*}{g-computation} & \multirow{4}{*}{Calibrated} 
& \multirow{2}{*}{30} & cluster-ATE & -2.6 (1.6) & 94.8 & 0.347 & 0.363 & 1.404 \\
&&& individual-ATE & -3.3 (1.3) & 94.8 & 0.358 & 0.373 & 1.517 \\
&& \multirow{2}{*}{60} & cluster-ATE & -2.3 (1.1) & 95.2 & 0.231 & 0.231 & 1.629 \\
&&& individual-ATE & -3.2 (0.9) & 94.2 & 0.240 & 0.242 & 1.738 \\

\midrule
\multirow{4}{*}{Model-robust} & \multirow{4}{*}{Uncalibrated} 
& \multirow{2}{*}{30} & cluster-ATE & -2.8 (1.6) & 38.4 & 0.346 & 0.117 & 1.408 \\
&&& individual-ATE & -3.4 (1.3) & 39.7 & 0.357 & 0.116 & 1.524 \\
&& \multirow{2}{*}{60} & cluster-ATE & -2.3 (1.1) & 27.1 & 0.230 & 0.041 & 1.649 \\
&&& individual-ATE & -3.1 (0.9) & 27.6 & 0.238 & 0.042 & 1.763 \\

\midrule
\multirow{4}{*}{Model-robust} & \multirow{4}{*}{Calibrated} 
& \multirow{2}{*}{30} & cluster-ATE & -2.8 (1.6) & 93.4 & 0.346 & 0.342 & 1.408 \\
&&& individual-ATE & -3.4 (1.3) & 93.6 & 0.357 & 0.349 & 1.524 \\
&& \multirow{2}{*}{60} & cluster-ATE & -2.3 (1.1) & 94.5 & 0.230 & 0.225 & 1.649 \\
&&& individual-ATE & -3.1 (0.9) & 93.6 & 0.238 & 0.233 & 1.763 \\
\bottomrule
\end{tabular}

}
  \label{tab:Antonelli_tsc3_nsim_1000}
\end{table}

\begin{table}[ht]
  \centering
    \caption{
    Results from the second simulation experiment based on $1,000$ replications for scenario (iii), estimating the cluster-average treatment effect, $\Delta_C=0.637$, and the individual-average treatment effect, $\Delta_I=0.754$. A linear mixed model, which adjusts for covariates with main effects and treatment-covariate interactions, is used as the fitted model. For the calibrated variance estimation method, $100$ resampled datasets are used within each simulation to compute posterior variance. $M$ denotes the number of clusters. ``Relative Bias'' refers to the relative bias of the estimated value expressed as a percentage, with standard errors shown in parentheses. ``Coverage'' refers to the $95$\% interval coverage across all simulated datasets expressed as a percentage. ``MCSD'' denotes Monte Carlo standard deviation. ``AESE'' denotes average of estimated standard error. ``RE'' refers to the relative efficiency of the proposed estimator (either the g-computation estimator or the model-robust estimator) with covariate adjustment, compared to the unadjusted moment estimator.}
    \begin{tabular}{ccrrrrrrr}
    \toprule
    Estimator & Posterior Inference & M & Estimand & Relative Bias (SE) & Coverage & MCSD & AESE & RE \\
    \midrule

    \multirow{4}{*}{g-computation} & \multirow{4}{*}{Uncalibrated} 
& \multirow{2}{*}{30} & cluster-ATE & -27.0 (23.9) & 85.0 & 4.810 & 3.831 & 1.050 \\
&&& individual-ATE & -40.1 (24.3) & 80.0 & 5.787 & 4.216 & 1.168 \\
&& \multirow{2}{*}{60} & cluster-ATE & 1.7 (17.1) & 83.8 & 3.446 & 2.670 & 1.193 \\
&&& individual-ATE & -9.4 (17.6) & 76.5 & 4.203 & 2.880 & 1.299 \\

\midrule
\multirow{4}{*}{g-computation} & \multirow{4}{*}{Calibrated} 
& \multirow{2}{*}{30} & cluster-ATE & -27.0 (23.9) & 87.1 & 4.810 & 3.995 & 1.050 \\
&&& individual-ATE & -40.1 (24.3) & 84.6 & 5.787 & 4.379 & 1.168 \\
&& \multirow{2}{*}{60} & cluster-ATE & 1.7 (17.1) & 86.8 & 3.446 & 2.733 & 1.193 \\
&&& individual-ATE & -9.4 (17.6) & 80.8 & 4.203 & 2.942 & 1.299 \\

\midrule
\multirow{4}{*}{Model-robust} & \multirow{4}{*}{Uncalibrated} 
& \multirow{2}{*}{30} & cluster-ATE & -28.0 (21.3) & 47.3 & 4.293 & 1.447 & 1.318 \\
&&& individual-ATE & -34.1 (22.7) & 40.0 & 5.419 & 1.532 & 1.331 \\
&& \multirow{2}{*}{60} & cluster-ATE & -0.7 (15.6) & 31.9 & 3.145 & 0.688 & 1.432 \\
&&& individual-ATE & -2.3 (16.9) & 26.6 & 4.030 & 0.719 & 1.413 \\

\midrule
\multirow{4}{*}{Model-robust} & \multirow{4}{*}{Calibrated} 
& \multirow{2}{*}{30} & cluster-ATE & -28.0 (21.3) & 93.0 & 4.293 & 4.154 & 1.318 \\
&&& individual-ATE & -34.1 (22.7) & 91.2 & 5.419 & 5.091 & 1.331 \\
&& \multirow{2}{*}{60} & cluster-ATE & -0.7 (15.6) & 93.4 & 3.145 & 2.921 & 1.432 \\
&&& individual-ATE & -2.3 (16.9) & 91.8 & 4.030 & 3.689 & 1.413 \\
\bottomrule
\end{tabular}

  \label{tab:Antonelli_tsc4_nsim1000}
\end{table}

\newpage
\clearpage
\section*{Appendix 4. BART simulation results with informative cluster size}
\begin{table}[ht]
  \centering
    \caption{
    BART-based results under varying scenarios with informative cluster sizes, based on $250$ simulations, estimating the cluster-average treatment effect, $\Delta_C$, and the individual-average treatment effect, $\Delta_I$. A BART model, which adjusts for covariates with main effects and treatment-covariate interactions, is used as the fitted model. For the calibrated variance estimation method, $100$ resampled datasets are used within each simulation to compute posterior variance. ``Relative Bias'' refers to the relative bias of the estimated value expressed as a percentage, with standard errors shown in parentheses. ``Coverage'' refers to the $95$\% interval coverage across all simulated datasets expressed as a percentage. ``MCSD'' denotes Monte Carlo standard deviation. ``AESE'' denotes average of estimated standard error.}
    \begin{tabular}{cccrrrrrr}

\toprule
Scenario & Estimator & Posterior Inference & M & Estimand & Relative Bias (SE) & Coverage & MCSD & AESE \\
\midrule
\multirow{12}{*}{(ii)} & \multirow{6}{*}{g-computation} & \multirow{6}{*}{Uncalibrated} & \multirow{2}{*}{30} 
& cluster-ATE & -8.4 (3.6) & 76.8 & 0.384 & 0.243 \\
&&&& individual-ATE & -12.7 (3.0) & 75.6 & 0.395 & 0.247 \\
&&& \multirow{2}{*}{60} 
& cluster-ATE & -0.3 (2.0) & 83.6 & 0.218 & 0.155 \\
&&&& individual-ATE & -5.1 (1.7) & 84.0 & 0.226 & 0.157 \\
&&& \multirow{2}{*}{90} 
& cluster-ATE & -2.8 (1.6) & 81.6 & 0.177 & 0.121 \\
&&&& individual-ATE & -5.3 (1.4) & 77.6 & 0.188 & 0.123 \\
\cmidrule{2-9}
& \multirow{6}{*}{Model-robust} & \multirow{6}{*}{Calibrated} & \multirow{2}{*}{30} 
& cluster-ATE & -7.9 (3.7) & 84.8 & 0.392 & 0.285 \\
&&&& individual-ATE & -10.3 (3.1) & 82.4 & 0.407 & 0.292 \\
&&& \multirow{2}{*}{60} 
& cluster-ATE & 0.0 (2.0) & 92.4 & 0.219 & 0.205 \\
&&&& individual-ATE & -3.3 (1.7) & 92.4 & 0.225 & 0.215 \\
&&& \multirow{2}{*}{90} 
& cluster-ATE & -2.7 (1.7) & 91.2 & 0.180 & 0.170 \\
&&&& individual-ATE & -3.5 (1.5) & 90.8 & 0.195 & 0.178 \\
\midrule
\multirow{12}{*}{(iii)} & \multirow{6}{*}{g-computation} & \multirow{6}{*}{Uncalibrated} & \multirow{2}{*}{30} 
& cluster-ATE & -54.6 (37.7) & 89.2 & 3.796 & 3.318 \\
&&&& individual-ATE & -59.8 (34.0) & 88.0 & 4.057 & 3.354 \\
&&& \multirow{2}{*}{60} 
& cluster-ATE & 35.1 (23.1) & 92.8 & 2.327 & 2.190 \\
&&&& individual-ATE & 22.3 (21.1) & 91.2 & 2.513 & 2.217 \\
&&& \multirow{2}{*}{90} 
& cluster-ATE & -12.3 (15.5) & 94.0 & 1.560 & 1.628 \\
&&&& individual-ATE & -20.4 (14.2) & 93.6 & 1.696 & 1.658 \\
\cmidrule{2-9}
& \multirow{6}{*}{Model-robust} & \multirow{6}{*}{Calibrated} & \multirow{2}{*}{30} 
& cluster-ATE & -40.1 (41.5) & 89.6 & 4.176 & 3.363 \\
&&&& individual-ATE & -44.8 (43.3) & 87.2 & 5.166 & 4.029 \\
&&& \multirow{2}{*}{60} 
& cluster-ATE & 39.8 (24.2) & 91.6 & 2.435 & 2.074 \\
&&&& individual-ATE & 36.9 (25.4) & 89.6 & 3.030 & 2.515 \\
&&& \multirow{2}{*}{90} 
& cluster-ATE & -6.6 (16.0) & 91.6 & 1.616 & 1.489 \\
&&&& individual-ATE & -7.0 (16.6) & 93.6 & 1.979 & 1.803 \\
\bottomrule
\end{tabular}
  \label{tab:BART_nsim250}
\end{table}

\newpage
\clearpage
\section*{Appendix 5. Simulation results without informative cluster size}

To guarantee there is no informative cluster sizes, we defined our data-generating process as follows. Let $(N_i, C_{i1}, C_{i2})$, for $i=1, \ldots, M$, be independent draws from distribution $\mathcal{P}^N \times \mathcal{P}^{C_1} \times \mathcal{P}^{C_2 | C_1}$, where $\mathcal{P}^N$ is uniform over support $(10,50)$, $\mathcal{P}^{C_1} = \mathcal{N}(3,4)$, $\mathcal{P}^{C_2 | C_1} = \mathcal{B}\left[\operatorname{expit}\left\{\log (3) \times C_1\right\}\right]$ is a Bernoulli distribution with $\operatorname{expit}(x) = \left(1+e^{-x}\right)^{-1}$. Next, for each individual, we generated the two individual-level covariates from $X_{ij1} \sim \mathcal{B}\left(0.6\right)$, and $X_{ij2} \sim \mathcal{N}\left\{X_{ij1} \times \left(2 C_{i2}-1\right), 9\right\}$. Potential outcomes $Y_{ij}(A_i)$ are generated from normal distribution $Y_{ij}(A_i) \sim \mathcal{N}\left(\psi_{ij}(A_i), 1\right)$, where
$\psi_{ij}(A_i) \sim \mathcal{N}\left(\eta_{ij}(A_i),\sigma_{\phi}^2\right)$, and $\eta_{ij}(A_i)$ is the fixed effect portion of the outcome-generating model, $\eta_{ij}(A_i)= -0.1 - \frac{0.3}{1 + \exp\left\{ -6 \times \left(30 + X_{ij1} + X_{ij2} \right) \right\}} + 0.2 X_{ij1} X_{ij2} + 0.3 \times \left( 3 + X_{ij1} \right) A_i - \frac{1}{1 + \exp\left\{ -4(X_{ij1} + X_{ij2}) \right\}} A_i$. Let $\sigma_{\phi}^2 = 0.25$, which results in a ICC of $0.2$. The observed outcomes are generated by independently sampling $A_i \sim \mathcal{B}(0.5)$ and defining them as $Y_{ij}=A_i Y_{ij}(1)+\left(1-A_i\right) Y_{ij}(0)$. Tables \ref{tab:LMM_NoICS} and \ref{tab:BART_NoICS} summarize the simulation results under this scenario, evaluating the performance of LMM and BART, respectively, in estimating both the cluster-ATE and individual-ATE.

\begin{table}[ht]
  \centering
    \caption{LMM-based results under the scenario with no informative cluster size, based on $250$ simulations, estimating the cluster-average treatment effect, $\Delta_C=0.54$, and the individual-average treatment effect, $\Delta_I=0.54$. A linear mixed model, which adjusts for covariates with main effects and treatment-covariate interactions, is used as the fitted model. For the calibrated variance estimation method, $100$ resampled datasets are used within each simulation to compute posterior variance. ``Relative Bias'' refers to the relative bias of the estimated value expressed as a percentage, with standard errors shown in parentheses. ``Coverage'' refers to the $95$\% interval coverage across all simulated datasets expressed as a percentage. ``MCSD'' denotes Monte Carlo standard deviation. ``AESE'' denotes average of estimated standard error.}
    \begin{tabular}{ccrrrrrr}

\toprule

Estimator & Posterior Inference & M & Estimand & Relative Bias (SE) & Coverage & MCSD & AESE \\
\midrule
\multirow{6}{*}{g-computation} & \multirow{6}{*}{Uncalibrated} & \multirow{2}{*}{30} 
& cluster-ATE & -3.6 (2.8) & 91.6 & 0.240 & 0.250 \\
&&& individual-ATE & -2.3 (3.0) & 93.6 & 0.253 & 0.261 \\
&& \multirow{2}{*}{60} 
& cluster-ATE & -1.4 (1.7) & 95.6 & 0.146 & 0.151 \\
&&& individual-ATE & -2.5 (1.9) & 95.2 & 0.158 & 0.157 \\
&& \multirow{2}{*}{90} 
& cluster-ATE & -0.9 (1.4) & 94.8 & 0.123 & 0.123 \\
&&& individual-ATE & -0.7 (1.5) & 95.6 & 0.126 & 0.129 \\
\midrule
\multirow{6}{*}{Model-robust} & \multirow{6}{*}{Calibrated} & \multirow{2}{*}{30} 
& cluster-ATE & -3.4 (2.8) & 91.6 & 0.238 & 0.233 \\
&&& individual-ATE & -2.3 (2.9) & 89.6 & 0.248 & 0.242 \\
&& \multirow{2}{*}{60} 
& cluster-ATE & -0.8 (1.7) & 95.2 & 0.144 & 0.146 \\
&&& individual-ATE & -2.1 (1.8) & 94.4 & 0.156 & 0.151 \\
&& \multirow{2}{*}{90} 
& cluster-ATE & -0.6 (1.4) & 94.0 & 0.123 & 0.120 \\
&&& individual-ATE & -0.4 (1.5) & 94.4 & 0.125 & 0.124 \\
\bottomrule
\end{tabular}
  \label{tab:LMM_NoICS}
\end{table}

\begin{table}[ht]
  \centering
    \caption{BART-based results under the scenario with no informative cluster size, based on $250$ simulations, estimating the cluster-average treatment effect, $\Delta_C=0.54$, and the individual-average treatment effect, $\Delta_I=0.54$. A BART model, which adjusts for covariates with main effects and treatment-covariate interactions, is used as the fitted model. For the calibrated variance estimation method, $100$ resampled datasets are used within each simulation to compute posterior variance. ``Relative Bias'' refers to the relative bias of the estimated value expressed as a percentage, with standard errors shown in parentheses. ``Coverage'' refers to the $95$\% interval coverage across all simulated datasets expressed as a percentage. ``MCSD'' denotes Monte Carlo standard deviation. ``AESE'' denotes average of estimated standard error.}
    \begin{tabular}{ccrrrrrr}

\toprule
Estimator & Posterior Inference & M & Estimand & Relative Bias (SE) & Coverage & MCSD & AESE \\
\midrule
\multirow{6}{*}{g-computation} & \multirow{6}{*}{Uncalibrated} & \multirow{2}{*}{30} 
& cluster-ATE & -2.0 (2.4) & 92.8 & 0.205 & 0.193 \\
&&& individual-ATE & -1.9 (2.4) & 93.2 & 0.206 & 0.193 \\
&& \multirow{2}{*}{60} 
& cluster-ATE & -1.7 (1.5) & 96.0 & 0.131 & 0.137 \\
&&& individual-ATE & -1.6 (1.5) & 96.0 & 0.131 & 0.137 \\
&& \multirow{2}{*}{90} 
& cluster-ATE & -0.3 (1.3) & 94.4 & 0.113 & 0.113 \\
&&& individual-ATE & -0.3 (1.3) & 94.8 & 0.113 & 0.113 \\
\midrule
\multirow{6}{*}{Model-robust} & \multirow{6}{*}{Calibrated} & \multirow{2}{*}{30} 
& cluster-ATE & -1.4 (2.4) & 91.6 & 0.207 & 0.192 \\
&&& individual-ATE & -0.9 (2.6) & 91.2 & 0.218 & 0.202 \\
&& \multirow{2}{*}{60} 
& cluster-ATE & -1.1 (1.5) & 95.6 & 0.131 & 0.137 \\
&&& individual-ATE & -2.4 (1.7) & 94.0 & 0.144 & 0.142 \\
&& \multirow{2}{*}{90} 
& cluster-ATE & -0.1 (1.3) & 93.2 & 0.113 & 0.114 \\
&&& individual-ATE & 0.3 (1.4) & 93.6 & 0.119 & 0.119 \\
\bottomrule
\end{tabular}
  \label{tab:BART_NoICS}
\end{table}

\newpage
\clearpage
\section*{Appendix 6. Comparison of relative efficiency from the second simulation experiment}

\begin{figure}[ht]
    \centering
    \includegraphics[width=1\textwidth]{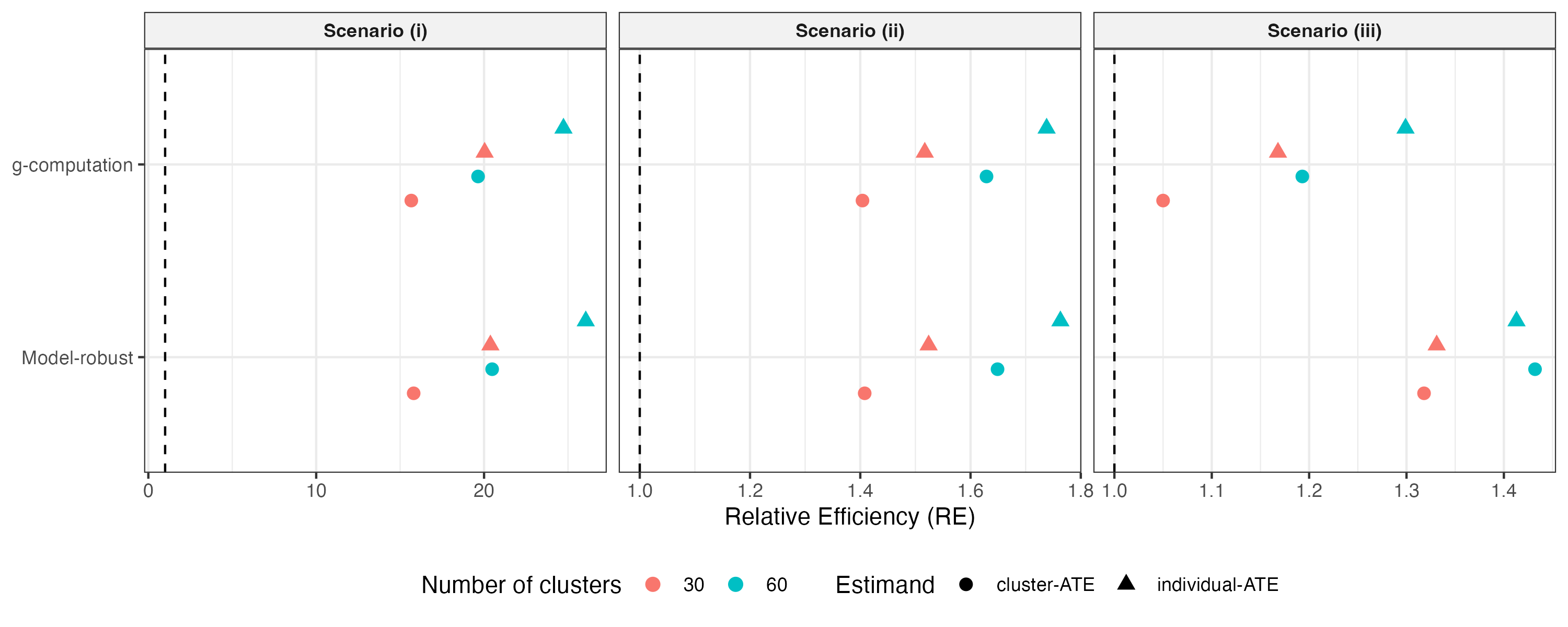}
    \caption{Relative efficiency (RE) from the second simulation experiment based on $1,000$ replications for scenario (i), (ii), and (iii), estimating the cluster-average treatment effect (cluster-ATE) and individual-average treatment effect (individual-ATE). The point estimation methods are categorized into g-computation and model-robust approaches. A linear mixed model, which adjusts for covariates with main effects and treatment-covariate interactions, is used as the fitted model. Shapes distinguish estimands (circles for cluster-ATE and triangles for individual-ATE), and colors indicate the number of clusters ($M=30$ in red, $M=60$ in blue). The horizontal dashed line at $\text{RE}=1$ marks the reference value where the proposed estimator (either the g-computation estimator or the model-robust estimator) with covariate adjustment has the same efficiency as the unadjusted moment estimator; values above $1$ indicate higher efficiency.}
    \label{fig:second simulation experiment RE}
\end{figure}

\newpage
\clearpage
\section*{Appendix 7. Comparison with frequentist model-robust standardization}

We compare our Bayesian approach against the frequentist model-robust standardization (MRS) estimator of \citet{li2025model}, applied to two different working outcome models: a generalized estimating equations (GEE) model with an exchangeable working correlation structure, and a linear mixed model (LMM) with a random intercept. In both cases, the MRS estimator standardizes the output of the fitted working model and uses a cluster jackknife for variance estimation, providing estimand-aligned inference that is consistent regardless of whether the working model is correctly specified. This comparison helps clarify what the Bayesian calibrated variance estimator offers relative to an established frequentist benchmark, and how the choice of working model affects performance.

For both the LMM and GEE working models, we adjust for all individual-level covariates ($X_{ij1}$, $X_{ij2}$), cluster-level covariates ($C_{i1}$, $C_{i2}$), and cluster size ($N_i$), matching the covariate set used by the Bayesian working model; the GEE working model additionally uses an exchangeable working correlation structure. We consider scenarios (ii) and (iii) and report results across $M=30$, $60$, and $90$ clusters. All aspects of the data-generating process are identical to those reported in Appendix 4.

\begin{table}[ht]
  \centering
    \caption{Frequentist model-robust standardization (MRS) results under scenarios (ii) and (iii), adjusting for all individual-level covariates, cluster-level covariates, and cluster size in the working model, based on $250$ simulations. ``GEE MRS'' refers to the MRS estimator of \citet{li2025model} applied to a GEE outcome model with an exchangeable working correlation structure; ``LMM MRS'' refers to the same MRS estimator applied to a linear mixed model with a random intercept. Both use a cluster jackknife variance estimator. ``Relative Bias'' refers to the relative bias of the estimated value expressed as a percentage, with standard errors shown in parentheses. ``Coverage'' refers to the $95$\% confidence interval coverage across all simulated datasets expressed as a percentage. ``MCSD'' denotes Monte Carlo standard deviation. ``AESE'' denotes average of estimated standard error.}
    \begin{tabular}{cccrrrrrr}
\toprule
Scenario & Estimator & Variance & M & Estimand & Relative Bias (SE) & Coverage & MCSD & AESE \\
\midrule
\multirow{8}{*}{(ii)} 
& \multirow{4}{*}{LMM MRS} & \multirow{4}{*}{Jackknife} & \multirow{2}{*}{30} 
& cluster-ATE & -9.1 (3.6) & 95.2 & 0.392 & 0.446 \\
&&&& individual-ATE & -7.6 (3.1) & 95.6 & 0.410 & 0.452 \\
&&& \multirow{2}{*}{60} 
& cluster-ATE & -0.3 (2.0) & 96.0 & 0.217 & 0.240 \\
&&&& individual-ATE & -2.4 (1.6) & 96.8 & 0.219 & 0.251 \\
\cmidrule(l){2-9}
& \multirow{4}{*}{GEE MRS} & \multirow{4}{*}{Jackknife} & \multirow{2}{*}{30}
& cluster-ATE & -9.1 (3.6) & 95.2 & 0.392 & 0.446 \\
&&&& individual-ATE & -7.6 (3.1) & 95.6 & 0.410 & 0.453 \\
&&& \multirow{2}{*}{60} 
& cluster-ATE & -0.3 (2.0) & 96.0 & 0.217 & 0.240 \\
&&&& individual-ATE & -2.4 (1.6) & 96.8 & 0.219 & 0.251 \\
\midrule
\multirow{8}{*}{(iii)} 
& \multirow{4}{*}{LMM MRS} & \multirow{4}{*}{Jackknife} & \multirow{2}{*}{30} 
& cluster-ATE & -22.9 (54.1) & 97.2 & 5.449 & 7.060 \\
&&&& individual-ATE & -34.5 (57.7) & 96.0 & 6.884 & 7.775 \\
&&& \multirow{2}{*}{60} 
& cluster-ATE & 37.5 (33.1) & 95.2 & 3.335 & 3.330 \\
&&&& individual-ATE & 35.8 (35.9) & 94.4 & 4.281 & 4.140 \\
\cmidrule(l){2-9}
& \multirow{4}{*}{GEE MRS} & \multirow{4}{*}{Jackknife} & \multirow{2}{*}{30} 
& cluster-ATE & -23.0 (54.1) & 97.2 & 5.449 & 7.062 \\
&&&& individual-ATE & -34.6 (57.7) & 96.0 & 6.883 & 7.775 \\
&&& \multirow{2}{*}{60} 
& cluster-ATE & 37.5 (33.1) & 95.2 & 3.334 & 3.329 \\
&&&& individual-ATE & 35.8 (35.9) & 94.4 & 4.280 & 4.139 \\
\bottomrule
\end{tabular}
  \label{tab:freq_compare_tsc23}
\end{table}

\newpage
\clearpage

\section*{Appendix 8. Sensitivity analysis: BART performance under scenario (iii) with covariates generated independently of cluster size}
To further probe the comparatively weaker performance of BART under scenario (iii), we investigate whether it is driven by the dependence of the covariate-generating process on cluster size. Specifically, we simplify the data-generating process so that the cluster-level covariates $C_{i1}$ and $C_{i2}$ and the individual-level covariates $X_{ij1}$ and $X_{ij2}$ are generated independently of the cluster size $N_i$, while the potential outcome model remains identical to that of scenario (iii). Because the potential outcomes continue to depend on $N_i$, informative cluster size is retained, and the cluster-ATE and individual-ATE remain distinct. Concretely, we let $N_i \sim U(10,50)$, $C_{i1} \sim \mathcal{N}(3,4)$, $C_{i2} \mid C_{i1} \sim \mathcal{B}[\operatorname{expit}\{\log(3) \times C_{i1}\}]$, $X_{ij1} \sim \mathcal{B}(0.6)$, and $X_{ij2} \sim \mathcal{N}\{X_{ij1} \times (2C_{i2}-1), 9\}$, with the potential outcome model unchanged from scenario (iii). All other aspects of the estimation procedure remain identical to those reported in Appendix 4. We focus exclusively on scenario (iii), as it is the most complex data-generating process, posing the greatest challenge for the working models. Under this simplified covariate-generating process, the true cluster-ATE and individual-ATE equal $0.625$ and $0.703$, respectively.

\begin{table}[ht]
  \centering
    \caption{BART-based results under scenario (iii) with covariates generated independently of cluster size, based on $250$ simulations. A BART model, which adjusts for all individual-level covariates, cluster-level covariates, and cluster size, is used as the fitted model. For the calibrated variance estimation method, $100$ resampled datasets are used within each simulation to compute posterior variance. ``Relative Bias'' refers to the relative bias of the estimated value expressed as a percentage, with standard errors shown in parentheses. ``Coverage'' refers to the $95$\% interval coverage across all simulated datasets expressed as a percentage. ``MCSD'' denotes Monte Carlo standard deviation. ``AESE'' denotes average of estimated standard error.}
    \begin{tabular}{ccrrrrrr}
\toprule
Estimator & Posterior Inference & M & Estimand & Relative Bias (SE) & Coverage & MCSD & AESE \\
\midrule
\multirow{6}{*}{g-computation} & \multirow{6}{*}{Uncalibrated} & \multirow{2}{*}{30} 
& cluster-ATE & -7.4 (35.4) & 92.8 & 3.504 & 3.301 \\
&&& individual-ATE & -17.5 (33.2) & 91.6 & 3.687 & 3.333 \\
&& \multirow{2}{*}{60} 
& cluster-ATE & -2.8 (19.9) & 94.8 & 1.965 & 2.248 \\
&&& individual-ATE & -13.4 (18.8) & 95.2 & 2.089 & 2.282 \\
&& \multirow{2}{*}{90} 
& cluster-ATE & -5.4 (16.4) & 94.8 & 1.622 & 1.623 \\
&&& individual-ATE & -11.6 (16.4) & 93.6 & 1.818 & 1.652 \\
\midrule
\multirow{6}{*}{Model-robust} & \multirow{6}{*}{Calibrated} & \multirow{2}{*}{30} 
& cluster-ATE & 5.6 (38.4) & 90.0 & 3.798 & 3.378 \\
&&& individual-ATE & 7.7 (42.5) & 86.4 & 4.729 & 4.078 \\
&& \multirow{2}{*}{60}
& cluster-ATE & 5.6 (20.6) & 94.8 & 2.038 & 2.154 \\
&&& individual-ATE & -5.0 (22.2) & 94.8 & 2.463 & 2.666 \\
&& \multirow{2}{*}{90} 
& cluster-ATE & 1.0 (16.8) & 92.0 & 1.656 & 1.489 \\
&&& individual-ATE & 1.0 (18.9) & 91.6 & 2.098 & 1.854 \\
\bottomrule
\end{tabular}
  \label{tab:BART_indepN_iii}
\end{table}

\newpage
\clearpage
\section*{Appendix 9. Arm-stratified resampling for the calibrated variance estimator}

In the main paper, the calibrated variance estimator generates bootstrap datasets by resampling clusters with replacement from the empirical distribution of observed clusters, reusing the posterior draws from the original data across bootstrap datasets. A natural question is whether treated and control clusters should be resampled separately so as to preserve the original treatment allocation. Resampling clusters without regard to arm allows the treated-to-control ratio to vary across bootstrap datasets and, in small samples, can occasionally yield datasets that are severely imbalanced in one arm. To respect the design of the cluster-randomized trial, we consider an \emph{arm-stratified} resampling scheme that fixes the number of treated and control clusters in every bootstrap dataset.

Let $M_1$ and $M_0$ denote the numbers of treated and control clusters in the observed data, with $M_1 + M_0 = M$. Each bootstrap dataset is constructed by (i) sampling $M_1$ clusters with replacement from the $M_1$ observed treated clusters, (ii) sampling $M_0$ clusters with replacement from the $M_0$ observed control clusters, and (iii) combining the two resampled sets. This preserves the observed treatment allocation $(M_1, M_0)$ in every bootstrap dataset, in contrast to the unstratified scheme used in the main text, which fixes only the total number of clusters $M$. As before, the posterior draws obtained from the original data are reused across bootstrap datasets, so no additional model fitting is required. We evaluate the arm-stratified scheme using the BART working model under the model-robust calibrated estimator, focusing on scenarios (ii) and (iii), the most complex data-generating processes and hence those for which the resampling scheme is most likely to matter. All other aspects of the data-generating process and estimation procedure are identical to those in Appendix 4. 

For CRTs with more structured randomization, the resampling may need to mirror the design used in the trial. In a stratified CRT, clusters would be resampled with replacement within each randomization stratum, preserving the arm allocation within strata. In a pair-matched CRT, the matched pair is the natural resampling unit. One should resample pairs with replacement and retain both clusters within each sampled pair, so that the matched structure is preserved. For covariate-constrained randomization, we need to account for the replication of the constrained design under resampling. A practical approach is to restrict resampling to allocations that satisfy the same balance criterion, i.e., those whose balance score, computed from the constraining covariates, falls within the acceptable range, so that the resampled datasets remain compatible with the design constraints. A detailed evaluation of these extensions is left to future work.

\begin{table}[ht]
  \centering
    \caption{
    BART-based model-robust calibrated results under the unstratified and arm-stratified resampling schemes, for scenarios (ii) and (iii), based on $250$ simulations. A BART model is used as the fitted model. For each calibrated variance estimate, $100$ resampled datasets are used within each simulation. The unstratified scheme resamples clusters with replacement while fixing only the total number of clusters $M$; the arm-stratified scheme resamples treated and control clusters separately, preserving the observed treatment allocation. ``Relative Bias'' refers to the relative bias of the estimated value expressed as a percentage, with standard errors shown in parentheses. ``Coverage'' refers to the $95$\% interval coverage across all simulated datasets expressed as a percentage. ``MCSD'' denotes Monte Carlo standard deviation. ``AESE'' denotes average of estimated standard error.}
    \begin{tabular}{cccrrrrr}
\toprule
Scenario & Resampling & M & Estimand & Relative Bias (SE) & Coverage & MCSD & AESE \\
\midrule
\multirow{12}{*}{(ii)} & \multirow{6}{*}{Unstratified} & \multirow{2}{*}{30} 
& cluster-ATE & -7.9 (3.7) & 84.8 & 0.392 & 0.285 \\
&&& individual-ATE & -10.3 (3.1) & 82.4 & 0.407 & 0.292 \\
&& \multirow{2}{*}{60} 
& cluster-ATE & 0.0 (2.0) & 92.4 & 0.219 & 0.205 \\
&&& individual-ATE & -3.3 (1.7) & 92.4 & 0.225 & 0.215 \\
&& \multirow{2}{*}{90} 
& cluster-ATE & -2.7 (1.7) & 91.2 & 0.180 & 0.170 \\
&&& individual-ATE & -3.5 (1.5) & 90.8 & 0.195 & 0.178 \\
\cmidrule{2-8}
& \multirow{6}{*}{Arm-stratified} & \multirow{2}{*}{30} 
& cluster-ATE & -7.9 (3.7) & 84.8 & 0.392 & 0.284 \\
&&& individual-ATE & -10.3 (3.1) & 82.0 & 0.407 & 0.291 \\
&& \multirow{2}{*}{60} 
& cluster-ATE & 0.0 (2.0) & 93.2 & 0.219 & 0.203 \\
&&& individual-ATE & -3.3 (1.7) & 91.2 & 0.225 & 0.214 \\
&& \multirow{2}{*}{90} 
& cluster-ATE & -2.7 (1.7) & 91.6 & 0.180 & 0.169 \\
&&& individual-ATE & -3.5 (1.5) & 90.8 & 0.195 & 0.177 \\
\midrule
\multirow{12}{*}{(iii)} & \multirow{6}{*}{Unstratified} & \multirow{2}{*}{30} 
& cluster-ATE & -40.1 (41.5) & 89.6 & 4.176 & 3.363 \\
&&& individual-ATE & -44.8 (43.3) & 87.2 & 5.166 & 4.029 \\
&& \multirow{2}{*}{60} 
& cluster-ATE & 39.8 (24.2) & 91.6 & 2.435 & 2.074 \\
&&& individual-ATE & 36.9 (25.4) & 89.6 & 3.030 & 2.515 \\
&& \multirow{2}{*}{90} 
& cluster-ATE & -6.6 (16.0) & 91.6 & 1.616 & 1.489 \\
&&& individual-ATE & -7.0 (16.6) & 93.6 & 1.979 & 1.803 \\
\cmidrule{2-8}
& \multirow{6}{*}{Arm-stratified} & \multirow{2}{*}{30} 
& cluster-ATE & -40.1 (41.5) & 91.2 & 4.176 & 3.375 \\
&&& individual-ATE & -44.8 (43.3) & 87.2 & 5.166 & 4.055 \\
&& \multirow{2}{*}{60} 
& cluster-ATE & 39.8 (24.2) & 91.2 & 2.435 & 2.083 \\
&&& individual-ATE & 36.9 (25.4) & 89.6 & 3.030 & 2.526 \\
&& \multirow{2}{*}{90} 
& cluster-ATE & -6.6 (16.1) & 92.8 & 1.616 & 1.492 \\
&&& individual-ATE & -7.0 (16.6) & 92.0 & 1.979 & 1.802 \\
\bottomrule
\end{tabular}
  \label{tab:arm_stratified}
\end{table}

\newpage
\clearpage
\section*{Appendix 10. Degeneracy of the uncalibrated posterior interval for the unadjusted model-robust estimator}

In the data application (Table 3 of the main paper), the uncalibrated posterior interval for the cluster-ATE under the unadjusted linear mixed model has zero width, with both endpoints coinciding with the point estimate. This appendix shows that the phenomenon is an exact algebraic property of the model-robust standardization estimator rather than a numerical artifact, and explains why it arises only for the cluster-ATE and only under the unadjusted working model.

Let $\pi_a = \pi^a(1-\pi)^{1-a}$, and let $\widehat{m}_{ai}$ denote the fitted conditional mean of the cluster-average outcome. For a generic weight $\omega_i$, the model-robust standardization formula is
\[
\widehat{\mu}_{\omega}^{\text{mr}}(a)
= \left(\sum_{i=1}^M \omega_i\right)^{-1} \sum_{i=1}^M \omega_i
\left\{ \frac{\mathbb{I}(A_i=a)\left(\overline{Y}_i - \widehat{m}_{ai}\right)}{\pi_a}
+ \widehat{m}_{ai} \right\}.
\]
Under the unadjusted linear mixed model, no covariates or cluster-size terms enter the working model, $\widehat{m}_{ai} \equiv \widehat{m}_a = \beta_0 + \beta_1 a$. Pulling $\widehat{m}_a$ out
of the sum gives
\begin{equation}
\widehat{\mu}_{\omega}^{\text{mr,unadj}}(a)
= \underbrace{\left(\sum_{i=1}^M \omega_i\right)^{-1}
\frac{1}{\pi_a}\sum_{i=1}^M \omega_i\,\mathbb{I}(A_i=a)\,\overline{Y}_i}_{\displaystyle \widehat{\mu}_{\omega}^{\text{np}}(a)}
\;+\; \widehat{m}_a\left(1 - \frac{q_a^{\omega}}{\pi_a}\right),
\label{eq:degeneracy}
\end{equation}
where
\[
q_a^{\omega} \;=\; \frac{\sum_{i=1}^M \omega_i\,\mathbb{I}(A_i=a)}{\sum_{i=1}^M \omega_i},
\]
and $\widehat{\mu}_{\omega}^{\text{np}}(a)$ is exactly the nonparametric unadjusted estimator defined in Section 4.2 of the main text, which is a function of the observed data alone. Equation~\eqref{eq:degeneracy} isolates the dependence of the estimator on the model parameters into the single term $\widehat{m}_a(1 - q_a^{\omega}/\pi_a)$. This term vanishes if and only if $q_a^{\omega} = \pi_a$, that is, when the realized weight in arm $a$ coincides exactly with the randomization probability. When this holds for both arms, $\widehat{\Delta}_{\omega}^{\text{mr,unadj}}$ reduces to the parameter-free quantity $\widehat{\mu}_{\omega}^{\text{np}}(1) - \widehat{\mu}_{\omega}^{\text{np}}(0)$, takes an identical value at every posterior draw of $\bm{\beta}$, and its posterior distribution is degenerate at a point.

For the cluster-ATE, $\omega_i = 1$ and $q_a^{C} = M_a/M$, the proportion of clusters assigned to arm $a$. In a CRT with $1{:}1$ randomization and an equal number of clusters per arm, $q_1^{C} = q_0^{C} = 1/2 = \pi_1 = \pi_0$, so the parameter-dependent term vanishes in both arms exactly. This is the situation in the PPACT trial, where $53$ of the $106$ clinics were randomized to each arm. To make the cancellation explicit in this case, expand the difference estimator directly:
\begin{equation*}
\begin{aligned}
\widehat{\Delta}_{C}^{\text{mr,unadj}}
& = \widehat{\mu}_C^{\text{mr,unadj}}(1) - \widehat{\mu}_C^{\text{mr,unadj}}(0)\\
&= \frac{1}{M}\sum_{i=1}^M \left\{ \frac{\mathbb{I}(A_i=1)\left(\overline{Y}_i - (\beta_0+\beta_1)\right)}{1/2} + (\beta_0 + \beta_1) \right\} - \frac{1}{M}\sum_{i=1}^M \left\{ \frac{\mathbb{I}(A_i=0)\left(\overline{Y}_i - \beta_0\right)}{1/2} + \beta_0 \right\} \\
&= \underbrace{\frac{1}{M}\sum_{i=1}^M \beta_1}_{\text{(I)}}
\;+\; \underbrace{\frac{1}{M}\sum_{i=1}^M \left( \frac{\mathbb{I}(A_i=1)\overline{Y}_i}{1/2} - \frac{\mathbb{I}(A_i=0)\overline{Y}_i}{1/2} \right)}_{\text{(II)}}
\;+\; \underbrace{\frac{1}{M}\sum_{i=1}^M \frac{-\mathbb{I}(A_i=1)\beta_1}{1/2}}_{\text{(III)}}.
\end{aligned}
\end{equation*}
Term (I) equals $\beta_1$ and term (II) is the nonparametric unadjusted moment estimator, which involves no model parameter. Because the two arms contain an equal number of clusters, $\sum_{i=1}^M \mathbb{I}(A_i=1) = M/2$, term (III) equals $\tfrac{1}{M}\cdot\tfrac{M}{2}\cdot\tfrac{-\beta_1}{1/2} = -\beta_1$ and cancels term (I) exactly, leaving
\begin{equation*}
\begin{aligned}
\widehat{\Delta}_{C}^{\text{mr,unadj}}
&= \underbrace{\beta_1}_{\text{(I)}} + \underbrace{\frac{1}{M}\sum_{i=1}^M \left( \frac{\mathbb{I}(A_i=1)\overline{Y}_i}{1/2} - \frac{\mathbb{I}(A_i=0)\overline{Y}_i}{1/2} \right)}_{\text{(II)}} + \underbrace{(-\beta_1)}_{\text{(III)}} \\
&= \underbrace{\frac{1}{M}\sum_{i=1}^M \left( \frac{\mathbb{I}(A_i=1)\overline{Y}_i}{1/2} - \frac{\mathbb{I}(A_i=0)\overline{Y}_i}{1/2} \right)}_{\text{(II)}},
\end{aligned}
\end{equation*}
a quantity determined entirely by the observed data. It therefore takes the same value at every posterior draw, and the resulting zero-width interval is the one reported in Table 3.

The cancellation is exact only because the cluster-ATE uses equal weights $\omega_i = 1$. For the individual-ATE, $\omega_i = N_i$ and $q_a^{I} = N_{(a)}/N$, where $N_{(a)} = \sum_{i=1}^M N_i\mathbb{I}(A_i=a)$ and $N = \sum_{i=1}^M N_i$, the proportion of individuals in arm $a$. Generally $q_a^{I} \neq \pi_a$. The parameter-dependent term is retained, and the interval has positive width, albeit a narrow one whenever the arms are close to balanced in total size. Under the covariate-adjusted LMM, the fitted conditional mean $\widehat{\mathbb{E}}(\overline{Y}_i \mid A_i, \bm{X}_i, N_i)$ contains covariate and cluster-size terms, so the parameters do not cancel and the estimator continues to depend on $\bm{\beta}$. Its posterior distribution is therefore non-degenerate, and the interval has positive width.

This is a symptom of the uncalibrated posterior inference, which propagates only parameter uncertainty and ignores the sampling variability of the data. When the estimator reduces to a data-only functional, the uncalibrated variance collapses to zero, even though the estimator clearly has non-trivial sampling variability. The calibrated variance estimator resolves this by adding the cluster-level resampling component, which recovers the variability of term (II), so that the calibrated interval has the expected positive width. This example thus provides a concrete illustration of why the calibrated variance estimator is necessary.

\label{lastpage}

\end{document}